\documentclass[acmsmall]{acmart}
\AtBeginDocument{\providecommand\BibTeX{{\normalfont B\kern-0.5em{\scshape i\kern-0.25em b}\kern-0.8em\TeX}}}

\setcopyright{acmcopyright}
\copyrightyear{2018}
\acmYear{2018}
\acmDOI{XXXXXXX.XXXXXXX}

\acmJournal{TOSEM}
\acmVolume{0}
\acmNumber{0}
\acmArticle{0}
\acmMonth{0}

\usepackage{soul}
\usepackage{url}

\usepackage[utf8]{inputenc}
\usepackage[T1]{fontenc}

\usepackage{balance}
\usepackage{esvect}
\usepackage{anyfontsize}
\usepackage[all]{nowidow}
\usepackage[normalem]{ulem}
\usepackage{xcolor}
\usepackage{colortbl}
\usepackage{enumitem}  \usepackage[switch]{lineno}

\usepackage{subcaption}

\usepackage{makecell}
\usepackage{multirow}

\usepackage{hyperref}
\usepackage{dblfloatfix}
\usepackage{float}

\usepackage[framemethod=TikZ]{mdframed}
\usepackage{framed}
\usepackage{tikz}
\usepackage{lipsum}

\usepackage{listings}
\usepackage[ruled,vlined,lined,commentsnumbered]{algorithm2e}
\usepackage{algorithmic}

 \definecolor{cocoabrown}{rgb}{0.82, 0.41, 0.12}
\definecolor{dimgray}{rgb}{0.41, 0.41, 0.41}
\definecolor{greenn}{rgb}{0.4, 0.69, 0.2}
\definecolor{darkred}{rgb}{0.75, 0.3,0.3}

\colorlet{rev}{cocoabrown}
\colorlet{rep}{red}
\colorlet{comment}{greenn}
\colorlet{revision}{blue}
\colorlet{revision2}{blue}

\newcommand\remove[1]{}

\newcolumntype{Y}{>{\centering\arraybackslash}X}
\newcolumntype{R}{>{\raggedleft\arraybackslash}X}

\definecolor{light-gray}{gray}{0.9}

\mdfsetup{skipabove=5pt,skipbelow=3pt}
\mdfdefinestyle{RQFrame}{linecolor=black,
	outerlinewidth=0.15pt,
	roundcorner=3pt,
	innertopmargin=2pt,
	innerbottommargin=2pt,
	innerrightmargin=4pt,
	innerleftmargin=4pt,
	backgroundcolor=light-gray,
	nobreak=true,
	}

\definecolor{javared}{rgb}{0.6,0,0} \definecolor{javagreen}{rgb}{0.25,0.5,0.35} \definecolor{javapurple}{rgb}{0.5,0,0.35} \definecolor{javadocblue}{rgb}{0.25,0.35,0.75} \definecolor{javagrey}{rgb}{0.46,0.45,0.48} 

\newcommand{\var}[1]{\mathit{#1}}
\newcommand{\fun}[1]{\mathit{#1}}

\lstdefinestyle{Alg}{
  basicstyle=\ttfamily\footnotesize,
  breaklines=true,
  tabsize=2,
  mathescape,
  numbers=left,
  xleftmargin=2.5em,
  xrightmargin=0.5em,
  frame=tb,
  framexleftmargin=2em,
  emph={Algorithm,Input,Output,for,each,do,if,else,Function,while,let,be,repeat,until,return,times,and,or,break,in,then,end,none},
  emphstyle={\textbf},
  escapechar=?,
  morecomment=[l][\color{javagreen}]{//},
  columns=flexible,
}

\lstdefinestyle{Alg2}{
  basicstyle=\ttfamily\footnotesize,
  tabsize=2,
  mathescape,
  numbers=left,
  xleftmargin=2.5em,
  xrightmargin=1em,
  frame=tb,
  framerule=0mm,
  framexleftmargin=2em,
  emph={Algorithm,Input,Output,for,each,do,if,else,Function,while,let,be,repeat,until,where,return,times,and,or,break,in,then,end,none},
  emphstyle={\textbf},
  escapechar=?,
  morecomment=[l][\color{javagreen}]{//},
  columns=flexible,
  breaklines=true,
  postbreak=\mbox{\hspace{3em}},
}

\hypersetup{
    colorlinks=true,
    linkcolor=blue,citecolor=blue,urlcolor=blue,}

\begin{document}

\title{Probabilistic Safe WCET Estimation for Weakly Hard Real-Time Systems at Design Stages}

\author{Jaekwon Lee}
\email{jaekwon.lee@uni.lu}
\affiliation{\institution{University of Luxembourg}
  \streetaddress{29 Avenue John F. Kennedy}
  \city{Luxembourg}
  \postcode{1859}
  \country{Luxembourg}
}
\affiliation{\institution{University of Ottawa}
  \streetaddress{800 King Edward Avenue}
  \city{Ottawa}
  \postcode{ON K1N 6N5}  
  \country{Canada}
}

\author{Seung Yeob Shin}
\email{seungyeob.shin@uni.lu}
\affiliation{\institution{University of Luxembourg}
  \streetaddress{29 Avenue John F. Kennedy}
  \city{Luxembourg}
  \postcode{1859}
  \country{Luxembourg}
}

\author{Lionel C. Briand}
\email{lionel.briand@uni.lu}
\affiliation{\institution{University of Luxembourg}
  \streetaddress{29 Avenue John F. Kennedy}
  \city{Luxembourg}
  \postcode{1859}  
  \country{Luxembourg}
}
\affiliation{\institution{University of Ottawa}
  \streetaddress{800 King Edward Avenue}
  \city{Ottawa}
  \postcode{ON K1N 6N5}  
  \country{Canada}
}

\author{Shiva Nejati}
\email{snejati@uottawa.ca}
\affiliation{\institution{University of Ottawa}
  \streetaddress{800 King Edward Avenue}
  \city{Ottawa}
  \postcode{ON K1N 6N5}  
  \country{Canada}
}

\renewcommand{\shortauthors}{J. Lee, S. Shin, L. Briand, S. Nejati}
 
\begin{abstract}
Weakly hard real-time systems can, to some degree, tolerate deadline misses, but their schedulability still needs to be analyzed to ensure their quality of service. 
Such analysis usually occurs at early design stages to provide implementation guidelines to engineers so that they can make better design decisions. 
Estimating worst-case execution times (WCET) is a key input to schedulability analysis. 
However, early on during system design, estimating WCET values is challenging and engineers usually determine them as plausible ranges based on their domain knowledge.
Our approach aims at finding restricted, safe WCET sub-ranges given a set of ranges initially estimated by experts in the context of weakly hard real-time systems.
To this end, we leverage (1)~multi-objective search aiming at maximizing the violation of weakly hard constraints in order to find worst-case scheduling scenarios and (2)~polynomial logistic regression to infer safe WCET ranges with a probabilistic interpretation.
We evaluated our approach by applying it to an industrial system in the satellite domain and several realistic synthetic systems. 
The results indicate that our approach significantly outperforms a baseline relying on random search without learning, and estimates safe WCET ranges
with a high degree of confidence in practical time (< 23h).
\end{abstract}



\maketitle

\section{Introduction}
\label{sec:intro}

Real-time systems are required to perform operations under time constraints, specifying execution deadlines~\cite{Cheng2003}. In real-world applications across many industry sectors, such as automotive and aerospace, real-time systems can often tolerate occasional deadline misses when their consequences are negligible with respect to achieving the system objectives and are not noticeable by users. The systems that are robust to occasional deadline misses are known as weakly hard real-time systems~\cite{Bernat2001}. Weakly hard deadline constraints specify the extent to which real-time tasks can tolerate deadline misses. For example, a control-loop task that computes throttle angles and sends throttle commands to an autonomous vehicle at a fixed rate (e.g., 20 Hz) can accept at most three deadline misses out of 20 task arrivals. When the task violates the weakly hard deadline constraint, e.g., four deadline misses within 20 task arrivals, it may result in the vehicle failing to arrive at the target destination on time or even colliding with objects.

While developing a weakly hard real-time system, estimating \emph{safe} worst-case execution times (WCETs) of real-time tasks is an important activity to ensure that the system meets its deadline constraints. Engineers deem tasks' WCETs to be safe when, under such specified execution times, task executions satisfy their (weakly hard) deadlines constraints; i.e., the tasks are schedulable~\cite{Bini2004}. In particular, safe WCET estimates are practically useful at early design stages when tasks' implementations are not yet completed. Such estimates provide development objectives to guide engineers in making appropriate design and implementation decisions and thus prevent deadline misses. For example, depending on the safe WCET estimated for a data-processing task, engineers may choose either an in-memory storage, a file system, or an external database system to store and access data.

At early design stages, engineers find it challenging to estimate safe WCETs and thus to guarantee that tasks always meet their deadlines~\cite{Gustafsson2009}. WCETs are determined based on a variety of factors such as task scheduling policies, task implementations, and hardware specifications. Regarding scheduling policies, advanced real-time operating systems (e.g., QNX Neutrino~\cite{QNXNeutrino}) applied in industry employ sophisticated scheduling policies to accommodate various systems' requirements in different domains, such as automotive and aerospace. For example, an adaptive partitioning scheduler (APS)~\cite{APS} developed by BlackBerry prevents unimportant tasks from monopolizing system resources (e.g., processing units) by using adaptive partitions. Such partitions separate tasks into virtual containers with their own resource-utilization budgets, which are adaptive depending on system performance. Due to the complexity of such scheduling policies, engineers face difficulties when applying existing WCET analysis techniques~\cite{Cucu2012,Santinelli2017} that are valid only when systems employ traditional scheduling policies, e.g., rate monotonic scheduling policy~\cite{Liu1973}. Furthermore, the problem of estimating safe WCETs becomes more challenging (i.e., computationally expensive) when real-time systems are constrained by weakly hard deadlines, which specify tolerable degrees for deadline misses. In addition, decisions regarding task implementations and hardware components are often not fully known at early design stages. Hence, engineers cannot determine exact WCET values ensuring that tasks are schedulable. Therefore, engineers usually resort to estimating WCET ranges that can ensure tasks are schedulable  with a high probability~\cite{Davis2019,Lee2022a}.

The problem of estimating WCET has been widely studied, relying mainly on measurements~\cite{Wenzel2005,Cucu2012,Santinelli2017} and static analysis~\cite{Ferdinand1998, Theiling2000, Mueller2000, Hardy2011}. Measurement-based approaches estimate WCETs by analyzing multiple executions on the target hardware or an accurate simulator using a set of worst-case inputs. In contrast, static analysis-based approaches estimate WCETs by investigating the longest path in source code and the cache hit ratio based on hardware specifications. There are approaches~\cite{Gustafsson2009, Altenbernd2016, Bonenfant2017} aiming at estimating WCETs at early stages of implementation. For example, \citet{Altenbernd2016} first create a timing model that predicts the execution times of machine instructions. Given source code to analyze, they then translate it to machine instructions captured in the timing model. The execution times of these instructions are used to approximate the source code's WCET.
In contrast to our work that aims at estimating safe WCET ranges, these prior approaches aim at estimating the WCET of a real-time task, without accounting for the task's schedulability (i.e., deadline constraints).
In addition, since these approaches rely on source code available only at implementation stages, they are not applicable at early design stages.
Recently, SAFE~\cite{Lee2022a} has been proposed to estimate with a probabilistic interpretation, at early design stages, safe WCET ranges that satisfy deadline constraints for real-time systems. SAFE utilizes task models instead of source code to simulate task executions and estimate safe WCET ranges using machine learning and meta-heuristic search. However, SAFE does not account for the specificities of weakly hard real-time systems accepting occasional deadline misses and advanced, sophisticated scheduling policies used in industry. Instead, SAFE targets real-time systems that do not tolerate any occurrence of a deadline miss and relies on a simple task model.
Hence, the problem addressed in our work is more complex than the problem tackled by SAFE.
Our work complements SAFE and extends it to probabilistically estimate safe WCET ranges for weakly hard real-time systems involving advanced industry scheduling policies.

\textbf{Contributions.} In this article, we propose SWEAK, a \underline{S}afe \underline{W}CET analysis method for w\underline{EAK}ly hard real-time systems. SWEAK searches for effective test cases that likely cause violations of weakly hard deadline constraints using a multi-objective search algorithm~\cite{Luke2013}. SWEAK then estimates safe WCET ranges with a probabilistic interpretation by using logistic regression~\cite{Hosmer2013}. SWEAK evaluates the schedulability of a set of real-time tasks by using an industrial scheduler APS that supports complex scheduling policies, accounting for multi-core platforms and adaptive partitions~\cite{Schaffer2011}. In a multidimensional WCET space defined by different tasks in a system, SWEAK identifies a \emph{safe WCET border} characterizing safe WCET ranges with a probability $p$ of violating weakly hard deadline constraints. Such a border allows engineers to investigate, for each task, suitable WCET values by analyzing trade-offs within the safe ranges.

We evaluated SWEAK with an industrial system from the satellite domain and several realistic synthetic systems that were created following guidelines provided by our industry partner, Blackberry. Experimental results show that SWEAK can efficiently and accurately estimate safe WCET ranges for various weakly hard real-time systems. Regarding the execution time of SWEAK, it takes at most 22.1h across a large number of synthetic systems, indicating that SWEAK is acceptable in practice as an offline analysis tool. All the details of our evaluation results are available online~\cite{Artifacts3}.

\textbf{Organization.} This article is organized as follows: Section~\ref{sec:motivation} motivates our work. 
Section~\ref{sec:problem} precisely defines the problem of estimating safe WCET ranges for weakly hard real-time systems.
Section~\ref{sec:approach} describes SWEAK. Section~\ref{sec:evaluation} empirically evaluates SWEAK. Section~\ref{sec:relatedworks} contrasts SWEAK against related work. Section~\ref{sec:conclusion} concludes this article. \section{Motivation}
\label{sec:motivation}

Our work is motivated by the practical needs identified in collaboration with our partner company, BlackBerry. They have developed a real-time operating system (RTOS), named QNX Neutrino~\cite{QNXNeutrino}, which satisfies the functional safety standard ISO-26262~\cite{ISO26262} with the highest automotive safety integrity level (ASIL-D). Due to the stringent assurance requirements, QNX Neutrino has been used in many safety-critical, real-time industries such as automotive and medical domains. 

\textbf{Adaptive partitioning scheduler (APS).}
QNX Neutrino employs a sophisticated scheduler named APS, which has been studied and applied in many systems~\cite{Bletsas2009,Massa2016,Abeni2020,Dasari2021,Dasari2022}, to support complex system requirements in managing real-time tasks. APS is based on a priority-driven preemptive scheduling policy, allocating tasks to processing cores based on the tasks' priorities for scheduling. The policy ensures that the highest priority task always has access to a processing core when required. APS also supports task partitions in which tasks are assigned. The time budgets of partitions, which impact task executions, are dynamically controlled depending on the system load. Such budget management not only scales up and down the budgets of partitions according to the tasks' demands, but also prevents tasks from monopolizing processors.
In addition, APS supports various scheduling policies, e.g., FIFO and Round-Robin, and multi-core platforms. These rich features of APS (and QNX Neutrino) have made it widely applicable in practice, but have also complicated schedulability analysis.

\textbf{Analysis needs.}
BlackBerry provides customers who develop real-time systems with an APS simulator, allowing them to design and evaluate real-time tasks running on QNX Neutrino in a realistic and scalable manner. In particular, the APS simulator emulates tasks' behaviors without requiring their implementations or hardware devices.
Hence, the simulator is applicable to analyze real-time tasks at early design stages.
However, the simulator cannot estimate safe WCET ranges at early stages, which is important to develop and assess real-time systems.

Many organizations~\cite{Akesson2020}, including BlackBerry customers, develop weakly hard real-time systems that can tolerate occasional deadline misses. However, existing WCET analysis techniques~\cite{Bernat2002, Altmeyer2008, Lee2022a} work on hard real-time systems with simpler scheduling policies than APS. In such contexts, using state-of-the-art techniques would therefore result in unnecessarily restricted WCET ranges compared to acceptable WCET ranges that would allow tasks to occasionally miss some deadlines. 
In practice, as early WCET estimates guide engineers to make appropriate implementation decisions and hardware resource choices, overly pessimistic WCET estimates may add unnecessary overhead for code optimization or over-provisioning of hardware resources.
To this end, BlackBerry is interested in developing an APS simulation-based solution to estimate safe WCET ranges for weakly hard real-time systems.

 \section{Problem definition}
\label{sec:problem}

This section introduces the notation we use in this article and our task model. The latter builds on our previous work~\cite{Lee2022a} and extends it with weakly hard deadline constraints, complex scheduling policies, and context switching times.
We then describe the problem of identifying safe WCET ranges that satisfy the deadline constraints with a certain level of confidence.

\textbf{Task model.}
We analyze a real-time system running $n$ tasks in parallel on a multi-core platform.
Each task $\tau_i$ ($1\leq i \leq n$) is identified as either periodic or aperiodic. 
A periodic task, which arrives at regular intervals, is characterized by the following temporal parameters: offset $O_i$, period $T_i$, WCET $C_i$, and relative deadline $D_i$. 
These parameters determine the arrival times of a periodic task and its absolute deadlines.
Specifically, the $k$th arrival of a periodic task $\tau_i$, denoted by $a_{i,k}$, is $O_i$ + $(k-1)$ $\times$ $T_i$. 
A periodic task arrived at $a_{i,k}$ is supposed to complete its execution, even in the worst case $C_i$, before the absolute deadline determined by $a_{i,k}$ + $D_i$.

An aperiodic task has irregular arrival times as it is activated by external stimuli. In general, there is no limit on the arrival times of an aperiodic task.
However, in real-time analysis, we typically specify a minimum inter-arrival time and maximum inter-arrival time to characterize irregular arrivals.
For an aperiodic task $\tau_j$, we define [$T^{min}_j$, $T^{max}_j$], indicating the minimum and maximum time intervals between two consecutive arrivals of $\tau_j$. 
Thus, an arrival time $a_{j,k}$ is determined by the $k{-}1$th arrival time of $\tau_j$ and its minimum and maximum arrival times as follows: [$T^{min}_{j}, T^{max}_{j}$] for $k = 1$ and [$a_{j,{k-1}}+T^{min}_{j}$, $a_{j,{k-1}}+T^{max}_{j}$] for $k > 1$.
We note that, in real-time analysis, sporadic tasks can be separately defined as they have irregular arrivals and hard deadlines~\cite{Liu2000}. However, in our task model, we do not introduce new notations for sporadic tasks because the deadline and period concepts defined above are sufficient to characterize them.

Under a priority-driven preemptive scheduling policy~\cite{Bini2004}, each task $\tau_i$ has its priority, denoted by $P_i$, which determines its order of execution. A task $\tau_i$ can preempt another task $\tau_j$ when the priority value of $\tau_j$ is less than the value of $\tau_i$, i.e., $P_j < P_i$.

\textbf{WCET ranges.}
As discussed earlier, engineers have a hard time estimating exact WCET values at early stages of development because there are many uncertain factors, for example related to hardware configuration, input data, and source code. In this study, we assume that engineers can provide WCET ranges instead of a single value. We denote by [$C^{min}_i$, $C^{max}_i$] the minimum and maximum WCET values of a task $\tau_i$.

\textbf{Weakly hard deadline constraints.}
A deadline $D_i$ is the time constraint of a task $\tau_i$. Our task model requires the deadline to be greater than or equal to $C_i$~\cite{Chen2018a}, i.e., $D_i \geq C_i$. For a task $\tau_j$ with a WCET range, the deadline value must be greater than or equal to the maximum WCET value $C^{max}_j$.
A task constrained by a hard deadline must meet its deadline in all executions of the task. However, when a task is subject to a weakly hard deadline constraint, which is also called a soft deadline, the task can occasionally miss its deadline. We adopt the formal definition of weakly hard deadline constraints introduced by \citet{Bernat2001}, i.e., ($m, K$)-constraint, which has been commonly used in weakly hard real-time studies~\cite{Alenawy2005,Gettings2015,Pazzaglia2021}.
The ($m_i, K_i$)-constraint of a task $\tau_i$ specifies that, within a time window $K_i$ (i.e., the number of consecutive arrivals of $\tau_i$), $m_i$ consecutive task arrivals are allowed to miss the deadline $D_i$. 
For instance, ($m_i, K_i$) $=$ ($2, 5$) specifies that two consecutive deadline misses of a task $\tau_i$ are acceptable within five consecutive arrivals of $\tau_i$. When $\tau_i$ is subject to a hard deadline constraint, $m_i$ $=$ 0.
We note that, in weakly hard real-time systems, analyzing consecutively missed deadlines is important as the consequence of a deadline miss at a task arrival is propagated to the next arrivals. However, non-consecutive deadline misses may not be noticeable to users as a task execution may complete before its deadline even after a deadline miss occurred at the previous arrival.

\begin{figure}[t]
\begin{center}
    \includegraphics[width=0.65\columnwidth]{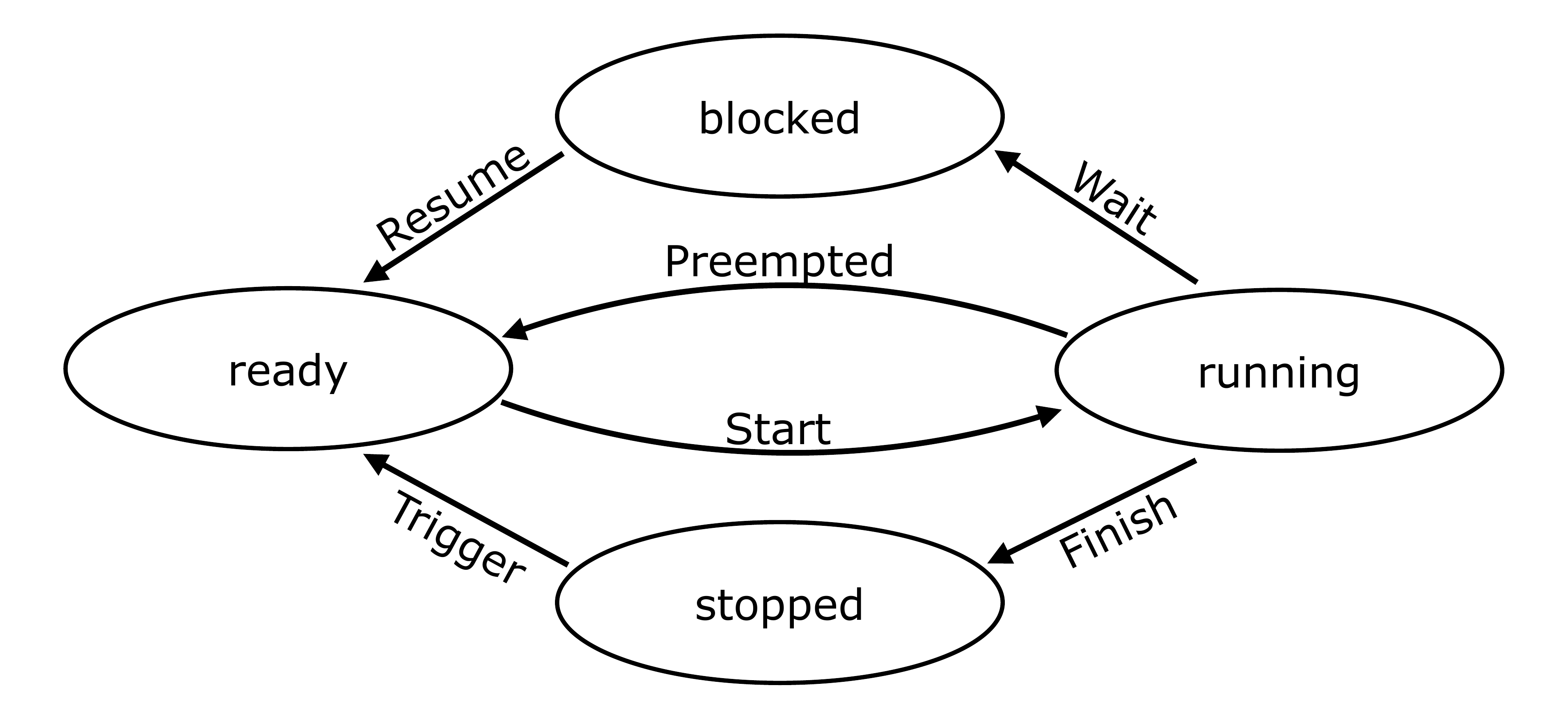}
    \caption{
    A task state transition model.
    }
    \label{fig:taskstate}
\end{center}
\end{figure}

\textbf{Context switching times.}
In addition to the task model described above, our study accounts for context switching time, which is the time required for a task to change state during scheduling.
Fig.~\ref{fig:taskstate} shows the task state transition model used in APS. 
According to this model, a task is \emph{ready} when it is prepared to execute on a processing core. APS executes the task by assigning it to an idle processing core and sets its state to \emph{running}.
A \emph{running} task can be \emph{blocked} when it requires resources used by other tasks or \emph{stopped} when it finishes its execution. APS can also preempt a \emph{running} task when a higher priority task is \emph{ready} or the partition budget that the task belongs to runs out. 
Each state transition requires time for exchanging data between memories and scheduling overheads. To account for such time, we define three types of context switching times. 
Start-up time, denoted by $\lambda_s$, is the time required to change the state of a task from \emph{ready} to \emph{running}. 
Exit time, denoted by $\lambda_x$, is the time required to change the state of a task from \emph{running} to other states, \emph{ready} or \emph{blocked}, or to finalize its execution.
Moreover, since APS deals with multi-core platforms, tasks may need to be assigned to different processing cores when their states are changed from \emph{ready} to \emph{running} depending on the availability of processing cores.
Inter-processor interrupt (IPI) time, denoted by $\lambda_p$, is required to transfer a task execution from one core to another in a multi-core platform. 
These context switching times are affected by hardware performance as well as scheduling overhead in a scheduler, which are uncertain. 
Hence, we specify them as ranges instead of single values.
For example, the start-up time $\lambda_s$ can be any value in the range [$0.012\mathit{ms}, 0.022\mathit{ms}$]. Note that representing these context switching times as ranges is consistent with the APS simulator provided by BlackBerry QNX.

\begin{figure}[t]
	\begin{center}
\begin{subfigure}[t]{\columnwidth}\centering
		\includegraphics[width=0.75\columnwidth]{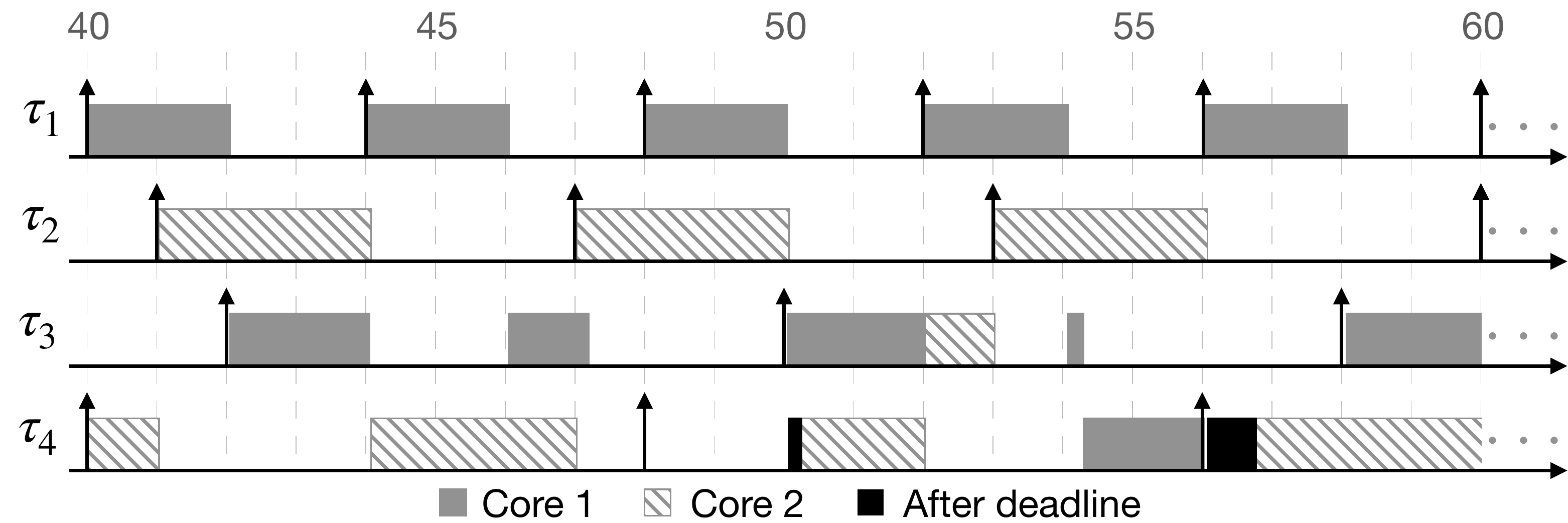}
		\label{fig:scheduling multi}
	\end{subfigure}
    \caption{
    A schedule scenario example that describes task executions of four tasks, $\tau_1$, $\tau_2$, $\tau_3$, and $\tau_4$, running on a two-core platform. This scenario includes context switching times.
    }
    \label{fig:scheduling}
	\end{center}
\end{figure}

\textbf{Schedule scenario}
Based on a scheduling result, we define a schedule scenario, denoted by $S$, describing the executions of all tasks in a system in terms of their start and end times. Specifically, we formulate a schedule scenario as a list of tuples $(\tau_i, a_{i,k}, e_{i,k})$, where $a_{i,k}$ and $e_{i,k}$ are, respectively, the arrival time and the end (or completion) time of the $k$th arrival of the task $\tau_i$.

For example, let $\Gamma$ be a set of $n$ tasks to be scheduled by a real-time scheduler. A scheduler then dynamically schedules the executions of tasks in $\Gamma$ over the scheduling period $\mathbb{T} = [0,\mathbf{t}]$ according to a scheduling policy (e.g., APS scheduling policy~\cite{APS}).
Fig.~\ref{fig:scheduling} describes a schedule result running four tasks $\tau_1$, $\tau_2$, $\tau_3$, and $\tau_4$ on a two-core platform.
The periodic task $\tau_1$ is characterized by: $O_1 = 0$, $T_1 = 4$, and $C_1^{min}$ $=$ $C_1^{max}$ $=$ 2. 
The aperiodic task $\tau_2$ is characterized by: [$T^{min}_2, T^{max}_2$] = [6, 12] and $C_2^{min}$ $=$ $C_2^{max}$ $=$ 3. The aperiodic task $\tau_3$ is characterized by: [$T^{min}_3, T^{max}_3$] $=$ [$8, 14$], and [$C_3^{min},C_3^{max}$] $=$ [$2, 3$].
The periodic task $\tau_4$ is characterized by: $O_4$ = 0, $T_4$ = 8, and [$C_4^{min}, C_4^{max}$] $=$ [$2, 4$]. 
All the task executions should be finished before the next task period or next minimum task arrival (i.e., $D_1$ $=$ $T_1$ $=$ $4$, $D_2$ $=$ $T_2^{min}$ $=$ $6$, $D_3$ $=$ $T_3^{min}$ $=$ $8$, and $D_4$ $=$ $T_4$ $=$ $8$).
The tasks' priorities are $P_1 > P_2 > P_3 > P_4$, which means that $\tau_1$ can preempt the processing cores at any time. All the context switching times in the example are 0.025, i.e., $\lambda_s = \lambda_x = \lambda_p = 0.025$.

As shown in Fig.~\ref{fig:scheduling}, during the scheduling period $\mathbb{T} = [40, 60]$, the schedule scenario $S$ is 
\{$(\tau_1, 40, 42.05)$, $\cdots$, 
$(\tau_2, 41, 44.075)$,  $\cdots$, 
$(\tau_3, 42, 47.2)$, $\cdots$, 
$(\tau_4, 40, 50.3)$, $\cdots$, $(\tau_4, 56, 60.825)$\}.
Due to randomness in task execution times, aperiodic task arrivals, and context switching times, a schedule scenario (i.e., scheduling result) can differ in each scheduler run.

\textbf{Schedulability.}
We analyze the schedulability of a given schedule scenario $S$ by checking the ($m_i, K_i$)-constraint of each task in $S$. If a schedule scenario shows a violation of any ($m_i, K_i$)-constraint, the schedule scenario is not schedulable.
For example, the schedule scenario in Fig.~\ref{fig:scheduling} has two deadline misses for task $\tau_4$ at the first and second arrivals. 
If the deadline constraint ($m_4, K_4$) of $\tau_4$ is ($1, 4$), the schedule scenario $S$ is not schedulable as the deadline constraint of $\tau_4$ only allows one deadline miss in four consecutive arrivals. 
However, the scenario $S$  becomes schedulable when ($m_4, K_4$) $=$ ($2, 4$), accepting two consecutive deadline misses. Note that a set $\Gamma$ of tasks is schedulable when every schedule scenario $S$ of $\Gamma$ is schedulable with respect to the tasks' ($m_i, K_i$)-constraints.

\textbf{Problem.} 
The effective design and assessment of real-time systems rely on the accurate evaluation of the task parameters. 
Among these parameters, WCET values are estimated as ranges, which is inevitable given the high uncertainty at early stages of development.
Upper WCET bounds are the worst-case WCET values that are most likely to have deadline misses, since larger WCET values increase the probability of deadline constraint violations. Lower WCET bounds are tasks' best-case WCET values but are harder to implement in practice.

Our work aims at determining the maximum upper bounds that allow tasks to be schedulable, under weakly hard deadline constraints, at a certain level of probability of violating deadline constraints.
Practitioners can use these upper bounds as an objective when implementing the tasks.
Specifically, for every task $\tau_i \in \Gamma$ to be analyzed, our approach computes a new upper bound value for the WCET range of $\tau_i$ (denoted by $C^\mathit{max*}_i$) by restricting it from $C^\mathit{max}_i$ to $C^\mathit{max*}_i$ such that, at a certain level of confidence, deadline constraint violations should not occur.
For instance, as we aforementioned, the schedule scenario in Fig.~\ref{fig:scheduling} is not schedulable under (1,4)-constraint for the task $\tau_4$. However, the tasks become schedulable when restricting the maximum WCET of $\tau_4$ from $C^{max}_4 = 4$ to $C^{max*}_4 = 3$ or WCET of $\tau_3$ from $C^{max}_3 = 3$ to $C^{max*}_3 = 2$.

 \section{Approach}
\label{sec:approach}

\begin{figure*}[t]
\begin{center}
\includegraphics[width=\textwidth]{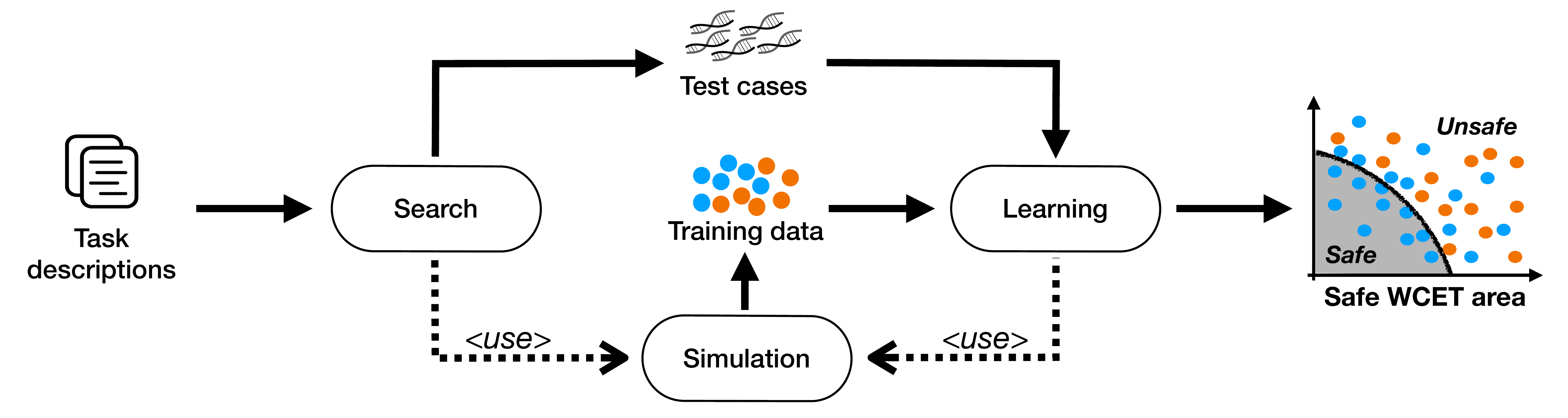}
\caption{An overview of our \underline{S}afe \underline{W}CET analysis method for w\underline{EAK}ly hard real-time system (SWEAK).}
\label{work3:fig:overview}
\end{center}
\end{figure*}

Fig.~\ref{work3:fig:overview} shows an overview of SWEAK, our \underline{S}afe \underline{W}orst-case execution time (WCET) analysis method for w\underline{EAK}ly hard real-time system.
Given task descriptions, SWEAK first finds test cases, which consist of sequences of task arrivals and context switching times, using meta-heuristic search to maximize the magnitude and consecutiveness degree of deadline misses
(Section~\ref{subsec:search}).
During search, SWEAK uses an industrial scheduling simulator, i.e., APSSimulator that simulates the APS policy, to evaluate the schedulability of test cases and produce a training dataset (Section~\ref{subsec:simulation}). 
Using the training data, SWEAK then builds a logistic regression model to distinguish between the \textit{safe} and \textit{unsafe} areas in the WCET space with respect to satisfying and violating weakly hard deadline constraints (Section~\ref{subsec:learning}). 
The model estimates, with a probabilistic interpretation, safe WCET ranges under which tasks are likely to be schedulable.
To improve the accuracy of the estimation model, SWEAK then augments the training dataset by running simulations with the test cases obtained from the search. In the next sections, we describe each step of SWEAK in detail.

We note that SWEAK is based on our past work (i.e., SAFE~\cite{Lee2022a}), which estimates probabilistic safe WCET ranges for (hard) real-time systems using a single-objective search algorithm, single-queue multi-core scheduling policy simulation, and logistic regression. 
In contrast to SAFE, which targets real-time systems with relatively simple scheduling policies, SWEAK aims at estimating probabilistic safe WCET ranges for weakly hard real-time systems involving advanced industrial scheduling policies.
Recall from Section~\ref{sec:problem} that the ($m_i, K_i$)-constraint (i.e., weakly hard constraint) of a task $\tau_i$ specifies that, within a time window $K_i$, $m_i$ consecutive task arrivals are allowed to miss the deadline of $\tau_i$. Such a weakly hard deadline constraint is more complex than a hard deadline constraint that does not allow any occurrence of a deadline miss.
Analyzing weakly hard real-time systems, therefore, requires different techniques, compared to SAFE, to track and record the tasks that have missed their deadlines and the frequency of such occurrences.
Regarding the underlying techniques of the search step (Section~\ref{subsec:search}) of SWEAK, it employs a multi-objective search algorithm to generate test cases that are likely to violate weakly hard deadline constraints and maximize the magnitude of deadline misses.
During search, SWEAK further accounts for the context switching times, i.e., start-up, exit, and IPI times, which have an impact on scheduling results (described in Section~\ref{sec:problem}). In contrast, SAFE does not consider these time aspects.
For the simulation step (Section~\ref{subsec:simulation}), SWEAK uses APSSimulator that simulates the APS policy.
Regarding the learning step (Section~\ref{subsec:learning}), SWEAK also opts to use logistic regression, similar to the learning step of SAFE, to provide a probabilistic interpretation for safe WCET ranges.
Hence, we adapt the learning step of SAFE to develop the learning step of SWEAK, which accounts for weakly hard deadline constraints and integrates with APSSimulator.

\subsection{Searching for effective test cases}
\label{subsec:search}

The search step of SWEAK aims to generate test cases that likely violate weakly hard deadline constraints and maximize the magnitude of deadline misses.
We apply a multi-objective search algorithm for finding test cases guided by the following two objectives: (1)~maximizing the magnitude of deadline misses and (2)~maximizing the consecutiveness degree of deadline misses. 
These two objectives enable the search step to generate test cases that cause larger deadline misses (in terms of distances between tasks' completion times and their deadlines) and more consecutive deadline misses.
To evaluate the test cases with respect to the objectives through simulations, SWEAK applies multiple sets of WCET values, that are randomly sampled within their specified ranges, since WCET values can lead to different schedule results with the same test case.
We describe our search-based approach by defining the solution representation, the fitness functions, and the computational search algorithm, as recommended in the checklists for search-based software engineering research~\cite{Ralph2020}.

\textbf{Representation.}
A feasible solution represents a test case for checking the schedulability of a set of tasks defined in the input task descriptions. 
Given a set $\Gamma$ of tasks to be scheduled, a solution $I$ consists of two parts: context switching times and sequences of task arrivals for all tasks in $\Gamma$. 
The context switching times are three scalar values, i.e., start-up $\lambda_s$, exit $\lambda_x$, and IPI $\lambda_p$ times, each of which is selected within their valid ranges (see Section~\ref{sec:problem}). 
The sequences of task arrivals are denoted by a set $A$ of tuples $(\tau_i, a_{i,k})$, where $\tau_i \in \Gamma$ and $a_{i,k}$ is the $k$th arrival time of $\tau_i$.
The number of arrivals of $\tau_i$ is restricted by the scheduling time period $\mathbb{T}=[0, \mathbf{t}]$. 
For example, if a task $\tau_i$ is periodic and its offset $O_i=0$, the number of $\tau_i$ arrivals is $\mathbf{t}/T_i$ where $T_i$ denotes the period of $\tau_i$ (see Section~\ref{sec:problem}). 
In the case of aperiodic tasks, the number of arrivals varies with changing inter-arrival times (see Section~\ref{sec:problem}).
Therefore, the size of $I$ varies across different solutions along with the size of $A$.

\textbf{Fitness.} To evaluate the fitness of each solution, we define two objective functions, which quantify the magnitude of deadline misses and the consecutiveness degree of deadline misses.
These objective functions compute fitness values using multiple simulations to account for uncertainty in WCETs. 
Specifically, given a solution $I$ for a set $\Gamma$ of tasks, SWEAK runs APSSimulator $ns$ times with WCET values for the tasks in $\Gamma$ that are randomly selected from their WCET ranges and thus obtains schedule scenarios $\mathbf{S}$ $=$ \{$S_1$, $S_2$, $\cdots$, $S_{ns}$\} (see Section~\ref{subsec:simulation}). 
Given the scenarios, SWEAK calculates the fitness values for a solution $I$ using the fitness functions described below.

\textit{Fitness for the magnitude of deadline misses.} 
We denote by $\mathit{fd}(I,\Gamma^\delta,ns)$ a fitness function that quantifies the magnitude of deadline misses regarding a solution $I$, a set $\Gamma^\delta \subseteq \Gamma$ of target tasks, and $ns$ simulations. 
We note that SWEAK provides the capability of selecting target tasks $\Gamma^\delta$ as practitioners often need to focus on a subset of critical tasks.
The function $\mathit{fd}(I,\Gamma^\delta,ns)$ calculates a fitness value using a distance function $\mathit{dist}(\tau_i,k)$ defined as follows:
\begin{equation*}
\mathit{dist}(\tau_i,k) = e_{i,k} - a_{i,k} + D_i
\end{equation*}
This function computes the distance between the end time and the deadline of the $k$th arrival of task $\tau_i$ in a schedule scenario (see notation in Section~\ref{sec:problem}). 
If an arrival $a_{i,k}$ misses its absolute deadline $a_{i,k} + D_i$, the value of $\mathit{dist}(\tau_i,k)$ is larger than 0. 
As a larger distance value leads to a higher probability of deadline misses, SWEAK finds the maximum distance among all the arrivals for each scenario by using the fitness function $fd(I, \Gamma^\delta, \mathit{ns})$ defined as follows:
\begin{equation*}
\mathit{fd}(I,\Gamma^\delta,ns) = \sum_{h =1}^{ns}\max_{\tau_i \in \Gamma^\delta,~ k \in [1,\mathit{lk}(\tau_i)]}\mathit{dist}_h(\tau_i, k) / {ns}
\end{equation*}
where $\mathit{lk}(\tau_i)$ is the number of $\tau_i$ arrivals in $I$. 
We denote by $\mathit{dist}_h({\tau_i, k})$ the distance function for each schedule scenario $S_h \in \mathbf{S}$.
SWEAK aims to maximize the fitness value computed by $fd(I, \Gamma^\delta, \mathit{ns})$.

\begin{figure}[t]
\begin{center}
\includegraphics[width=0.6\columnwidth]{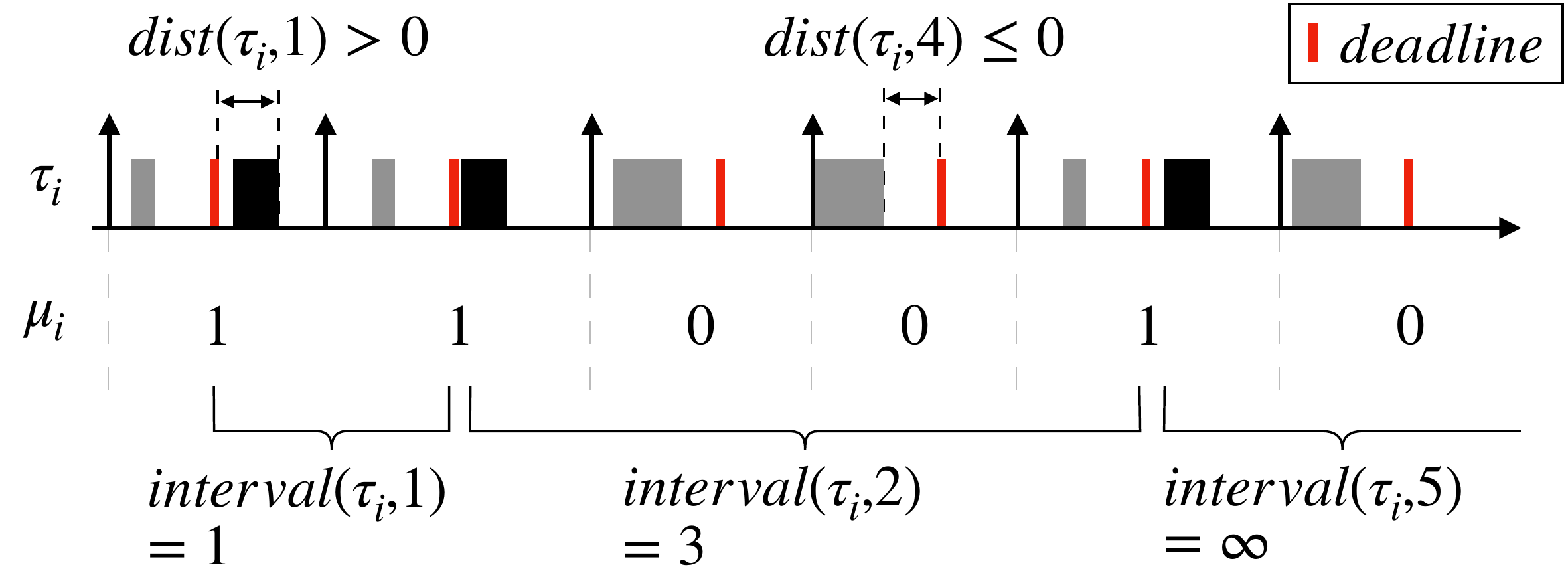}
\caption{An example of a $\mu$-pattern for a task $\tau_i$.}
\label{fig:interval}
\end{center}
\end{figure}

\textit{Fitness for the consecutiveness degree of deadline misses.}
We denote by $\mathit{fc}(I,\Gamma^\delta,ns)$ the fitness function that quantifies the consecutiveness degree of deadline misses regarding a solution $I$, a set $\Gamma^\delta \subseteq \Gamma$ of target tasks, and $ns$ simulations. 
To compute the consecutiveness degree of deadline misses, SWEAK converts a schedule scenario into $\mu$-patterns~\cite{Bernat2001} by checking whether task arrivals in a schedule scenario meet their deadlines or not.
Specifically, given a schedule scenario $S$, a $\mu$-pattern $\mu_i$ for a task $\tau_i$ is a sequence of $(\mu_i(1), \ldots, \mu_i(k), \ldots, \mu_i(\mathit{lk}(\tau_i)))$ where $k$ is the $k$th arrival of $\tau_i$, $\mathit{lk}(\tau_i)$ is the number of $\tau_i$ arrivals in $S$, and $\mu_i(k)$ is defined as follows:
\begin{equation*}
\mathit{\mu_i}(k) =
    \begin{cases}
    1 & , dist(\tau_i, k) > 0 \\
    0 & , otherwise\\
    \end{cases}
\end{equation*}
Fig.~\ref{fig:interval} shows an example converting task arrivals of $\tau_i$ into a $\mu$-pattern $\mu_i$. The task $\tau_i$ has three deadline misses at the first, second, and fifth arrivals, resulting in $\mu_i$ to be equal to (1,1,0,0,1,0).

Based on a $\mu$-pattern, we calculate the interval, denoted by $\mathit{interval}(\tau_i, k)$, between the $k$th and the $k^\prime$th arrivals of a task $\tau_i$, where $k^\prime > k$, the $k$th and $k^\prime$th arrivals miss their deadlines, and all arrivals between the $k$th and $k^\prime$th arrivals meet their deadlines.
For example, in Fig.~\ref{fig:interval}, $\mathit{interval}(\tau_i, 2)$ is equal to 3 because, after the 2nd arrival, the next deadline miss occurs at the 5th arrival; hence, $\mathit{interval}(\tau_i, 2) = 5-2 = 3$.
Note that the $\mathit{interval}(\tau_i, 5)$ in Fig.~\ref{fig:interval} is defined as $\infty$ because, after the 5th arrival of $\tau_i$, the next deadline miss is unknown. When $\mu_i(k) = 0$, we define $\mathit{interval}(\tau_i, k) = 0$.

Given the function $\mathit{interval}(\tau_i, k)$ and a $\mu$-pattern $\mu_i$, 
we denote by $\mathit{consec}(\tau_i,k)$ the consecutiveness degree of deadline misses regarding the $k$th arrival of a task $\tau_i$. The function $\mathit{consec}(\tau_i,k)$ is defined as follows:
\begin{equation*}
\mathit{consec}(\tau_i,k) =
    \begin{cases}
    10^\frac{1}{\mathit{interval}(\tau_i, k)} & , \mu_i(k)=1 \\ 0 & , \mu_i(k)=0\\
    \end{cases}
\end{equation*}
To reward small intervals and penalize large intervals between consecutive deadline misses, we invert $\mathit{interval}(\tau_i,k)$ and use an exponential function as shown in the $\mathit{consec}(\tau_i, k)$ definition. Hence, a consecutiveness degree $\mathit{consec}(\tau_i, k)$ exponentially decreases with the increasing value of $\mathit{interval}(\tau_i,k)$.
For example, given the $\mu$-pattern $\mu_i$ in Fig.~\ref{fig:interval}, the value of $\mathit{consec}(\tau_i, k)$ decreases when the value of $\mathit{interval}(\tau_i,k)$ increases, i.e., 
$\mathit{consec}(\tau_i, 1)$ = $\sqrt[1]{10}$ = 10,
$\mathit{consec}(\tau_i, 2)$ = $\sqrt[3]{10}$ = 2.15, and
$\mathit{consec}(\tau_i, 5)$ = $\sqrt[\infty]{10}$ = 1, where $1/\infty$ $=$ $0$.

To compute the fitness function $fc(I,\Gamma^\delta,ns)$, SWEAK runs APSSimulator $ns$ times for $I$ and obtains $ns$ schedule scenarios $S_1, S_2, \ldots, S_{ns}$. 
For each schedule scenario $S_h$, we denote by $\mathit{consec}_h({\tau_i, k})$ the consecutiveness degree of deadline misses regarding the $k$th arrival of a task $\tau_i$ observed in each schedule scenario $S_h$.
SWEAK aims to maximize the $\mathit{fc}(I,\Gamma^\delta,ns)$ fitness defined as follows:
\begin{equation*}
\mathit{fc}(I,\Gamma^\delta,ns) =
    \sum_{h =1}^{ns}{\Big(
        \max_{\tau_i \in \Gamma^\delta}{
            \sum_{k=1}^{\mathit{lk}(\tau_i)}\mathit{consec}_h(\tau_i, k)
        \Big)}
    }\big/{ns}
\end{equation*}

\begingroup
\begin{algorithm}[!t]
\parbox{0.95\columnwidth}{
	\caption{An algorithm for searching test cases, aiming at maximizing (1)~the magnitude of deadline misses and (2)~the consecutiveness degree of deadline misses, based on NSGA-II.}
 \label{alg:search}
}
\begin{lstlisting}[style=Alg2]
Input $\Gamma$: a set of tasks
Input $\var{ns}$: number of samples
Input $\var{np}$: population size
Input $\var{pc}$: crossover probability
Input $\var{pm}$: mutation probability
Output $\mathbf{P}_{\alpha}$: population of test cases

$\mathbf{P}_{\alpha} \leftarrow \{\}$
$\mathbf{P} \leftarrow \fun{CreatePopulation}(\Gamma, np)$

repeat
	?\vrule?	// calculating fitness for $\color{javagreen}{\mathbf{P}}$
	?\vrule?	for each $I \in \mathbf{P}$ do
	?\vrule?		?\vrule?	$\mathbf{W} \leftarrow \fun{SampleWCET}(ns)$
	?\vrule?		?\vrule?	$\mathbf{S} \leftarrow \fun{RunSimulation}(I, \mathbf{W})$
	?\vrule?		?\vrule?	$\mathit{fd}(I,\Gamma^\delta,ns) = \sum_{h =1}^{ns}\max_{\tau_i \in \Gamma^\delta,~ k \in [1,\mathit{lk}(\tau_i)]}\mathit{dist}_h(\tau_i, k) / {ns}$
	?\vrule?		?\vrule?	$\mathit{fc}(I,\Gamma^\delta,ns) =
    \sum_{h =1}^{ns}{\Big(
        \max_{\tau_i \in \Gamma^\delta}{
            \sum_{k=1}^{\mathit{lk}(\tau_i)}\mathit{consec}_h(\tau_i, k)
        \Big)}
    }\big/{ns}$
	?\vrule?	end for
	?\vrule?
	?\vrule?	// update archive
	?\vrule?	$\mathbf{P}_\alpha \leftarrow \mathbf{P_\alpha} \cup \mathbf{P}$
	?\vrule?	$\fun{ComputeFrontRanks}(\mathbf{P}_\alpha)$
	?\vrule?	$\fun{ComputeSparsities}(\mathbf{P}_\alpha)$
	?\vrule?	$\mathbf{P}_\alpha \leftarrow \fun{SelectArchive}(\mathbf{P}_\alpha, np)$
	?\vrule?	$BestFront \leftarrow \fun{ParetoFront}(\mathbf{P}_\alpha)$
	?\vrule?
	?\vrule?	//creating a new population
	?\vrule?	$\mathbf{P} \leftarrow \fun{Breed}(\mathbf{P}_\alpha, np, pc, pm)$
until we have run out of time or $BestFront$ is the ideal Pareto front
return $\mathbf{P}_{\alpha}$
\end{lstlisting}
\end{algorithm}
\endgroup

\textbf{Computational search.} SWEAK employs the NSGA-II algorithm~\cite{Luke2013} as shown in Algorithm~\ref{alg:search}. 
It generates an initial population $\mathbf{P}$ (line 9) and iterates to evolve the population until finding the ideal Pareto front or exhausting the execution budget (lines 11--29).
At each iteration, the algorithm first evaluates individuals in $\mathbf{P}$ with the fitness functions defined above (lines 13--18) and adds them to the archive $\mathbf{P}_{\alpha}$ (line 21).
The algorithm then calculates Pareto front rankings and sparsities of the solutions in the archive $\mathbf{P}_{\alpha}$ using the fitness values (lines 22-23). The calculation is used for determining the $\mathit{np}$ individuals to be kept in the archive and the best Pareto front (lines 24--25).
Based on the archive, the algorithm breeds a new population $\mathbf{P}$ to produce the next generation's population using the following genetic operators:
(1)~\emph{Selection} chooses candidate solutions as parents using a tournament selection technique, with the tournament size equal to two, which is the most common setting~\cite{Gendreau2010}. 
(2)~\emph{Crossover} creates offspring from the selected parents using a modified version of the one-point crossover. 
(3)~\emph{Mutation} changes the offspring according to a mutation rate and strategy.
After completing the evolution process, the algorithm returns the latest archive $\mathbf{P}_{\alpha}$ that contains the best found Pareto front. We describe our crossover and mutation approaches in detail.

\emph{Crossover.} 
A crossover operator produces offspring from two parent solutions by inheriting their characteristics. 
Our crossover operator, named SWEAKCrossover, modifies the standard one-point crossover operator~\cite{Luke2013} that selects a random crossover point among all genes and swaps them between parent solutions based on the crossover point. 
However, in our context, as the size of two parents can differ, such random selection may produce invalid offspring. 
To prevent it, SWEAKCrossover selects a crossover point among the context switching times, i.e., $\lambda_s$, $\lambda_x$, and $\lambda_p$, or the first arrivals of the aperiodic tasks in $\Gamma$. 
As the size of $\Gamma$ and context switching times are fixed for all solutions, SWEAKCrossover can crossover two solutions with different sizes.

\begin{figure*}[t]
\begin{center}
\includegraphics[width=\textwidth]{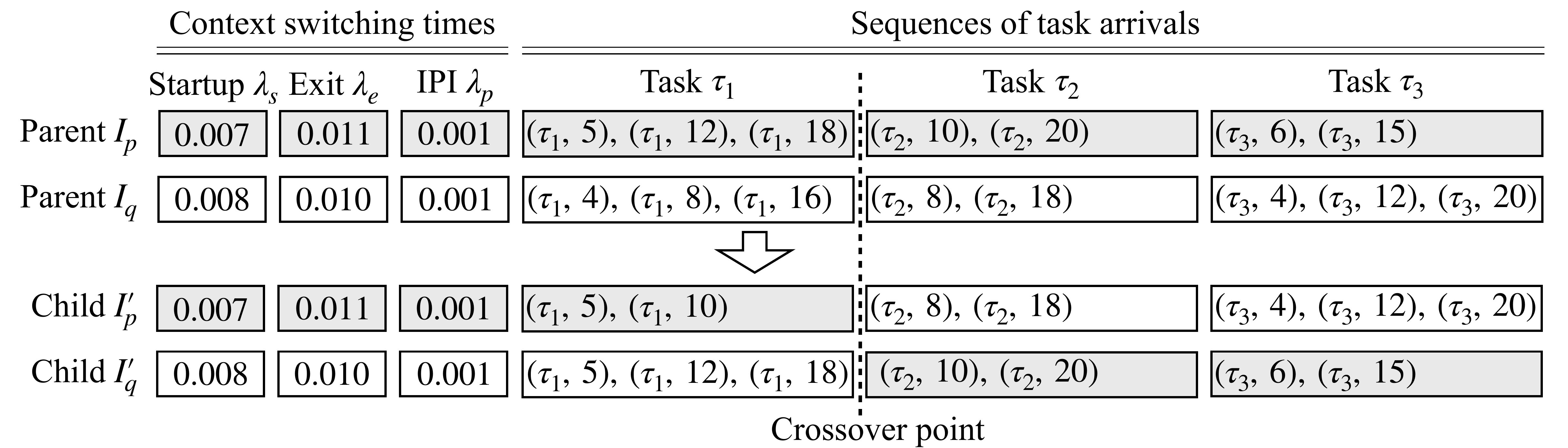}
\caption{An example of SWEAK's crossover operation. It swaps all context switching times and all task arrivals of task $\tau_1$ between two parent solutions $I_p$ and $I_q$ to produce offspring $I^\prime_p$ and $I^\prime_q$.}
\label{fig:crossover}
\end{center}
\end{figure*}

Fig.~\ref{fig:crossover} shows an example operation of SWEAKCrossover using a system with three aperiodic tasks, $\tau_1$, $\tau_2$, and $\tau_3$. 
Let two parent solutions $I_p$ and $I_q$ be as follows: $I_p$ $=$ ($0.007, 0.011, 0.001, (\tau_1,5)$, $\ldots$, $(\tau_2,10)$, $\ldots$, $(\tau_3,6)$, $(\tau_3,15)$) and $I_q$ $=$ ($0.008, 0.010, 0.001, (\tau_1,4)$, $\ldots$, $(\tau_2,8)$, $\ldots$, $(\tau_3,4)$, $\ldots$, $(\tau_3,20))$, where $(\tau_i,t)$ states that task $\tau_i$ arrives at time $t$.
Given the two parents $I_p$ and $I_q$, SWEAKCrossover randomly selects a point--- the first arrival of $\tau_2$ in this example---and then it swaps the context switching times and all the arrivals of $\tau_1$ between $I_p$ and $I_q$. 
As shown in Fig.~\ref{fig:crossover}, SWEAKCrossover then generates the offspring $I^\prime_p$ and $I^\prime_q$ as follows: $I^\prime_p$ $=$ ($0.007, 0.011, 0.001, (\tau_1,5)$, $\ldots$, $(\tau_2,8)$, $\ldots$, $(\tau_3,4)$, $\ldots$, $(\tau_3,20)$) 
and $I^\prime_q$ $=$ ($0.008, 0.010, 0.001, (\tau_1,4)$, $\ldots$, $(\tau_2,10)$, $\ldots$, $(\tau_3,6)$, $(\tau_3,15)$).
The shaded (resp. unshaded) cells in Fig.~\ref{fig:crossover} indicate which context switching times and task arrivals in child $I^\prime_q$ (resp. $I^\prime_p$) come from which parent.

\emph{Mutation.} 
SWEAK uses a heuristic mutation algorithm called SWEAKMutation. 
For a solution $I$, SWEAKMutation mutates the context switching times or the $k$th task arrival time $a_{i,k}$ of an aperiodic task $\tau_i$ with a mutation probability.
Regarding the context switching times, SWEAKMutation chooses a new time value from the range of each context switching time, i.e., startup $\lambda_s$, exit $\lambda_x$, and IPI $\lambda_p$ times.
Regarding arrivals of an aperiodic task $\tau_i$, 
SWEAKMutation chooses a new arrival time value $a_{i,k}$ based on the $[T^{min}_i, T^{max}_i]$ inter-arrival time range of $\tau_i$. 
If a mutation of the $k$th arrival time of $\tau_i$ does not affect the validity of the $k{+}1$th arrival time, the mutation operation ends.
Specifically, let $a^*_{i,k}$ be a mutated value of $a_{i,k}$. 
In case $a_{i,k+1} \in [a^*_{i,k} + T^{min}_i, a^*_{i,k} + T^{max}_i]$, SWEAKMutation returns the mutated $I$ solution.

After mutating the $k$th arrival time $a_{i,k}$ of a task $\tau_i$ in a solution $I$, if the $k{+}1$th arrival becomes invalid, SWEAKMutation corrects the remaining arrivals of $\tau_i$. 
We denote by $a^*_{i,k}$ the mutated $k$th arrival time of $\tau_i$. 
For all the arrivals of $\tau_i$ after $a^*_{i,k}$, SWEAKMutation first updates their original arrival time values by adding the difference $a^*_{i,k}-a_{i,k}$. 
Let $\mathbb{T} = [0,\mathbf{t}]$ be the scheduling period. SWEAKMutation then removes some arrivals of $\tau_i$ if they are mutated to arrive after $\mathbf{t}$ or adds new arrivals of $\tau_i$ while ensuring that all tasks arrive within $\mathbb{T}$.

Given the offspring presented in Fig.~\ref{fig:crossover}, SWEAKMutation, for example, mutates a child solution $I^\prime_q$ $=$ ($0.008, 0.010, 0.001, (\tau_1,4)$, $(\tau_1,8)$, $(\tau_1,16)$, $\ldots$, $(\tau_3,15)$). 
Let $[T^{min}_1, T^{max}_1]$ $=$ $[2,8]$ be the inter-arrival time range of task $\tau_1$, let $\mathbb{T} = [0,22)$ be the time period during which APSSimulator receives task arrivals, and let us assume SWEAKMutation selects the second arrival of task $\tau_1$, i.e., $(\tau_1,8)$ in Fig.~\ref{fig:crossover}, to mutate. 
Based on the inter-arrival time range of $\tau_1$, SWEAKMutation randomly chooses a new arrival time, e.g., $6$, for the second arrival of $\tau_1$. 
The third arrival $(\tau_1,16)$ of $\tau_1$ then becomes invalid due to the mutated second arrival $(\tau_1,6)$, i.e., $\tau_1$ cannot arrive at time $16$ because $16$ $\notin$ $[6 + 2, 6 + 8]$, where $[T^{min}_1, T^{max}_1]$ $=$ $[2,8]$. 
According to the correction procedure described above, the third arrival of $\tau_1$ is modified to $(\tau_1,14)$ as $14$ $=$ $16 + (6-8)$, where $16$, $6$, and $8$ are, respectively, the original third arrival time of $\tau_1$, the mutated second arrival time of $\tau_1$, and the original second arrival time of $\tau_1$. 
As APSSimulator can receive new arrivals of $\tau_1$ after time $14$, SWEAKMutation may add new arrivals of $\tau_1$ based on its inter-arrival time range.

Note that for a system that consists of only periodic tasks, SWEAK will search for effective test cases by varying context-switching times without changing sequences of task arrivals since periodic tasks will follow the same arrival patterns (see Section~\ref{sec:problem}).

 \subsection{Simulation}
\label{subsec:simulation}

The objective of the simulation step is to produce schedule scenarios and a labeled dataset (training dataset).
SWEAK uses a scheduling simulation technique to produce schedule scenarios since such a simulation technique can generate a large number of schedule scenarios at a lower cost compared to the cost required to run an actual system. Further, simulation enables analyzing the tasks based on task descriptions at early design stages when their actual code is not yet available. 
Hence, simulation techniques have been used in many prior studies~\cite{le2004generic,bini2005measuring,moallemi2013modeling,Lee2022a,Lee2022b}.
Based on simulation results, we generate a labeled data set for the learning step (described in Section~\ref{subsec:learning}).

\textbf{APSSimulator.}
A schedule simulator, named APSSimulator, simulates the behavior of APS according to the characteristics described in Section~\ref{sec:motivation}.
We note that APSSimulator is our extension of the industrial APS simulator provided by BlackBerry.
Our extension is mainly about applying the adapter pattern~\cite{gamma1994design} to integrate the industrial APS simulator into SWEAK.
Hence, below, we describe the interfaces (i.e., inputs and outputs) of APSSimulator, which need to be considered when adapting SWEAK to integrate with another scheduling simulator, which simulates, for example, different APS policies~\cite{Bletsas2009,Massa2016,Abeni2020}.
For more details about APS developed by BlackBerry, we refer readers to the APS user's guide~\cite{APS}.
As input, APSSimulator takes a feasible solution $I$, which contains sequences $A$ of task arrivals for a set $\Gamma$ of tasks, context switching times (startup $\lambda_s$, exit $\lambda_x$, IPI $\lambda_p$), and a set $W$ of WCET values of the tasks.
For a sequences $A$ of task arrivals, APSSimulator calculates when each task arrival will be completed given the context switching times in a solution $I$ and a set $W$ of WCET values, as well as APS configurations, e.g., the time window for partitioning and the timeslice for Round-Robin.
We set the APS configuration values following the guidelines provided by BlackBerry.
As output, a simulation result is encoded into a schedule scenario $S$ (see Section~\ref{sec:problem}).

\indent\textbf{Generating a labeled dataset.}
SWEAK requires a labeled dataset as it uses a supervised learning technique~\cite{Russell2010} to infer a model that predicts safe WCET ranges. Importantly, engineers want to have a certain level of confidence about our prediction results, i.e., safe WCET ranges. To this end, SWEAK applies logistic regression to enable a probabilistic interpretation of the prediction results. In our context, a prediction model, inferred from a given labeled dataset, captures the relationship between tasks' WCET values and the schedulability of the tasks. The detailed learning step is explained in Section~\ref{subsec:learning}.

A labeled dataset, denoted by $\vv{L}$, is a list of tuples $(W, \ell)$, where $W$ is a set of WCET values, and $\ell$ is the label indicating the schedulability of a schedule scenario resulting from $W$. 
SWEAK generates a tuple $(W, \ell)$ for each APSSimulator run.
For example, when SWEAK evaluates a feasible solution $I$ during search (Section~\ref{subsec:search}), it appends $ns$ tuples $(W, \ell)$ to the labeled dataset $\vv{L}$. 
To evaluate a solution $I$, SWEAK runs APSSimulator $ns$ times with the sampled sets of WCET values, i.e., \{$W_1, W_2, ...W_{ns}$\}. 
Each set $W_h$ consists of a set of tuples $(\tau_i, C_i)$, where $C_i$ is a randomly selected WCET value within the range [$C^{min}_i$, $C^{max}_i$] of $\tau_i \in \Gamma$.
Given a feasible solution $I$ and a set $W_h$ of WCET values, APSSimulator produces a schedule scenario $S_h$. 
SWEAK then labels $\ell$ as $\mathit{safe}$ when the schedule scenario $S_h$ satisfies all the deadline constraints of the target tasks in $\Gamma^\delta$; otherwise, it labels $\ell$ as $\mathit{unsafe}$. 
Since schedule scenarios vary across test cases, the labeled dataset $\vv{L}$ can contain tuples that have different labels for the same set $W$ of WCET values.

 \subsection{Learning logistic regression model}
\label{subsec:learning}

\begin{figure}[t]
\begin{center}
\includegraphics[width=0.85\textwidth]{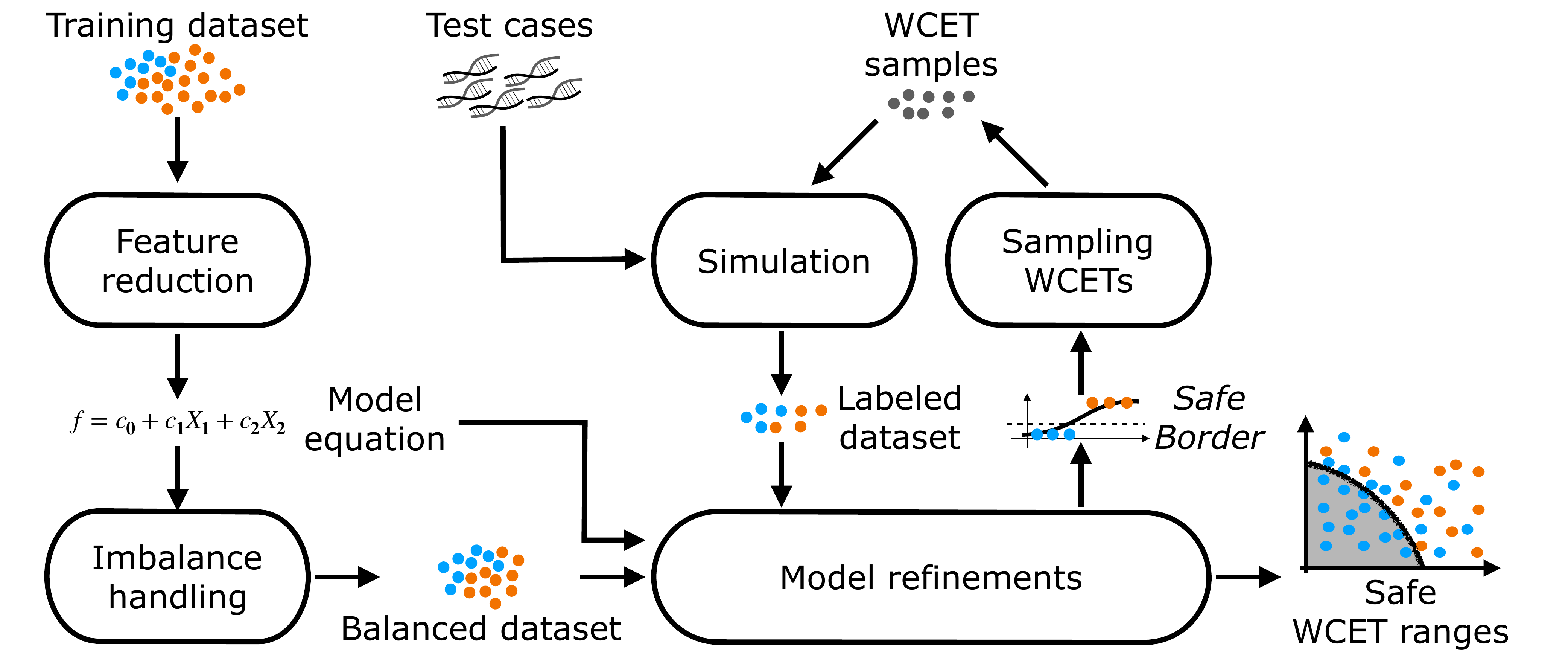}
\caption{An overview of the learning step.}
\label{fig:learning_overview}
\end{center}
\end{figure}

The objective of the learning step is to estimate safe ranges of WCET values under which target tasks are likely to be schedulable.
To achieve the objective, SWEAK builds a model to predict safe WCET ranges using logistic regression~\cite{Hosmer2013}.
This technique provides a probabilistic interpretation and enables trade-off analysis when making implementation decisions about safe WCET ranges. 
Fig.~\ref{fig:learning_overview} shows the overall process of the learning step. We note that SWEAK's learning step adapts the learning step of our previous work~\cite{Lee2022a} to account for weakly hard deadline constraints and an industrial simulator, i.e., APSSimulator.
We describe the learning step of SWEAK in the following order: feature reduction, imbalance handling, model refinements (including sampling and simulation), and selecting WCET ranges.

\begin{figure}[t]
\begin{center}
\includegraphics[width=0.7\columnwidth]{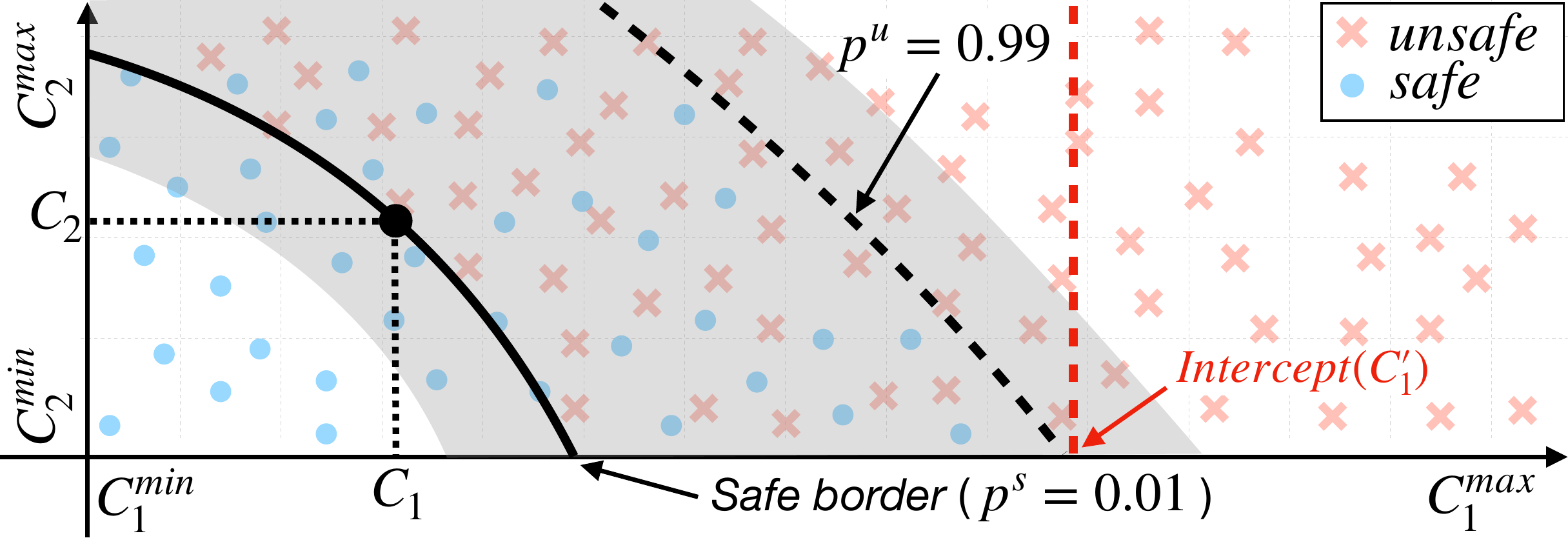}
\caption{A logistic regression model in the WCET space of two tasks $\tau_1$ and $\tau_2$.}
\label{fig:model}
\end{center}
\end{figure}

\textbf{Feature reduction.}
Given the training data $\vv{L}$ obtained from the search step, the feature reduction procedure generates an equation $f$ for logistic regression. 
Logistic regression builds a prediction model by inferring coefficients of a given equation $f$. The equation $f$ is formulated with the WCET variables of the tasks in $\Gamma$ recorded in our dataset $\vv{L}$.
Some WCET variables have significant effects in predicting whether the label is \emph{safe} or \emph{unsafe}, while other variables do not. 
Therefore, eliminating insignificant variables is needed to reduce computational complexity and increase model accuracy.

SWEAK applies a feature reduction technique based on random forest that has widely been used for dimensionality reduction~\cite{Hideko2012, Nguyen2015}. 
Given the labeled dataset $\vv{L}$, random forest builds a large number of decision trees to predict a label, i.e., either \emph{safe} or \emph{unsafe} in our case, using a randomly selected subset of WCET variables.
The technique then derives the importance of each variable based on Gini impurity~\cite{Breiman2001}.
SWEAK selects a set $V$ of important variables that are above a particular threshold.
Note that we describe the parameter values for our feature reduction in Section~\ref{subsec:param}.
Given important variables in $V$, SWEAK formulates an equation $f$ for logistic regression using a second-order polynomial response surface model (RSM)~\cite{Khuri2010} as follows: \begin{equation*}
\log \frac{p}{1-p} = c_0 + 
\sum_{i=1}^{|V|}{c_iv_i} + 
\sum_{i=1}^{|V|}{c_{ii}v_i^2} + 
\sum_{i=1}^{|V|-1}{\sum_{j=i+1}^{|V|}{c_{ij}v_iv_j}
}
\end{equation*}
\noindent where $v_i, v_j \in V$, $p$ is the probability of violating deadline constraints, and $c_0$, $c_i$, $c_{ii}$, and $c_{ij}$ are the coefficients that will be inferred by logistic regression.
Hence, the probability $p$ of violating deadline constraints is defined as follows:
\begin{equation*}
p = \frac{1}{
1+e^{-(
    c_0 + 
    \sum_{i=1}^{|V|}{c_iv_i} + 
    \sum_{i=1}^{|V|}{c_{ii}v_i^2} + 
    \sum_{i=1}^{|V|-1}{\sum_{j=i+1}^{|V|}{c_{ij}v_iv_j}}
    )
}}
\end{equation*}

In addition, SWEAK applies stepwise AIC (Akaike Information Criterion)~\cite{Yamashita2007} to the equation $f$ to eliminate terms that do not significantly help predict the label.
This enables logistic regression of SWEAK to predict only the coefficients of significant explanatory terms in $f$.
Since SWEAK requires building logistic regression models multiple times within a time budget, and these models are computationally expensive, stepwise AIC allows SWEAK to execute more efficiently.

\textbf{Imbalance handling.}
The performance of supervised machine learning highly depends on the training dataset $\vv{L}$. 
As $\vv{L}$ is generated by the search step that aims to find effective test cases (see Section~\ref{subsec:search}) with respect to violating deadline constraints, 
it tends to be imbalanced, containing more sets of WCET values that result in violating deadline constraints. 
In general, imbalanced data likely lead to unsatisfactory results when relying on supervised machine learning. 
SWEAK handles the imbalance problem as described below.

SWEAK builds an initial model $m$ from the training dataset $\vv{L}$ and the equation $f$. Logistic regression estimates a probability of violating deadline constraints for given tasks' WCET values.
For example, Fig.~\ref{fig:model} shows a model $m$ in the WCET space of two tasks $\tau_1$ and $\tau_2$. 
The gray area in Fig.~\ref{fig:model} represents the model area where the probability of violating deadline constraints is within the range [0.0001, 0.9999].
Given WCET ranges, a border can be defined by selecting a probability of violating deadline constraints dividing \emph{safe} and \emph{unsafe} areas.
SWEAK automatically selects a probability $p^u$ that maximizes the \emph{unsafe} area, while ensuring that all the data instances in the \emph{unsafe} area are classified as \emph{unsafe}, i.e., no false negative
(see the area above the border indicated by $p^u = 0.99$ in Fig.~\ref{fig:model}). 
SWEAK then calculates reduced WCET ranges [$C^{min}_i$, $C^{\prime}_i$], where $C^{\prime}_i$ is the intercept between the WCET axis of $\tau_i$ and the WCET border determined by the probability $p^u$ (see the red dashed line in Fig.~\ref{fig:model}).
The more balanced dataset $\vv{L}^b$ is produced by pruning the data instances outside the reduced WCET ranges.
Note that $C^{\prime}_i$ is equal to $C^{max}_i$ when there is no intercept for a task $\tau_i$.

\textbf{Model refinements.}
Given the balanced dataset $\vv{L}^b$ and the equation $f$, SWEAK builds a logistic regression model $m$.
SWEAK then finds a probability $p^s$ that maximizes the \emph{safe} area, while ensuring that all the data instances within the \emph{safe} area are classified as \emph{safe} with no false positive.
Note that $m$ and $p^s$ determine a \emph{safe border} that distinguishes safe and unsafe areas (see the solid line in Fig.~\ref{fig:model}).
More precisely, a safe border is defined by the equation $1 / (1 + e^{-f_m}) = p^s$, where $f_m$ denotes the function $f$ with the coefficients' values determined by $m$ (see the RSM coefficients described earlier). The safe area defined by $1 / (1 + e^{-f_m}) < p^s$ contains only safe WCETs, while for any probability $p^o > p^s$, the area defined by $1 / (1 + e^{-f_m}) < p^o$ contains both safe and unsafe WCETs.
To improve the safe border, SWEAK refines it using a distance-based sampling method that adds more WCET samples around the safe border (described in our previous work~\cite{Lee2022a}).
The sampled WCET values are evaluated and labeled by the simulation step using the test cases obtained from the search step.
These simulation results are included into a new labeled dataset $\vv{L}^{new}$. 
SWEAK then rebuilds the safe border after merging $\vv{L}^b$ with $\vv{L}^{new}$ using logistic regression. 
This refinement is repeated until either reaching the specified number of refinements (i.e., assigned analysis budget) or reaching an acceptable level of precision of the safe border using a standard $k$-fold cross-validation~\cite{Witten2011}.
In the cross-validation process, SWEAK first partitions the merged dataset into $k$ equal-size folds. SWEAK builds and evaluates logistic regression models $k$ times. Each time, SWEAK uses a different fold as the test dataset and the remaining $k-1$ folds as the training dataset.
From the $k$ evaluations, SWEAK computes the accuracy of the safe border determined by a logistic regression model $m$ and a probability $p^s$.

\textbf{Selecting WCET ranges.}
Given a safe border, safe WCET ranges are then determined by selecting one point on that border. 
A safe border represents a set of points that represents the upper bounds of safe WCET ranges. 
Engineers thus can find safe WCET ranges by choosing one appropriate point on a safe border depending on their system requirements. 
For example, if a black dot on the safe border in Fig.~\ref{fig:model} is the selected point, i.e., $[C_1, C_2]$, the safe WCET ranges become $[C^{min}_1, C_1]$ and $[C^{min}_2, C_2]$. 
However, engineers may not have proper contextual information to select a point at early design stages. 
Hence, SWEAK suggests a point, named \emph{best-size point}, on a safe border that maximizes the volume of the WCET ranges. 
The best-size point with the largest volume provides engineers with more relaxed objectives for tasks' execution times when no domain-specific information can guide such decision.
In addition, the inferred safe border provides engineers with the ability to select another point through trade-off analysis between tasks' WCET values.
  \section{Evaluation}
\label{sec:evaluation}

In this section, we evaluate SWEAK by answering three research questions below.
We do so by applying SWEAK to an industrial study subject from the satellite domain and several synthetic study subjects. 
All experiment results can be found online~\cite{Artifacts3}.

\subsection{Research questions}
\label{subsec:rqs}

\textbf{RQ1. (baseline comparison):} \textit{How does SWEAK fare against a baseline relying on random search?}
In general, comparing a search-based approach against a baseline, i.e., random search, is important to determine if the search problem indeed requires sophisticated search algorithms~\cite{Harman2012, Arcuri2014}. 
In addition, SWEAK is the first attempt to automatically estimate probabilistic safe WCET ranges for weakly hard real-time systems.
Hence, we compare SWEAK and a baseline built on random search to see if SWEAK can infer significantly better WCET options than the baseline with respect to their confidence levels and best-size points (defined in Section~\ref{subsec:learning}).
Note that due to the complexity introduced by uncertainties in task arrivals, context switching, adaptive partitioning, and multiple cores, finding optimal solutions in reasonable time is infeasible.
Hence, the baseline approach built on random search serves as our best alternative solution.

\textbf{RQ2. (probabilistic interpretation):} \textit{Can we rely on predicted probabilities from logistic regression?}
During the design stage, engineers tend to be conservative when determining safe WCET ranges and probabilities of violating deadline constraints as these determine objectives for tasks' implementations. Recall from Section~\ref{subsec:learning} that SWEAK employs logistic regression to infer such probabilities. 
We investigate the probabilities predicted by SWEAK by comparing them with those obtained from a large number of simulations with different WCET values within the estimated WCET ranges.
Given the same WCET ranges, our conjecture is that SWEAK infers higher or similar probabilities of violating deadline constraints compared to simulation-based probabilities, as SWEAK relies on fine-tuning of the logistic regression step.

\textbf{RQ3. (scalability):} \textit{Can SWEAK find safe WCET ranges for large-scale systems within a practical time budget? }
It is challenging to estimate acceptable WCET ranges for large-scale systems because of complex task interactions caused by combinations of arrival sequences, priorities, context switching times, and WCET values.
To assess the scalability of SWEAK in terms of execution time, we use a large number of synthetic systems that are generated with various characteristics.

\subsection{Synthetic systems}
\label{subsec:synthetic}

\begingroup
\begin{algorithm}[t]
\parbox{0.95\columnwidth}{
	\caption{An algorithm for generating synthetic systems including weakly hard real-time tasks and APS partitions.}
	\label{alg:synthetic}
}
\begin{lstlisting}[style=Alg2]
Input ${n}$: number of real-time tasks
Input $u^t$: target utilization of the system
Input $T^\mathit{min}$: minimum task period
Input $T^\mathit{max}$: maximum task period
Input $g$: granularity of task periods
Input $\theta$: maximum offset value
Input $\gamma$: ratio of aperiodic tasks
Input $\mu$: range factor to determine inter-arrival ?times?
Input $\omega$: number of tasks having WCET ranges
Input $\lambda$: range factor to determine WCET ranges
Input $\rho$: number of partitions
Input $(m,K)$: deadline constraint
Input ${nw}$: number of weakly hard real-time tasks
Output $\Gamma$: set of tasks

$\Gamma \leftarrow$ $\{\}$, $\mathbf{C} \leftarrow$ $\{\}$
// synthesize a set of periodic tasks
$\mathbf{U} \leftarrow$ $\fun{UUniFast\_discard}(n, u^t)$ // task utilizations 
$\mathbf{T} \leftarrow$ $\fun{generate\_task\_set}(n, T^\mathit{min}, T^\mathit{max}, g)$ //task periods
// determine WCETs
for each $i \in [1, n]$ do
?\vrule?    $\mathbf{C}$ $\leftarrow$ $\mathbf{C} \cup \{U_i {\cdot} T_i\}$, where $U_i \in \mathbf{U}$ and $T_i \in \mathbf{T}$
end for
$\Gamma \leftarrow$ $\fun{generate\_task\_periods}(\mathbf{T}, \mathbf{C}, \theta, g)$
// select weakly hard real-time tasks
$\Gamma \leftarrow$ $\fun{set\_deadline\_constraints}(\Gamma, {nw}, (m,K))$
// convert some tasks to aperiodic tasks
$\Gamma \leftarrow$ $\fun{convert\_to\_aperiodic\_tasks}(\Gamma, \gamma, \mu)$
// convert some WCET values to WCET ranges
// default argument $\color{javagreen}\lambda=\mathit{undefined}$
$\Gamma \leftarrow$ $\fun{convert\_to\_WCET\_ranges}(\Gamma, \omega, \lambda)$
// assign partitions and partition budgets
$\Gamma \leftarrow$ $\fun{assign\_partitions}(\Gamma, \rho)$
return $\Gamma$
\end{lstlisting}
\end{algorithm}
\endgroup

A synthetic system is an artificially generated system accounting for the characteristics of real-time tasks that reflect actual systems in the real world.
Synthetic systems are used in many real-time system studies~\cite{Davis2008,Zhang2009,Davis2011,Grass2018,Vonder2018,Durr2019} to evaluate approaches while being able to fully control and vary systems' characteristics, thus assessing their impact on results.
Algorithm~\ref{alg:synthetic} describes a procedure for generating a synthetic system by varying the key task parameters (lines 1-13).
Note that the algorithm is developed based on our prior approach for evaluating SAFE and the guidelines provided by BlackBerry to account for weakly hard deadline constraints and adaptive partitions.
Briefly, the algorithm synthesizes a set of periodic tasks (lines 18-24) and sets some tasks to have weakly hard deadline constraints (lines 25-26).
The algorithm then modifies the system as follows: 
(1)~converting some tasks to aperiodic tasks (lines 27-28), 
(2)~transforming some tasks' WCET values into WCET ranges (lines 29-31), and 
(3)~configuring partitions and assigning tasks to each partition (lines 32-33).

As shown in line 18 of Algorithm~\ref{alg:synthetic}, the algorithm first creates a set $\mathbf{U}$ of task utilization values using the UUniFast-Discard algorithm~\cite{Davis2011}, which is devised to generate an unbiased distribution of task utilization values. The UUniFast-Discard algorithm takes as input the number $n$ of tasks to be synthesized and a target utilization value $u^t$ of the system. It then outputs $n$ utilization values, \{$U_1$, $\ldots$, $U_n$\}, where $0 < U_i < 1$ for all $U_i$ and $\sum_{i=1}^{n}{U_i} = u^t$. The maximum value of target utilization $u^t$ relies on the number of processing cores, i.e., the maximum target utilization is equal to the number of processing cores. For example, if a system uses two processing cores, the maximum value of $u^t$ is 2.

On line 19, the algorithm generates $n$ task periods, $T_1$ $\ldots$ $T_n$ according to a log-uniform distribution within a range [$T^{min}, T^{max}$], i.e., $\mathit{log}~T_i$ follows a uniform distribution. For example, when a period range [$T^{min}, T^{max}$] is [10ms, 1000ms], the algorithm generates approximately an equal number of tasks in the period ranges [10ms, 100ms] and [100ms, 1000ms]. The parameter $g$ is used to determine the granularity of period values to be multiples of $g$. Lines 21-23 of Algorithm~\ref{alg:synthetic} describe how the algorithm synthesizes tasks' WCET values $\mathbf{C}$. Specifically, for each task $\tau_i$, the algorithm computes its WCET value $C_i$ as $U_i \cdot T_i$.

Given the task periods $\mathbf{T}$ and the WCET values $\mathbf{C}$, line 24 of Algorithm~\ref{alg:synthetic} synthesizes a set $\Gamma$ of periodic tasks with their offsets, priorities, and deadlines. Recall from Section~\ref{sec:problem} that a periodic task $\tau_i$ is characterized by a period $T_i$, a WCET $C_i$, an offset $O_i$, a priority $P_i$, and a deadline $D_i$. A task offset $O_i$ is randomly selected from an input range $[0, \theta]$ of offset values. 
The algorithm applies a rate-monotonic scheduling policy~\cite{Liu1973} to assign task priorities, in which tasks that have longer periods are given lower priorities. 
This policy assumes that task deadlines are equal to their periods.

Given the system $\Gamma$, line 26 assigns the $(m,K)$-constraint (defined in Section~\ref{sec:problem}) to $nw$ tasks to make them weakly hard real-time tasks. Note that real-time tasks in a system can have different deadline constraints, supported by SWEAK. 
However, to investigate the impact of different $(m,K)$-constraints in a controlled setting, we set the $nw$ tasks to be subjected to the same deadline constraint (see Section~\ref{subsec:design}). We select the $nw$ tasks from the lowest priority tasks as they have higher chances of missing deadlines compared to higher priority tasks.

Line 28 selects some periodic tasks and converts them into aperiodic tasks according to the ratio $\gamma$ of aperiodic tasks. 
The algorithm then uses a range factor $\mu$ to determine the minimum and maximum inter-arrival times of the aperiodic tasks. Specifically, for a task $\tau_i$ to be converted, the algorithm computes a range  $[T^{min}_i, T^{max}_i]$ of inter-arrival times as  $[T^{min}_i, T^{max}_i]$ $=$ $[T_i \times (1-\mu), T_i \times (1+\mu)]$, where $\mu \in (0,1)$. For example, if $\mu=0.45$ and $T_i=50$ for a task $\tau_i$ to be converted, $[T^{min}_i, T^{max}_i]$ $=$ $[27.5, 72.5]$.
For the converted aperiodic tasks, we set their offsets to 0, i.e., $O_i$=0 as the offsets are replaced by the tasks' inter-arrival times (see Section~\ref{sec:problem}).

To synthesize tasks' WCET ranges, line 31 randomly selects $\omega$ tasks in $\Gamma$ to convert their WCET point values into WCET ranges.
When the range factor $\lambda$ is defined, the algorithm  computes the WCET ranges by applying $\lambda$ to the selected tasks. More precisely, for a selected task $\tau_i$ $\in$ $\Gamma$, the algorithm computes a WCET range $[C^{min}_i, C^{max}_i]$ as $[C^{min}_i, C^{max}_i]$ $=$ $[C_i \times (1-\lambda), C_i \times (1+\lambda)]$, where 0 < $\lambda < 1$.
When the range factor $\lambda$ is undefined, the algorithm selects a range factor $\lambda_i$ for each task $\tau_i$ $\in$ $\Gamma$ from a log-uniform distribution in the range (0, 1). 
Each WCET range [$C_i^{min}$, $C_i^{max}$] for $\tau_i$ is then determined according to $[C_i \times (1 - \lambda_i),$ $C_i \times (1 + \lambda_i)]$. For example, if $\lambda_1=0.25$ and $C_1=10$ for a task $\tau_1$, $[C^{min}_1, C^{max}_1]$ $=$ $[7.5, 12.5]$. This procedure results in a system having a small number of tasks with large WCET ranges and a large number of tasks with small WCET ranges. Note that we discard the invalid cases where the minimum WCET is equal to 0 or the maximum WCET is greater or equal to its deadline, i.e., $C_i^{min}$ $=$ $0$ or $C_i^{max}$ $>=$ $D_i$.

Regarding APS partitioning, line 33 assigns tasks to partitions. The algorithm creates $\rho$ partitions and assigns evenly distributed partition budgets. For example, when $\rho$=2, the budget distribution is [50\%, 50\%]. If $\rho$=3, the budget distribution is [34\%, 33\%, 33\%]. 
The algorithm then randomly assigns tasks to partitions. 
Each partition must have at least one task, and a task can be assigned to only one partition.

 \subsection{Study subjects}
\label{subsec:subjects}

To address RQ1 and RQ2, we use four case study subjects: ESAIL~\cite{Lee2022a} (an industrial real-time system) and three synthetic systems. 
We note that, due to confidentiality, we were not able to obtain access to the actual weakly hard real-time systems developed by BlackBerry's customers, who use QNX Neutrino with APS.
However, the synthetic systems used in our experiments are realistic and representative as they are based on BlackBerry's guidelines and employ the industrial APS policy.
Regarding RQ3, we use a large number of synthetic systems generated by controlling parameters in Algorithm~\ref{alg:synthetic} (See Section~\ref{subsec:design}).
The full descriptions of the systems are available online~\cite{Artifacts3}.
We further describe the details of the systems below.

ESAIL is a commercial microsatellite developed by LuxSpace that tracks the movements of ships over the entire globe. 
The ESAIL management system is made up of 12 periodic tasks and 13 aperiodic tasks working on a single core platform with one partition. 
During the design stages, these tasks were analyzed their WCETs as ranges.
Regarding deadline constraints, five aperiodic tasks are considered weakly hard real-time tasks while the other tasks are hard real-time tasks.

To generate three synthetic systems for RQ1 and RQ2, we first create a base system using Algorithm~\ref{alg:synthetic}.
The base system is generated with the following parameter values: 
(1)~the number of tasks $n$ $=$ 25, the ratio of aperiodic tasks $\gamma$ $=$ 0.5, the range factor to determine inter-arrival times $\mu$ $=$ 0.25, and the maximum offset $\theta$ $=$ 0. These settings were decided based on the characteristics of ESAIL. 
(2)~the minimum task period $T^{min}$ $=$ 10ms, the maximum task period $T^{max}$ $=$ 1s, and the granularity $g$ = 10ms. These are commonly used in real-time systems~\cite{Baruah2011}.
(3)~the target utilization $u^t$ = 0.9 for a single-core platform.
This was decided to ensure the tasks sometimes miss their deadlines~\cite{Emberson2010}. 
(4)~Regarding the number of APS partitions, we set $\rho$ = 1.
This was decided for the base system to be simple so that it can easily be converted to other synthetic systems. 
(5)~Regarding deadline constraints, we set the 10 lowest priority tasks to be weakly hard real-time tasks (i.e., $nw$ = 10).  
The $(m,K)$-constraint of these 10 tasks was set to $(0,10)$, i.e., hard deadline constraint. It will subsequently be changed in our experiments (see Section~\ref{subsec:design}) to account for weakly hard deadline constraints, i.e., $m$ $>$ $0$.
(6)~For WCET ranges, we set the number $\omega$ of tasks with WCET ranges to 25 and the range factor $\lambda$ for determining WCET ranges to $\mathit{undefined}$ (see Algorithm~\ref{alg:synthetic}). Recall from Section~\ref{subsec:synthetic} that this configuration creates a system with a small number of tasks having large WCET ranges and a large number of tasks having small WCET ranges, aligning with the WCET characteristics of ESAIL.

Given the base system $\Gamma$, we synthesize the three systems described below by modifying $\Gamma$ to account for the characteristics of APS following the guidelines from BlackBerry. These synthetic systems enable us to evaluate SWEAK in different operational settings of APS.

\begin{itemize}[leftmargin=1.1em]
    \setlength\itemsep{0em}
    \item PARTITION: This system has two APS partitions with 60\% and 40\% budgets. For efficient scheduling, a partition budget should be enough to execute all tasks in the partition.
    We thus assign tasks to each partition so that each partition budget is close to the total utilization of the tasks in the partition.
    In our evaluation, 19 tasks with high priorities in $\Gamma$ are assigned to the first partition. The remaining six tasks are assigned to the second partition. 
    \item POLICY: This system contains two pairs of tasks with the same priority. 
    To make a pair, we randomly select two tasks from the given system $\Gamma$ and assign the same priority and scheduling policy to the selected tasks.
	One pair applies FIFO. The other pair uses Round-Robin.

    \item MULTICORE: This system works on a two-core platform and assigns core affinities to some tasks. To make this system, we multiply WCET values (as well as the WCET ranges) by two for all tasks in $\Gamma$ to make the total utilization $\approx$1.8. Recall that the maximum total utilization of a system is equal to the number of cores.
	We then assign core affinity to tasks using a random selection. For this system, we assign core 1 affinity to eight tasks, core 2 affinity to the other eight tasks, and no core affinity to the remaining nine tasks.
\end{itemize}

 \subsection{Experimental design}
\label{subsec:design}
To answer the three RQs described in Section~\ref{subsec:rqs}, we design three experiments EXP1, EXP2, and EXP3, respectively. We conduct EXP1 and EXP2 with the four case study subjects described in Section~\ref{subsec:subjects}: ESAIL, PARTITION, POLICY, and MULTICORE. For EXP3, we experiment with 600 synthetic systems with different parameter settings. We describe each experiment in detail below.

\textbf{EXP1.}
To answer RQ1, we implement a baseline approach (named Baseline) that uses random search without the learning step in SWEAK.
Baseline's random search is a variant of the search step in SWEAK that does not use genetic operators, i.e., selection, crossover, and mutation, to breed offspring (see Section~\ref{subsec:search}). 
Instead, Baseline generates offspring randomly for the next generation and evaluates them with the same multi-objective fitness functions as SWEAK. 
During search, a labeled dataset $\vv{L}$ is produced by Baseline, which contains tuples $(W,\ell)$ where $W$ is a set of tasks' WCET values and $\ell$ is the label that classifies the simulation result with $W$ as safe or unsafe.
Once the labeled dataset $\vv{L}$ is obtained, Baseline retrieves all tuples from $\vv{L}$ to select a specific tuple $(W_s,\ell_s)$ that is safe (i.e., $\ell_x=\mathit{safe}$) and maximizes the volume of the hyperbox defined by $W_s$. 
Note that $W_s$ should satisfy the condition that any tuple ($W_x$, $\ell_x$) contained in the hyperbox defined by $W_s$ be safe, i.e, $\ell_x=\mathit{safe}$.

EXP1 compares the results obtained from SWEAK against Baseline. 
Recall from Section~\ref{subsec:learning} that SWEAK suggests safe WCET ranges on the safe border by selecting a best-size point that maximizes its volume of the hyperbox. 
Since both SWEAK and Baseline return best-size points, the comparison can be done by measuring the best-size volumes. 
To analyze the relationship between deadline constraints and best sizes, we apply both approaches to the subjects with different deadline constraints $(m_i, K_i)$, where $m_i$ is the number of tolerable deadline misses and $K_i$ is the time window to check the deadline constraint (see Section~\ref{sec:problem}).
To do this, we vary $m_i$ from 0 to 4, with a fixed $K_i$ (10) by assuming that all tasks in a subject are subjected to the same deadline constraint; hence, EXP1 uses 4 $\times$ 5 synthetic systems (i.e., ESAIL, PARTITION, POLICY, and MULTICORE with five different deadline constraints). 
Note that we do not vary $K_i$ because it does not affect the results.

\textbf{EXP2.} To answer RQ2, EXP2 calculates the empirical probability of violating deadline constraints for the safe WCET ranges obtained from SWEAK.
To this end, we first randomly sample multiple test cases, defining task arrivals and context switching times, and execution times within the safe WCET ranges obtained from each approach. 
We then simulate many combinations of the test cases and execution times using APSSimulator and check for the presence of violating deadline constraints in each simulation result. 
The empirical probability is calculated as the number of simulations that violate deadline constraints over the number of all simulation runs.
We simulate 40000 times to compute the empirical probability of safe WCET ranges obtained by SWEAK and Baseline.
In addition, we conduct EXP2 by varying the number $m_i$ from 0 to 4 tolerable deadline misses to investigate the impact of deadline constraints.

\textbf{EXP3.} To answer RQ3, we conduct EXP3 to assess the execution time of SWEAK using 600 synthetic systems with different system characteristics using  Algorithm~\ref{alg:synthetic}. EXP3 varies each system parameter value while fixing the other parameters' values so that we can analyze correlations between the execution time of SWEAK and the varying parameter. We generate the synthetic systems by changing the following six parameters:
(a)~number of tasks, $n$ $\in$ \{5, 10, $\cdots$, 50\},
(b)~ratio of aperiodic tasks, $\gamma$ $\in$ \{0.05, 0.10, $\cdots$, 0.50\},
(c)~number of WCET ranges, $\omega$ $\in$ \{1, 2, $\cdots$, 10\},
(d)~number of processing cores, $\epsilon$ $\in$  \{1, 2, $\cdots$, 10\},
(e)~number of APS partitions, $\rho$ $\in$ \{1, 2, $\cdots$, 10\},
and (f)~simulation time, $\mathbf{t}$ $\in$ \{5s, 10s, 15s, $\cdots$, 50s\}.

The number of all tasks $n$, the ratio of aperiodic tasks $\gamma$, the number of WCET ranges $\omega$, and the number of processing cores $\epsilon$ are selected because they are the main factors when designing real-time systems.
The number of APS partitions $\rho$ is selected as it is adjustable by APS. We also include the simulation time $\mathbf{t}$ as the execution time of SWEAK obviously depends on simulation time.
We note that the total utilization of a generated synthetic system changes according to the number of processing cores.
For example, when the target utilization $u^t$ $=$ 0.9 and the number of cores $\epsilon$ $=$ 2, the total utilization of the system becomes 1.8 (see Section~\ref{subsec:synthetic}).

When varying a parameter's value, to enable controlled experiments, we fix the other parameters' values as follows:
(1)~We set the number of all tasks $n$ $=$ 25, the ratio of aperiodic tasks $\gamma$ $=$ 0.50, and the maximum offset $\theta$ $=$ 0. We set these values according to our industrial subject, ESAIL.
(2)~Regarding the task periods, we set the range [$T^{min}$, $T^{max}$] of minimum and maximum periods to [10ms, 1s] with granularity $g$ = 10ms. These values are commonly used in many real-time subjects~\cite{Baruah2011}.
(3)~We set the range factor to determine inter-arrival times of aperiodic tasks $\mu$ = 0.25, the number of WCET ranges $\omega$ = 2, the range factor to determine WCET ranges $\lambda$ = 0.25, and the target utilization per processing core $u^t$ = 0.9.
The parameters' values are determined based on our preliminary experiments. 
They ensure that the executions of the synthetic systems examined in EXP3 can sometimes violate their deadline constraints, i.e., they contain both safe and unsafe data instances (see Section~\ref{subsec:learning}).
(4)~We set the number of processing cores $\epsilon$ and the number of APS partitions $\rho$ equal to 1. These values are selected to build simple baseline systems.
(5)~For the simulation time of APSSimulator (see Section~\ref{subsec:simulation}), we assign the minimum simulation time of 5s to guarantee that any aperiodic task arrives at least once and all possible arrivals of periodic tasks can be analyzed during that period.
Additionally, with regard to the deadline constraint, we set $(m,K)$ to (2, 10) for the ten lowest priority tasks in each system. 

Due to the randomness of SWEAK, we conduct our experiments multiple times, i.e., 50 times for EXP1 and EXP2 and 10 times for each parameter configuration of EXP3. To compare the results, we perform a statistical comparison using the non-parametric Mann-Whitney U-test~\cite{Mann1947} and Vargha and Delaney's $\hat{A}_{12}$~\cite{Vargha2000}. The Mann-Whitney U-test is used to compare statistical differences between two independent sample groups. We set the level of significance $\alpha = 0.05$. 
Vargha and Delaney’s $\hat{A}_{12}$ is a measure of effect size to assess the practical significance of differences between two search algorithms. If $\hat{A}_{12}$ is equal to 0.5, the two algorithms are equivalent. If $\hat{A}_{12}$ is close to 1, the first algorithm is largely superior to the second algorithm.
 \subsection{Implementation and parameter tuning}
\label{subsec:param}

We set the following parameter values for running SWEAK and Baseline. 
For the NSGA-II search in SWEAK, we set the population size to 10, the crossover rate to 0.7, and the mutation rate to 0.2 based on existing guidelines~\cite{Haupt1988}. 
We set the number of iterations to 1000 since we observed that the fitness values reached a plateau after that in our preliminary experiment. 
To calculate the fitness values, we ran APSSimulator 20 times for each solution ($I$ in Section~\ref{subsec:search}). 
Based on preliminary experiments, we found that this number was sufficient to compute the fitness values of a solution within a reasonable time period (i.e., less than 1m).

For random search in Baseline, we set the number of iterations to 1500 to ensure that Baseline produces a dataset $\vv{L}$ of the same size as SWEAK. We used the same values as SWEAK for the population size and the number of APSSimulator runs.

To simulate study subjects, we set the simulation time to 60s for the ESAIL subject and 5s for other synthetic subjects. Such values are determined by the following rules:
(1)~If a system is composed of only periodic tasks, the simulation time is the least common multiple (LCM) of the period values for all tasks~\cite{Wang2017}. 
(2)~If a system contains aperiodic tasks, the simulation time is determined as the larger value of the following two values: the LCM of the period values of the periodic tasks and the maximum value among the maximum inter-arrival times of the aperiodic tasks. This simulation time allows us to simulate all possible patterns of arrivals of periodic tasks including at least one arrival of aperiodic tasks.

For each run of APSSimulator, we set the time window for partitioning to 100ms, which is the default value of APS. 
The timeslice for Round-Robin is set to 4ms, and the processor tick interval is 1ms. 
According to the guidelines from BlackBerry, the timeslice is usually set to 4 times a processor tick interval.

SWEAK has some parameters in the learning step for tuning feature reduction and model refinement.
Regarding the former, we employed the random forest algorithm that includes the following parameters: (1)~The tree depth was set to $\sqrt{|F|}$, where $|F|$ is the number of features, following the guidelines~\cite{Hastie2009}. For example, since the ESAIL subject contains 25 features (i.e., WCET ranges), we assigned $\sqrt{|25|}$ to the tree depth of the subject (see Section~\ref{subsec:learning}).
(2)~The number of trees was set to 100 as we found that learning more than 100 trees did not provide further benefits for reducing the number of features in our preliminary experiments.
Regarding model refinement, we set the number of WCET samples to 100 and the number of model updates to 100. 
We observed that the precision of the model reaches an acceptable level with these parameters in our preliminary experiments.

We note that all the parameters can be further tuned to improve the performance of SWEAK. However, we were able to clearly and convincingly support our conclusions with the current parameter settings mentioned above; hence, we do not report further experiments on tuning the parameters' values.

All experiments have been performed on nodes in the high-performance computing cluster~\cite{Varrette2014} at the University of Luxembourg. Each run of experiments was executed on a node by assigning 10 cores running at 2.6GHz and 16GB of memory.

\subsection{Results}
\label{subsec:results}

\begin{figure*}[t]
\begin{center}
	\begin{subfigure}[t]{0.48\textwidth}\centering
	    \includegraphics[width=\columnwidth]{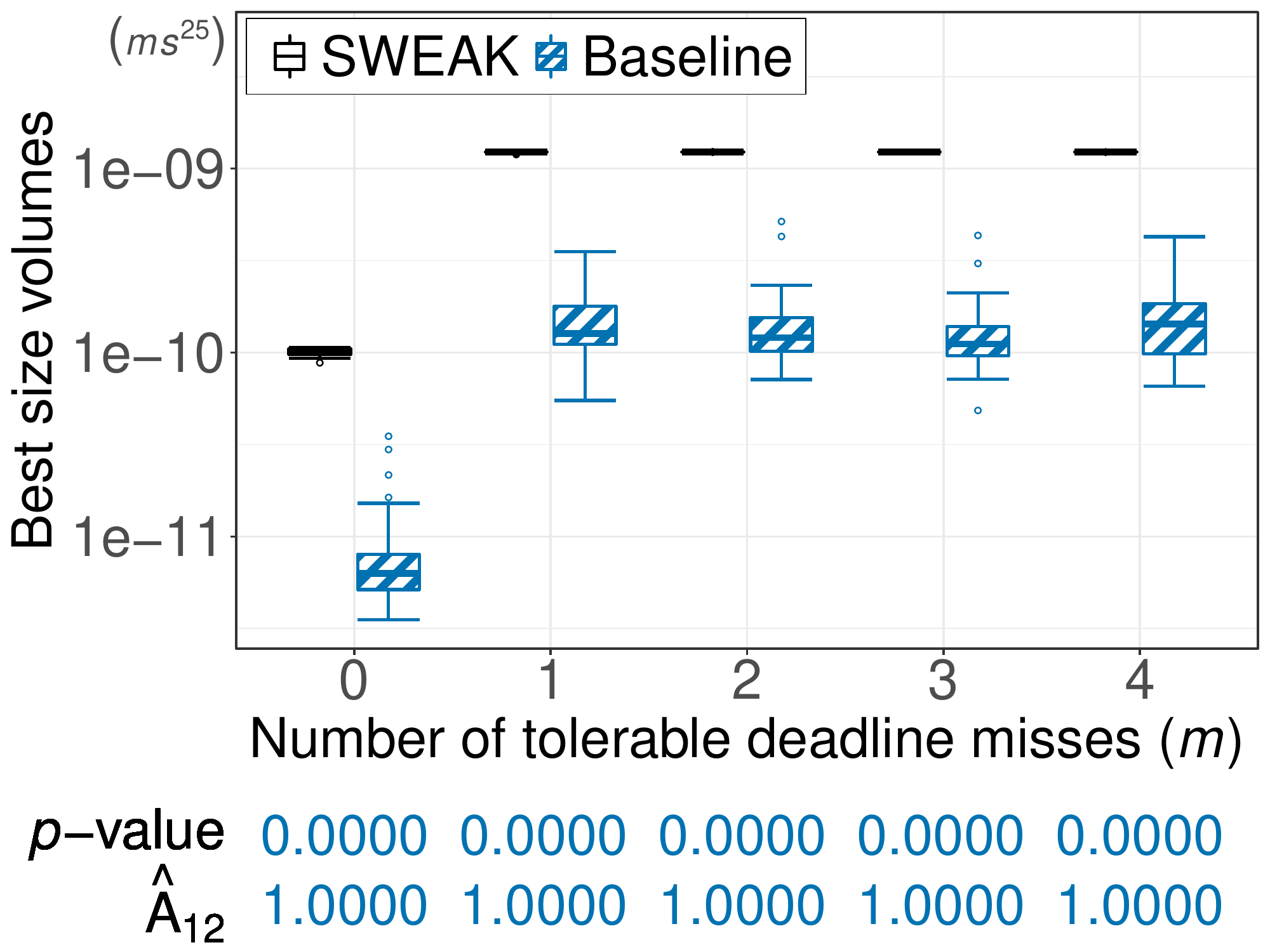}
		\caption{ESAIL}
		\label{fig:rq1 ESAIL}
	\end{subfigure}
	\hfill
	\begin{subfigure}[t]{0.48\textwidth}\centering
	    \includegraphics[width=\columnwidth]{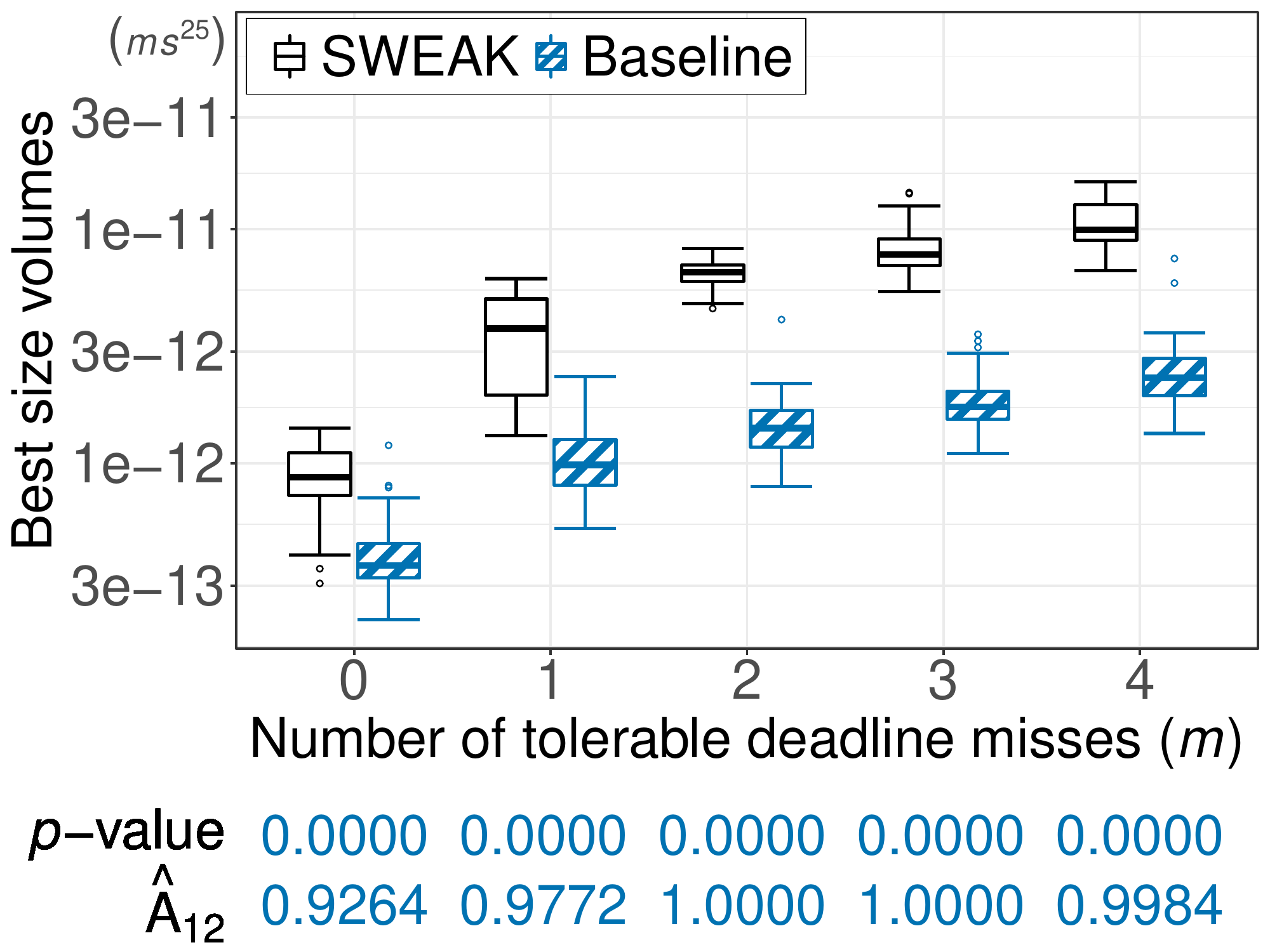}
		\caption{PARTITION}
	    \label{fig:rq1 PARTITION}
	\end{subfigure}

	\bigskip

	\begin{subfigure}[t]{0.48\textwidth}\centering
	    \includegraphics[width=\columnwidth]{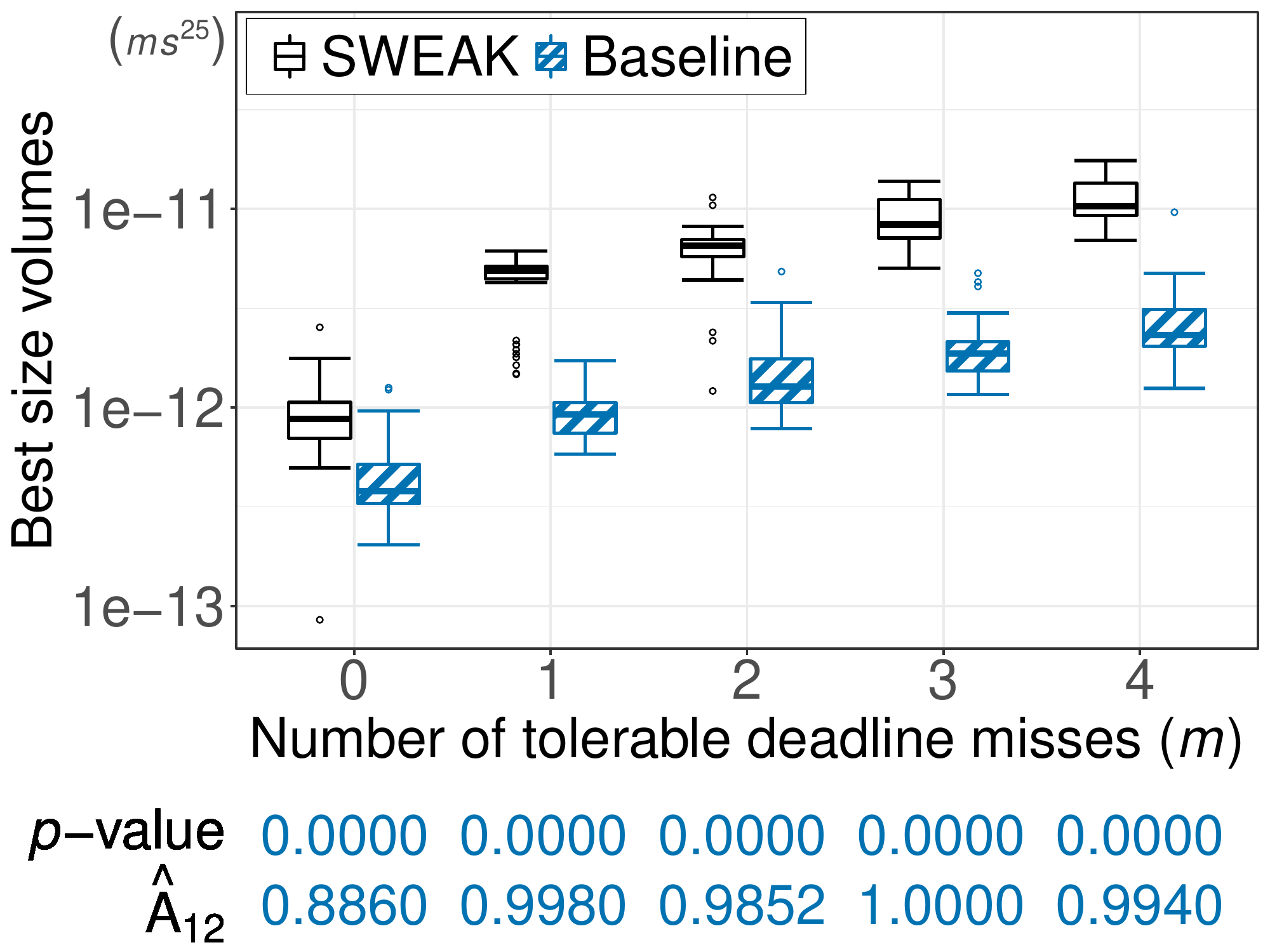}
		\caption{POLICY}
	    \label{fig:rq1 POLICY}
	\end{subfigure}
	\hfill
	\begin{subfigure}[t]{0.48\textwidth}\centering
	    \includegraphics[width=\columnwidth]{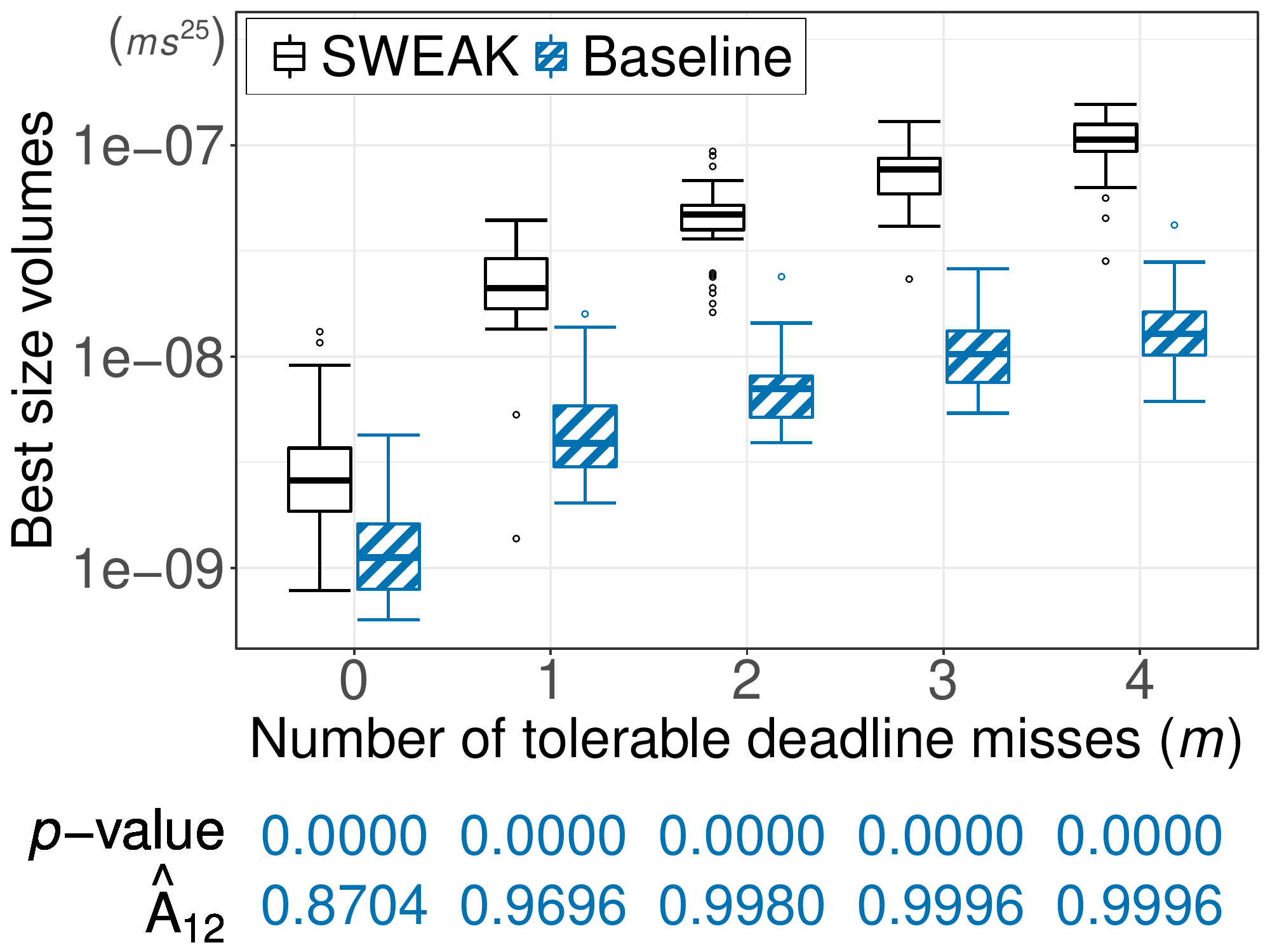}
		\caption{MULTICORE}
	    \label{fig:rq1 MULTICORE}
	\end{subfigure}
\end{center}
\caption{Distributions of the hyperboxes' volumes that are defined by the safe WCET ranges obtained from SWEAK and Baseline for (a)~ESAIL, (b)~PARTITION, (c)~POLICY, and (d)~MULTICORE. The boxplots (25\%-50\%-75\%) show the hyperboxes' volumes obtained from 50 runs of SWEAK and Baseline.}
\label{fig:rq1}
\end{figure*}

\textbf{RQ1.}
Fig.~\ref{fig:rq1} shows the results of EXP1, which compare the volumes of the hyperboxes defined by the safe WCET ranges computed by SWEAK and Baseline. The comparisons are carried out with five different numbers of tolerable deadline misses, 0 to 4, in deadline constraints of the following four study subjects: ESAIL (Fig.~\ref{fig:rq1 ESAIL}), PARTITION (Fig.~\ref{fig:rq1 PARTITION}), POLICY (Fig.~\ref{fig:rq1 POLICY}), and MULTICORE (Fig.~\ref{fig:rq1 MULTICORE}). Each boxplot in the figures shows a distribution (25\%-50\%-75\% quartiles) obtained from 50 runs of SWEAK and Baseline. The figures also report $p$-values and $\hat{A}_{12}$ values from comparing 50 runs of SWEAK and Baseline. Note that the unit of the volumes is $\mathit{ms}^{25}$, where the number of tasks with WCET ranges $\omega$ $=$ 25. Since the minimum time unit in our experiments is 0.1ms, the minimum volume of the study subjects is $1\times10^{-25}$$\text{ms}^{25}$.

As shown in Fig.~\ref{fig:rq1}, SWEAK produces larger volumes of hyperboxes compared to Baseline across all the subjects. 
Note that a larger hyperbox volume provides greater flexibility in selecting appropriate WCET values, as such a hyperbox has wider WCET ranges.
Regarding deadline constraints, for both SWEAK and Baseline, the larger $m$, the larger the volume of the hyperbox for the following subjects: PARTITION, POLICY, and MULTIFORE.
In particular, when $m$ = 1, the hyperbox volume becomes much larger than that obtained when $m$ = 0. The increase in volume is smaller when $m$ further increases to 2, 3, and 4.
This trend implies that when a deadline miss occurs, the subjects are likely to have more consecutive deadline misses. 
Unlike the three subjects discussed above, the hyperbox volumes of ESAIL are similar when the system is subjected to weakly hard deadline constraints ($m$ $\geq$ 1), which are significantly larger than the hyperbox volumes when the hard deadline constraint ($m$ $=$ 0) is applied. 
This trend is caused by the characteristics of ESAIL, which works on a single core platform with one partition (see Section~\ref{subsec:subjects}).
Hence, the lowest priority task in ESAIL can easily starve when a deadline violation occurs due to high priority tasks having long execution times, which continuously occupy the processing core of ESAIL.
Regarding the other study subjects, the results show that the different operational settings of APS, i.e., adaptive partitions (Fig.~\ref{fig:rq1 PARTITION}), multiple policies (Fig.~\ref{fig:rq1 POLICY}), and multiple cores (Fig.~\ref{fig:rq1 MULTICORE}), could alleviate the problem of starvation.
Across all the subjects and deadline constraints, the experiment results obtained from SWEAK are significantly superior to those obtained from Baseline (i.e., $p$-value $<$ 0.05 and large effect sizes of $\hat{A}_{12}$~$\approx$~1.0) with regard to the best-sizes of safe WCET ranges.
The average execution times of SWEAK and Baseline are 9.37h and 7.55h, respectively, when both methods produce datasets of the same size.

\begin{table*}[h!t]
\caption{Summary of the numbers of simulation runs that violate any deadline constraints in (a)~ESAIL (b)~PARTITION, (c)~POLICY, and (d)~MULTICORE. EXP1 ran 40000 simulation runs for each subject with different test cases and WCET values within the best-size WCET ranges obtained from SWEAK and Baseline. Each table shows the max, median, min, and average simulation runs under different deadline constraints, i.e., the number of tolerable deadline misses $m$ $=$ 0, 1, 2, 3, or 4.}
\label{tab:ndeadlines}
\fontsize{10}{10}\selectfont
\def\arraystretch{1.5}

\newcolumntype{U}[1]{@{\hspace{#1}}r}
\newcommand{\cs}{0.5em}  

\newcommand\tbColTitle{
    \parbox{10em}{
        \centering\textbf{Number of tolerable deadline misses ($m$)}
    }
}
\newcommand\tbApprI{\rotatebox{90}{\parbox{4em}{\centering{\textbf{SWEAK}}}}}
\newcommand\tbApprII{\rotatebox{90}{\parbox{4em}{\centering{\textbf{Baseline}}}}}

\begin{center}
    \begin{subtable}[t]{0.49\columnwidth}\centering
        \resizebox{\columnwidth}{!}{\begin{tabular}[t]{m{0.2em}U{1em}U{\cs}U{\cs}U{\cs}U{\cs}U{\cs}}
			& & \multicolumn{5}{c@{}}{\tbColTitle} \\
			& & 0 & 1 & 2 & 3 & 4 \\ \toprule
			\multirow{4}{*}{\tbApprI}
				& Max & 19.00 & 0.00 & 0.00 & 0.00 & 0.00 \\
				& Median & 2.00 & 0.00 & 0.00 & 0.00 & 0.00 \\
				& Min & 0.00 & 0.00 & 0.00 & 0.00 & 0.00 \\
				& Average & 3.80 & 0.00 & 0.00 & 0.00 & 0.00 \\
                \midrule
			\multirow{4}{*}{\tbApprII}
				& Max & 574.00 & 0.00 & 0.00 & 0.00 & 0.00  \\
				& Median & 0.00 & 0.00 & 0.00 & 0.00 & 0.00 \\
				& Min & 0.00 & 0.00 & 0.00 & 0.00 & 0.00 \\
				& Average & 19.24 & 0.00 & 0.00 & 0.00 & 0.00  \\
                \midrule
			\multicolumn{2}{r}{$p$-value} & 0.1269 & 1.0000 & 1.0000 & 1.0000 & 1.0000  \\
			\multicolumn{2}{r}{$\hat{A}_{12}$}  & 0.5830 & 0.5000 & 0.5000 & 0.5000 & 0.5000  \\
	        \bottomrule
		\end{tabular}
        }
        \vspace{0.1em}
        \caption{ESAIL}
        \label{tab:ndeadlines ESAIL}
    \end{subtable}
    \hfill
    \begin{subtable}[t]{0.49\columnwidth}\centering
        \resizebox{\columnwidth}{!}{\begin{tabular}[t]{m{0.2em}U{1em}U{\cs}U{\cs}U{\cs}U{\cs}U{\cs}}
			& & \multicolumn{5}{c@{}}{\tbColTitle} \\
			& & 0 & 1 & 2 & 3 & 4 \\ \toprule
	        \multirow{4}{*}{\tbApprI}
			 		& Max 		& 17.00 & 13.00 & 2.00 & 2.00 & 3.00 \\
			        & Median 	& 0.00 & 0.00 & 0.00 & 0.00 & 0.00 \\
			        & Min 		& 0.00 & 0.00 & 0.00 & 0.00 & 0.00 \\
			        & Average 	& 2.60 & 0.90 & 0.06 & 0.08 & 0.10 \\
                \midrule
	        \multirow{4}{*}{\tbApprII}
			 		& Max 		& 409.00 & 333.00 & 128.00 & 180.00 & 345.00 \\
					& Median 	& 76.00 & 10.50 & 7.00 & 2.00 & 1.50 \\
					& Min 		& 0.00 & 0.00 & 0.00 & 0.00 & 0.00 \\
					& Average 	& 111.36 & 42.00 & 15.68 & 13.68 & 15.94 \\
                \midrule
	        \multicolumn{2}{r}{$p$-value}      & 0.0000 & 0.0000 & 0.0000 & 0.0000 & 0.0000  \\
	        \multicolumn{2}{r}{$\hat{A}_{12}$} & 0.0196 & 0.1604 & 0.1280 & 0.1840 & 0.2378 \\
	        \bottomrule
		\end{tabular}
        }
        \vspace{0.1em}
        \caption{PARTITION}
        \label{tab:ndeadlines PARTITION}
    \end{subtable}

    \vspace{1.5em}

    \begin{subtable}[t]{0.49\columnwidth}\centering
        \resizebox{\columnwidth}{!}{\begin{tabular}[t]{m{0.2em}U{1em}U{\cs}U{\cs}U{\cs}U{\cs}U{\cs}}
			& & \multicolumn{5}{c@{}}{\tbColTitle} \\
			& & 0 & 1 & 2 & 3 & 4 \\ 
			\toprule
			\multirow{4}{*}{\tbApprI}
				& Max & 29.00 & 20.00 & 1.00 & 1.00 & 0.00 \\
				& Median & 1.50 & 0.00 & 0.00 & 0.00 & 0.00 \\
				& Min & 0.00 & 0.00 & 0.00 & 0.00 & 0.00 \\
				& Average & 3.48 & 1.14 & 0.04 & 0.08 & 0.00 \\
                \midrule
			\multirow{4}{*}{\tbApprII}
				& Max & 705.00 & 199.00 & 137.00 & 46.00 & 216.00 \\
				& Median & 56.00 & 7.00 & 2.50 & 1.00 & 0.00 \\
				& Min & 0.00 & 0.00 & 0.00 & 0.00 & 0.00 \\
				& Average & 125.70 & 18.46 & 15.44 & 7.18 & 15.16 \\
                \midrule
			\multicolumn{2}{r}{$p$-value}      & 0.0000 & 0.0000 & 0.0000 & 0.0000 & 0.0000  \\
			\multicolumn{2}{r}{$\hat{A}_{12}$} & 0.1024 & 0.2130 & 0.1988 & 0.2316 & 0.3000  \\
	        \bottomrule
		  \end{tabular}
        }
        \vspace{0.1em}
        \caption{POLICY}
        \label{tab:ndeadlines POLICY}
    \end{subtable}
    \hfill
    \begin{subtable}[t]{0.49\columnwidth}\centering
        \resizebox{1.022\columnwidth}{!}{\begin{tabular}[t]{m{0.2em}U{1em}U{\cs}U{\cs}U{\cs}U{\cs}U{\cs}}
			& & \multicolumn{5}{c@{}}{\tbColTitle} \\
			& & 0 & 1 & 2 & 3 & 4 \\ 
			\toprule
			\multirow{4}{*}{\tbApprI}
				& Max & 146.00 & 73.00 & 7.00 & 21.00 & 5.00 \\
                    & Median & 27.00 & 0.00 & 0.00 & 0.00 & 0.00 \\
                    & Min & 0.00 & 0.00 & 0.00 & 0.00 & 0.00 \\
                    & Average & 35.60 & 2.68 & 0.34 & 1.08 & 0.20 \\
                \midrule
			\multirow{4}{*}{\tbApprII}
				& Max & 1495.00 & 848.00 & 339.00 & 444.00 & 220.00 \\
                    & Median & 192.50 & 13.50 & 12.00 & 6.50 & 2.50 \\
                    & Min & 0.00 & 0.00 & 0.00 & 0.00 & 0.00 \\
                    & Average & 266.60 & 58.04 & 52.46 & 25.98 & 27.66 \\
                \midrule
			\multicolumn{2}{r}{$p$-value}      & 0.0000 & 0.0000 & 0.0000 & 0.0000 & 0.0000  \\
			\multicolumn{2}{r}{$\hat{A}_{12}$} & 0.1320 & 0.0958 & 0.1556 & 0.2026 & 0.1972  \\
	        \bottomrule
		  \end{tabular}
        }
        \vspace{0.1em}
        \caption{MULTICORE}
        \label{tab:ndeadlines MULTICORE}
    \end{subtable}
\end{center}
\end{table*}
 
In addition, EXP1 evaluates the best-size WCET ranges obtained from 50 runs of SWEAK and Baseline using 40000 simulation runs by varying test cases (i.e., task arrivals and context switching times) and WCET values within the best-size WCET ranges. 
Table~\ref{tab:ndeadlines} shows the (maximum, median, minimum, and average) number of simulation runs (out of 40000 runs) in which any violation of deadline constraints occurred in the following subjects:
ESAIL~(Table~\ref{tab:ndeadlines ESAIL}), PARTITION~(Table~\ref{tab:ndeadlines PARTITION}), POLICY~(Table~\ref{tab:ndeadlines POLICY}), and MULTICORE~(Table~\ref{tab:ndeadlines MULTICORE}).
Once again, we vary the number of tolerable deadline misses $m$ in the experiments from 0 to 4.
The $p$-values and $\hat{A}_{12}$ values report the differences between the results obtained from 50 runs of both approaches.

As shown in Table~\ref{tab:ndeadlines}, SWEAK is significantly better ($p$-values are less than 0.05) with respect to violating deadline constraints than Baseline across all values of $m$ for the PARTITION, POLICY, and MULTICORE subjects.
The $\hat{A}_{12}$ values are also much lower than 0.5. Specifically, SWEAK shows smaller variation than Baseline (i.e., the differences between maximum and minimum values) in the number of simulation runs that violate deadline constraints. 
Regarding the ESAIL subject, both SWEAK and Baseline show similar results when $m$ $=$ 1, 2, 3, and 4 due to the starvation of the lowest priority task in ESAIL as discussed above. 
However, Baseline has a larger variation in the number of simulation runs that violate deadline constraints when ESAIL is subjected to the hard deadline constraint (i.e., $m$ $=$ 0). 
The results indicate that the best-size WCET ranges computed by SWEAK are more reliable in terms of violating deadline constraints than those computed by Baseline.

\begin{mdframed}[style=RQFrame]
	\emph{The answer to {\bf RQ1} is that} SWEAK significantly outperforms the baseline approach with respect to maximizing the hyperbox volume of the best-size WCET ranges under hard and weakly hard deadline constraints. 
	The best-size WCET ranges obtained by SWEAK have a significantly smaller chance of violating deadline constraints than the baseline approach. SWEAK takes on average 9.37h to compute the safe WCET ranges, while the baseline takes on average 7.53h, which indicates that SWEAK is acceptable for use in practice as an offline analysis tool.
\end{mdframed}

\begin{figure*}[t]
\begin{center}
\begin{subfigure}[t]{0.48\textwidth}\centering
		\includegraphics[width=\columnwidth]{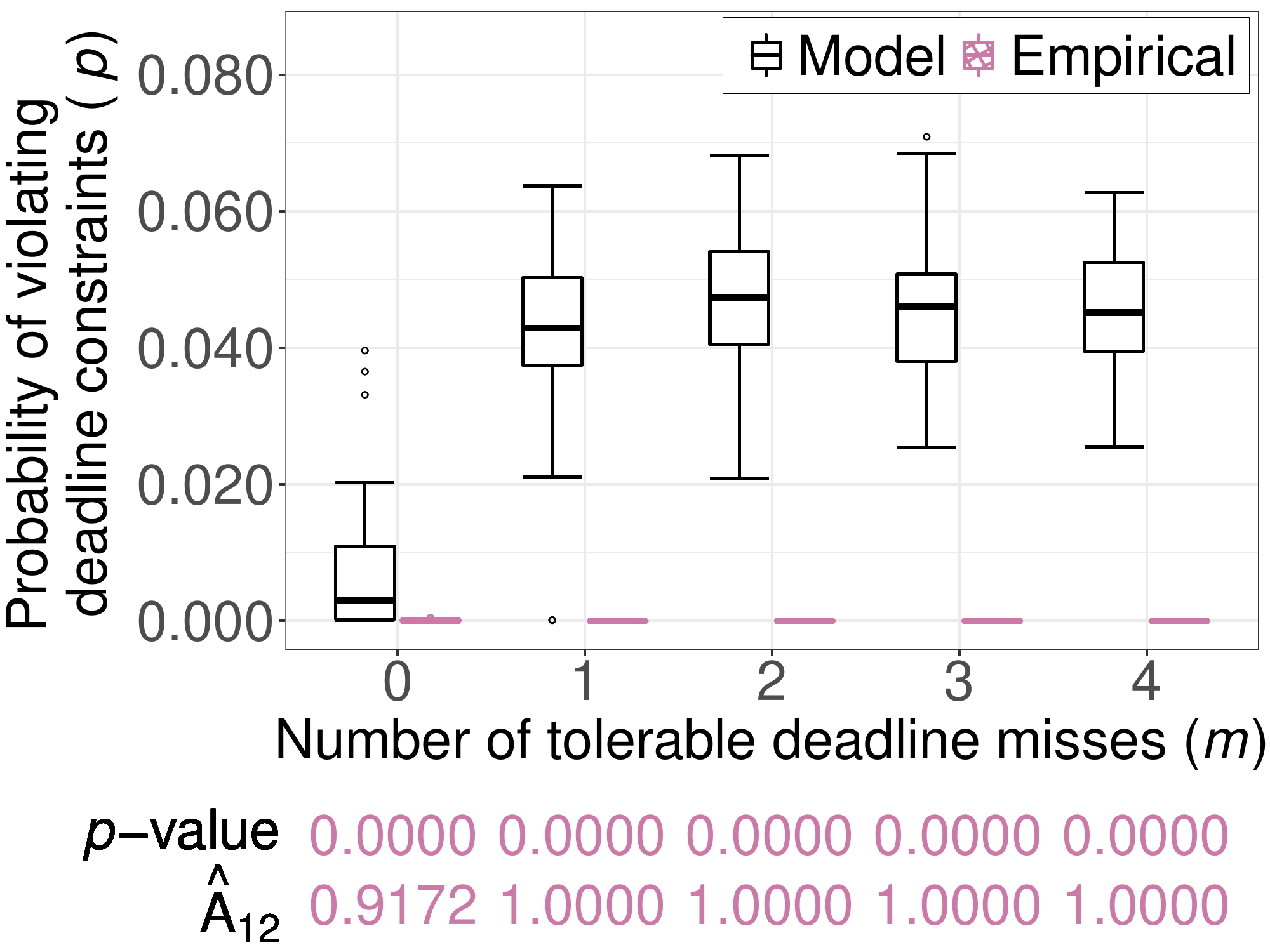}
		\caption{ESAIL}
		\label{fig:rq2 ESAIL}
	\end{subfigure}
	\hfill
	\begin{subfigure}[t]{0.48\textwidth}\centering
		\includegraphics[width=\columnwidth]{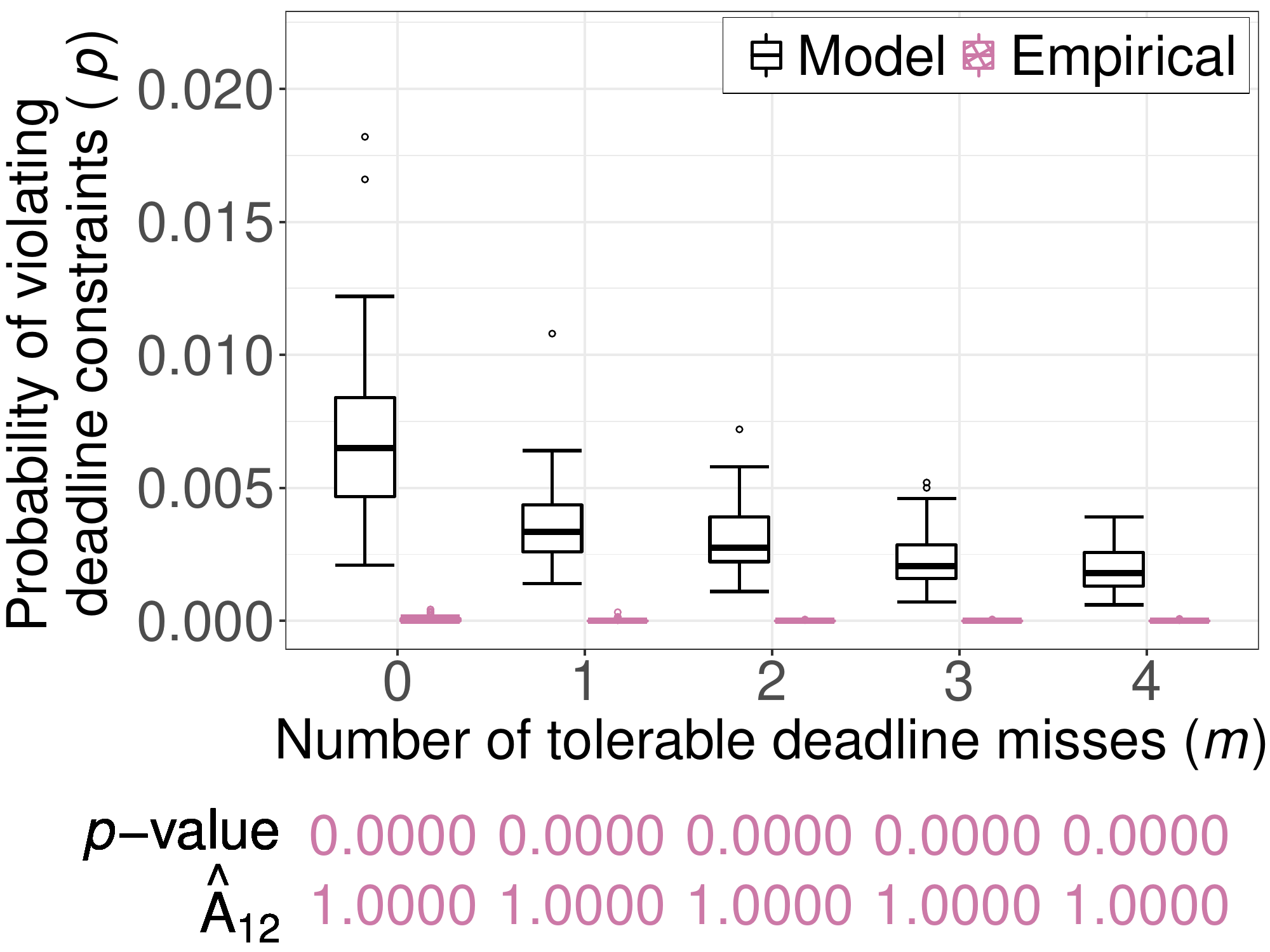}
		\caption{PARTITION}
		\label{fig:rq2 PARTITION}
	\end{subfigure}

	\bigskip

	\begin{subfigure}[t]{0.48\textwidth}\centering
		\includegraphics[width=\columnwidth]{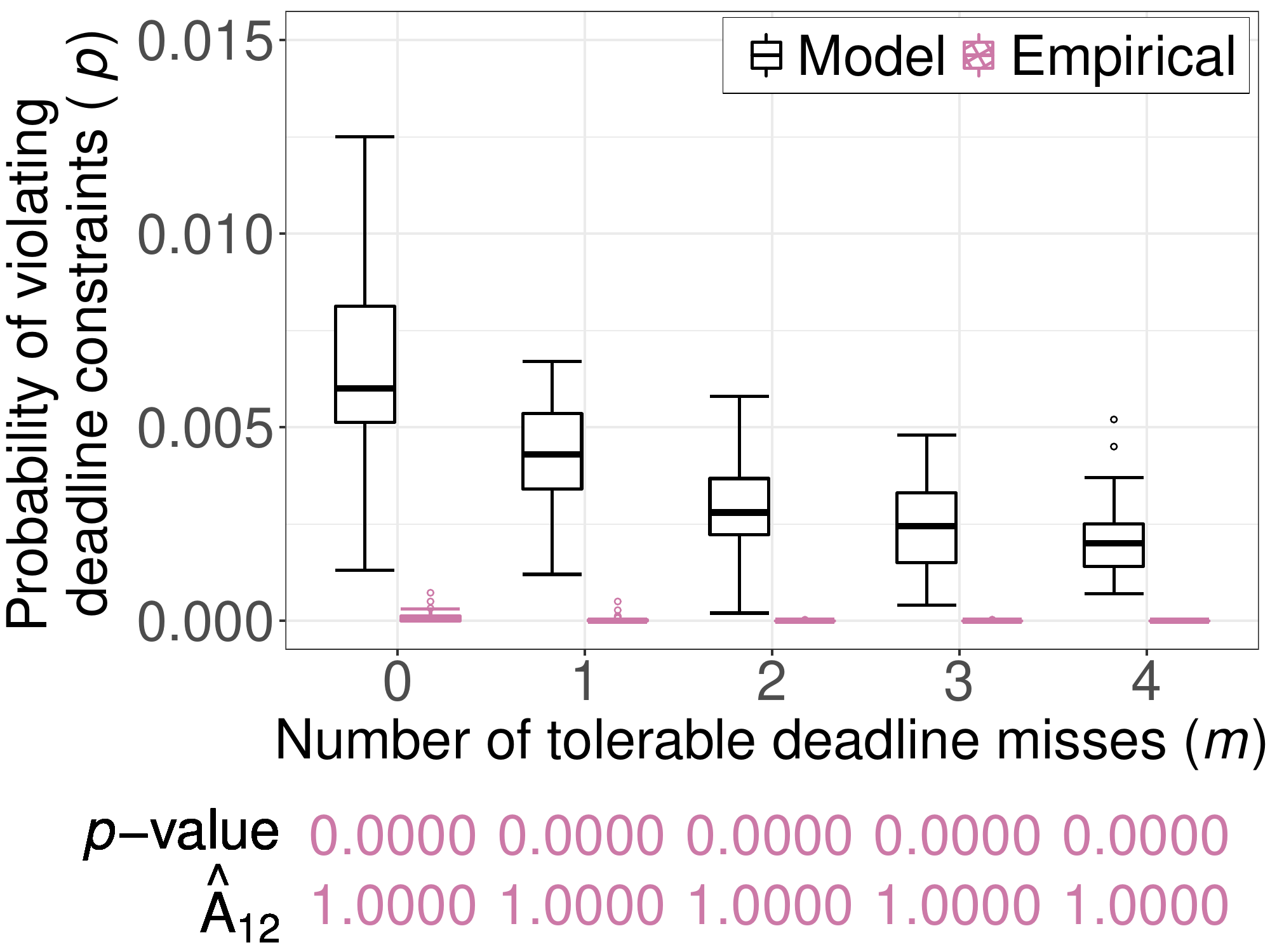}
		\caption{POLICY}
		\label{fig:rq2 POLICY}
	\end{subfigure}
	\hfill
	\begin{subfigure}[t]{0.48\textwidth}\centering
		\includegraphics[width=\columnwidth]{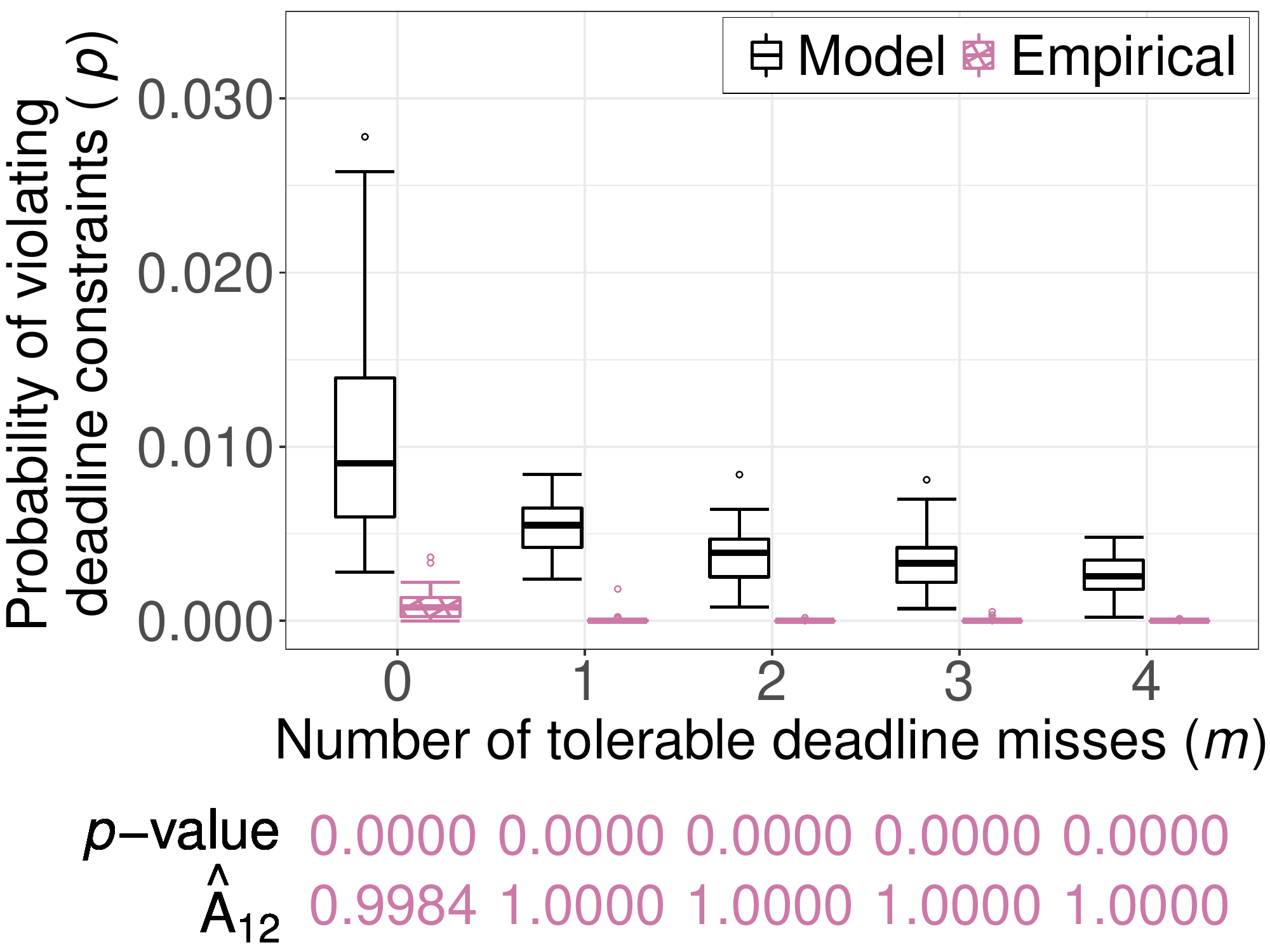}
		\caption{MULTICORE}
		\label{fig:rq2 MULTICORE}
	\end{subfigure}
\end{center}
\caption{Distributions of empirical probabilities and model probabilities across different numbers of tolerable deadline misses $m$ in (a)~ESAIL, (b)~PARTITION, (c)~POLICY, and (d)~MULTICORE. The boxplots (25\%-50\%-75\%) show distributions of probabilities for the best-size WCET ranges obtained from 50 runs of SWEAK.}
\label{fig:rq2}
\end{figure*}

\textbf{RQ2.}
Fig.~\ref{fig:rq2} depicts the results of EXP2 for all subjects: ESAIL (Fig.~\ref{fig:rq2 ESAIL}), PARTITION (Fig.~\ref{fig:rq2 PARTITION}), POLICY (Fig.~\ref{fig:rq2 POLICY}), and MULTICORE (Fig.~\ref{fig:rq2 MULTICORE}).
Each sub-figure compares the model probability (computed by SWEAK's logistic regression) and empirical probability (computed by simulations) of violating deadline constraints by varying the number of tolerable deadline misses $m$. 
Each boxplot in the figures shows the distributions (25\%-50\%-75\% quartiles) of model probabilities and empirical probabilities for the best-size WCET ranges obtained from 50 runs of SWEAK. 
As shown in Fig.~\ref{fig:rq2}, the empirical probabilities across all values of $m$ and all the subjects are significantly smaller than the model probabilities. Statistical comparisons show that all the $p$-values are 0 and all the $\hat{A}_{12}$ values are approximately 1.
Recall from Section~\ref{sec:approach} that SWEAK infers a logistic regression model with a probability of violating deadline constraints based on the labeled dataset that is generated by evaluating the worst-case task arrivals and context switching times, i.e., test cases. SWEAK thus infers the model that fits the worst-case test cases. 
However, SWEAK shows higher probabilities than the empirical probabilities, which are computed by running simulations with random test cases and random WCET values within the best-point WCET ranges obtained by SWEAK.
Hence, a logistic regression model produced by SWEAK allows engineers to probabilistically interpret the safe WCET ranges in a more conservative manner than evaluating the WCET ranges by simulations. Such results indicate that we can expect the actual probability of violating deadline constraints to be lower than the model probability determined by SWEAK. Note that such conservative interpretations of WCET ranges are desirable in practice.

Regarding the trend for model probabilities over the number $m$ of tolerable deadline misses, ESAIL (Fig~\ref{fig:rq2 ESAIL}) shows similar probability distributions when $m > 0$.
The probabilities in these distributions are higher than that obtained when $m = 0$.
This trend contrasts with that of the other three subjects (Figs.~\ref{fig:rq2 PARTITION}, \ref{fig:rq2 POLICY}, and \ref{fig:rq2 MULTICORE}), where model probabilities decrease with an increasing  number of tolerable deadline misses.
In ESAIL, this particular trend is caused by its characteristics (Section~\ref{subsec:subjects}), which make it prone to starvation, as described earlier (see the RQ1 results).
Hence, when high-priority tasks with long execution times continuously occupy the processing core of ESAIL, low-priority tasks with short execution times will likely encounter consecutive deadline misses due to starvation.
However, regarding the other three subjects that operate under the different settings of APS, i.e., adaptive partitions (Fig.~\ref{fig:rq2 PARTITION}), multiple policies (Fig.~\ref{fig:rq2 POLICY}), and multiple cores (Fig.~\ref{fig:rq2 MULTICORE}), we did not observe starvation problems that are likely to lead to consecutive deadline misses.

\begin{mdframed}[style=RQFrame]
	\emph{The answer to {\bf RQ2} is that} SWEAK estimates higher probabilities of violating deadline constraints for the safe WCET ranges than empirical probabilities computed by simulation-based evaluations for the ranges. SWEAK, therefore, enables conservative probabilistic interpretations of safe WCET ranges. 
\end{mdframed}

\begin{figure*}[ht]
\begin{center}
\begin{subfigure}[t]{0.4\textwidth}\centering
	    \includegraphics[width=\columnwidth]{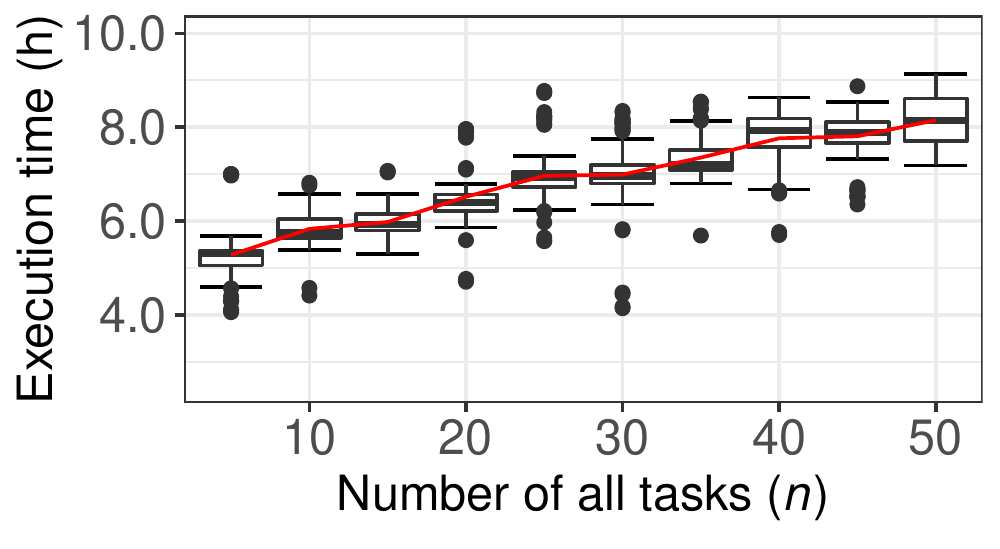}
		\caption{Number of all tasks ($n$)}
	    \label{fig:rq3 w1}
	\end{subfigure}
	\hspace{3em}
	\begin{subfigure}[t]{0.4\textwidth}\centering
	    \includegraphics[width=\columnwidth]{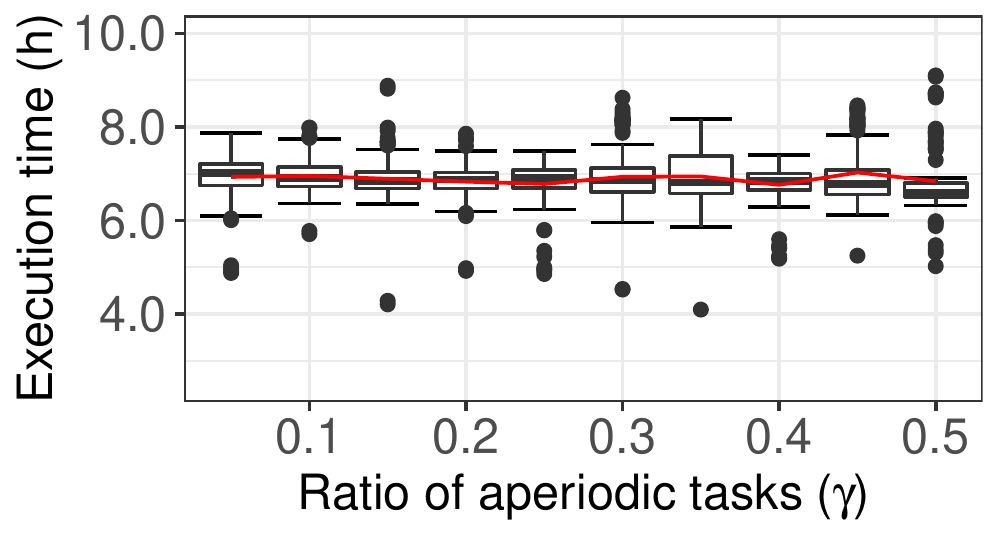}
		\caption{Ratio of aperiodic tasks ($\gamma$)}
	    \label{fig:rq3 w2}
	\end{subfigure}

	\bigskip

	\begin{subfigure}[t]{0.4\textwidth}\centering
	    \includegraphics[width=\columnwidth]{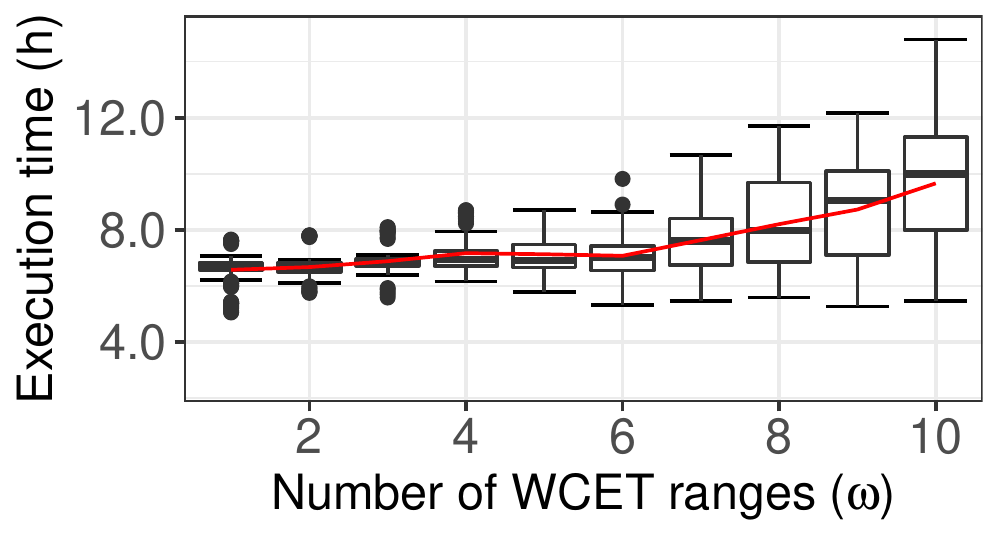}
		\caption{Number of WCET ranges ($\omega$)}
	    \label{fig:rq3 w3}
	\end{subfigure}
	\hspace{3em}
	\begin{subfigure}[t]{0.4\textwidth}\centering
	    \includegraphics[width=\columnwidth]{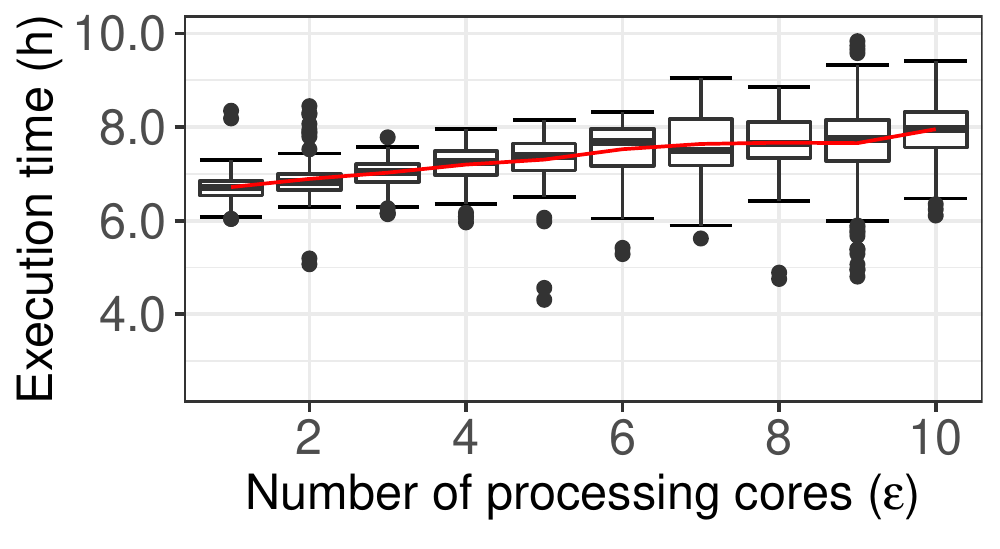}
		\caption{Number of processing cores ($\epsilon$)}
	    \label{fig:rq3 w4}
	\end{subfigure}

	\bigskip

	\begin{subfigure}[t]{0.4\textwidth}\centering
	    \includegraphics[width=\columnwidth]{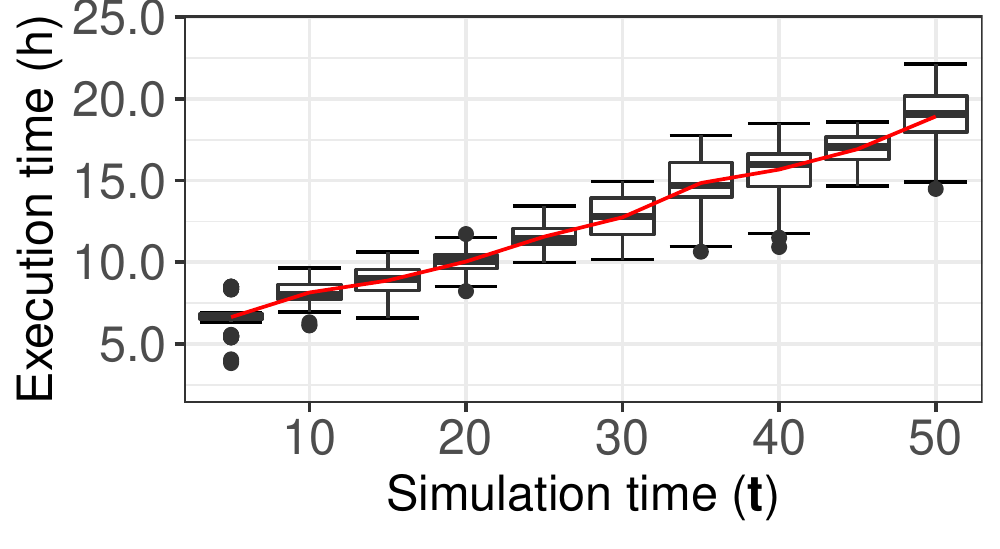}
		\caption{Simulation time ($\mathbf{t}$)}
	    \label{fig:rq3 w5}
	\end{subfigure}
	\hspace{3em}
	\begin{subfigure}[t]{0.4\textwidth}\centering
	    \includegraphics[width=\columnwidth]{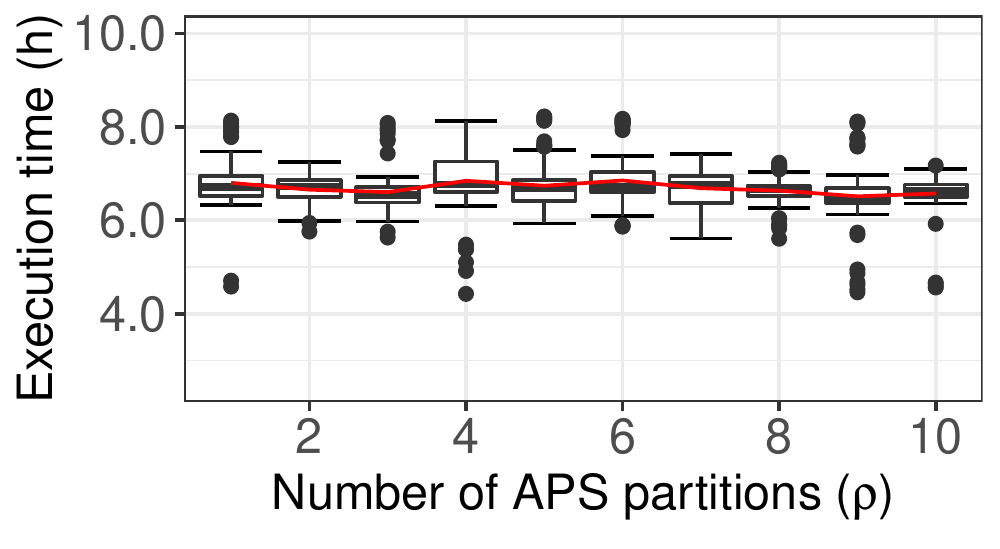}
		\caption{Number of APS partitions ($\rho$)}
	    \label{fig:rq3 w6}
	\end{subfigure}
\end{center}
\caption{Execution times of SWEAK when varying the values of the following parameters: (a)~number of all tasks $n$, (b)~ratio of aperiodic tasks $\gamma$, (c)~number of WCET ranges $\omega$, (d)~number of processing cores $\epsilon$, (e)~simulation time $\mathbf{t}$, and (f)~number of APS partitions $\rho$.
Each boxplot (25\%-50\%-75\%) shows the distributions of 100 execution time values measured from 10 runs of SWEAK for 10 synthetic systems with the same configuration. The red line in each figure represents the trend of mean values of the execution times over the parameter values.}
\label{fig:rq3}
\end{figure*}

\textbf{RQ3.} Fig.~\ref{fig:rq3} shows the results of EXP3 that present the distributions (boxplots) of execution times obtained from 10 $\times$ 10 runs of SWEAK, i.e., 10 runs for each synthetic system with the same experimental setting (see Section~\ref{subsec:design}). 
The red solid lines represent the changes in mean values of the execution times while varying the following control parameters: 
(a)~number of tasks $n$, 
(b)~ratio of aperiodic tasks $\gamma$, 
(c)~number of WCET ranges $\omega$, 
(d)~number of processing cores $\epsilon$, 
(e)~simulation time $\mathbf{t}$, and 
(f)~number of APS partitions $\rho$. 
The experiments in EXP3 took 22.1 hours at most, which is acceptable as an offline analysis technique.
As shown in Fig.~\ref{fig:rq3 w1}, Fig.~\ref{fig:rq3 w4}, and Fig.~\ref{fig:rq3 w5}, there are positive linear relationships between SWEAK's execution times and the parameters, $n$, $\epsilon$, and $\mathbf{t}$. 
Thus, we expect SWEAK to scale well as the number of tasks, the number of processing cores, and the simulation time increase.
Further, for parameters such as $\gamma$ (Fig.~\ref{fig:rq3 w2}) and $\rho$ (Fig.~\ref{fig:rq3 w6}), there is no correlation with execution time.

In contrast, parameter $\omega$ (Fig.~\ref{fig:rq3 w3}, number of WCET ranges) is quadratically related with the execution time of SWEAK. Recall from Section~\ref{subsec:learning} that, to build the equation used in logistic regression, we leverage the second-order polynomial RSM that contains linear, quadratic, and two-way interaction terms. The total number of terms is determined by the number of WCET ranges $\omega$, which is $1 + \omega + \omega + \binom{\omega}{2}$.
The logistic regression algorithm infers the coefficients for each term of RSM and is used during the execution of SWEAK. 
As the execution time of logistic regression is linearly related to the number of coefficients~\cite{Hosmer2013}, the execution time of SWEAK is quadratically correlated with the number of WCET ranges.
Moreover, as the number of WCET ranges $\omega$ increases, the variance of execution time becomes larger, as visible in Fig.~\ref{fig:rq3 w3}. 
We found that this phenomenon occurs due to feature reduction and stepwise regression in the learning step (see Section~\ref{subsec:learning}). 
These techniques output an equation consisting of terms that are considered highly related to the violation of deadline constraints.
Depending on the system characteristics, the synthetic systems generated by setting $\omega = 10$ have more diverse equations than those generated with $\omega = 1$.
This output affects the execution time for sampling WCET values (distance-based) and building logistic regression models.

\begin{mdframed}[style=RQFrame]
	\emph{The answer to {\bf RQ3} is that} SWEAK's execution time is linearly related to the number of tasks, the number of processing cores, and the simulation time. 
	However, execution time has a quadratic correlation with the number of tasks whose WCETs are defined as ranges. 
	Overall, SWEAK is applicable in practice as an offline analysis technique since it took at most 22.1h in our experiments, which is generally acceptable.
\end{mdframed}

\textbf{Usefulness of SWEAK from the perspective of practitioners.}
To understand the practical usefulness of SWEAK, we discussed it with two practitioners at Blackberry with whom we are closely collaborating.
The feedback we received is as follows: (1)~practitioners perceived that the test cases, i.e., worst-case task arrivals and context switching times, produced by SWEAK help them conduct further analysis of their systems' schedulability, (2)~practitioners agreed that SWEAK enables trade-off analysis by using a logistic regression model, and (3)~practitioners perceived that SWEAK can be applied to their customers' systems since it uses an industry-strength simulator (APS simulator)  and supports weakly hard real-time tasks.

As a company that develops a real-time operating system, Blackberry has a long-term goal of providing its customers with a schedulability analysis tool for systems with heterogeneous characteristics such as multiple policies, partitions, and weakly hard deadline constraints. SWEAK is a practical candidate solution as it is not only an industrial simulation-based approach but it also addresses both tasks with hard and weakly hard deadline constraints. 
Although we have not conducted user studies, given the positive feedback from Blackberry, we believe SWEAK can be practically applicable and is worthy of further research.

 \subsection{Threats to validity}
\label{subsec:threats}

\textbf{Internal validity.}
An internal validity threat is the randomness of SWEAK.
To mitigate this threat, we compared SWEAK against a baseline method under identical parameter settings, and ran both approaches 50 times for each experiment setting. We then performed statistical comparisons using the Mann-Whitney U-test and Vargha and Delaney’s $\hat{A}_{12}$.

Another threat to internal validity is related to the configuration of the experiments.
As SWEAK uses a multi-objective search algorithm, there are many parameters that need to be optimized to find best solutions.
In our experiments, we configured the parameters' values based on guidelines~\cite{Haupt1988} and preliminary experiments (see Section~\ref{subsec:param}).
Even though, to improve the performance of SWEAK, these values could be further tuned for different study subjects, our results clearly show that SWEAK is a promising solution.

Regarding NSGA-II, which SWEAK employs for searching worst-case test cases, \citet{ByersCD15} found that when the number of solution elements can freely evolve (i.e., variable-length genome), the crowding distance operator of NSGA-II prioritizes solutions with fewer elements. 
Hence, the search of NSGA-II is biased towards regions containing such solutions.
However, in SWEAK, the size of a test case cannot freely evolve during search. In fact, the size of a test case is bounded within a range determined by the minimum and maximum task arrivals during the given simulation time.
Furthermore, \citet{ByersCD15} observed this finding from the application of NSGA-II to address the remote data mirroring problem, which differs significantly from the problem addressed by SWEAK.
Hence, further study is needed to investigate the extent to which bounded-variable-length genomes impact the performance of NSGA-II in general application contexts.
In addition, considering other multi-objective search algorithms~\cite{Luke2013} can be beneficial to enhance the performance of SWEAK. 
Even though these research directions are interesting (but outside the scope of this article), our results nevertheless indicate that SWEAK significantly outperforms the baseline approach and estimates safe WCET ranges with a high degree of confidence in practical time.

\textbf{External validity.}
The main threat to external validity is that our results may not be generalizable to other contexts.
SWEAK explicitly targets WCET ranges that are estimated at early design stages when actual executions of the systems are not feasible.
Hence, SWEAK was evaluated with an industrial system (i.e., ESAIL) from the satellite domain, using its task design descriptions.
In addition, we applied SWEAK to a large number of synthetic systems generated by following BlackBerry's guidelines to ensure they were realistic and representative.
We precisely described the method used to generate synthetic systems (see Section~\ref{subsec:synthetic}). We have also made these systems available online~\cite{Artifacts3}.
Even though our results, based on simulations, show that SWEAK is applicable at early design stages for probabilistically estimating safe WCET ranges for weakly hard real-time systems, further studies remain necessary to evaluate whether or not the estimated safe WCET ranges are useful for engineers during later development stages, when tasks' implementations are available.  
Indeed, SWEAK is also applicable at later stages of development for testing the schedulability of systems and for providing more precise WCET ranges.
Furthermore, additional study systems in other domains should be further investigated to evaluate the general usefulness of SWEAK.

  \section{Related works}
\label{sec:relatedworks}

This section discusses previous studies on WCET estimation in real-time systems, schedulability analysis of real-time systems with weakly hard real-time tasks, and industry-strength task schedulers (e.g., APS).
In addition, we discuss existing work that relies on search techniques for analyzing real-time systems.
To our knowledge, no previous work investigated the probabilistic estimation of WCET ranges accounting for weakly hard real-time tasks and industry-strength task schedulers.

\textbf{WCET estimation in real-time systems.} 
Measurement-based methods to estimate WCETs are widely studied~\cite{Burns2000,Wenzel2005,Hansen2009,Cucu2012,Berezovskyi2014,Abella2017} and commonly used in practice~\cite{Akesson2020}. 
The basic idea of such methods is to run several task executions on the targeted hardware using various input data. 
To obtain tasks' input data, \citet{Wenzel2005} proposed a method that analyzes the execution paths of source code. 
\citet{Burns2000} proposed a probabilistic WCET estimation approach that applies statistical analysis to find worst-case input data. 
However, as they need executable source code and target hardware, these approaches are only applicable for systems at later stages of development. 

To not rely on target hardware when estimating WCET values, static analysis approaches~\cite{Ferdinand1998, Theiling2000, Mueller2000, Hardy2011} have been proposed. 
These approaches estimate WCET values based on an abstract model of the target hardware and software structure analysis. 
For example, some prior studies in these research strands rely on models of cache behaviors~\cite{Theiling2000, Mueller2000, Hardy2011} or timing models of hardware instructions~\cite{Gustafsson2009, Altenbernd2016, Bonenfant2017}.
However, these approaches still require source code; hence, they are not applicable at early design stages. 

In contrast to the previous studies that aim at estimating the WCETs of tasks regardless of their schedulability, SWEAK targets early stages of development, taking estimated, conservative WCET ranges as input. 
SWEAK then finds restricted safe WCET sub-ranges corresponding to a probability of violating deadline constraints, relying on a multi-objective search algorithm, simulation, feature reduction, a dedicated sampling strategy, and logistic regression.
SWEAK enables trade-off analysis of tasks' WCET values and allows practitioners to select an appropriate violation probability depending on the context.

\textbf{Schedulability analysis with weakly hard real-time tasks.}
To ensure quality of service (QoS) in real-world systems that can tolerate occasional deadline misses, \citet{Bernat2001} introduced the concept of weakly hard real-time systems and precisely defined the concept of weakly hard deadline constraints adopted in SWEAK.
The follow-up studies~\cite{Xu2015,Pazzaglia2021} on weakly hard real-time systems introduced analytical methods to analyze the schedulability of weakly hard real-time systems.

\citet{Xu2015} proposed a deadline miss model for computing the number of potential deadline misses within consecutive task arrivals when a task is under unexpected overload caused by task arrivals triggered through external events.
In particular, they introduced task arrival models that capture arrival patterns of both periodic and sporadic tasks, enabling their analysis method to account for possible overload situations.
Their analysis method relies on integer linear programming (ILP) and works under the assumption that tasks are running on a single-core platform with the fixed-priority preemptive or non-preemptive scheduling policy.

\citet{Pazzaglia2021} presented a schedulability analysis method for weakly hard real-time systems by accounting for free offsets and release jitters using a mixed integer linear programming (MILP) formulation.
Their analysis method checks whether or not a task satisfies its ($m, K$)-constraint to ensure that the task has no more than $m$ deadline misses out of any $K$ consecutive arrivals.
This MILP-based method works under the assumption that a real-time system is composed of periodic tasks scheduled by fixed-priority with preemption on a single-core platform.

In contrast to these analytical schedulability analysis methods that takes fixed WCETs as input, SWEAK addresses the problem of probabilistically estimating safe WCET ranges that satisfy weakly-hard deadline constraints.
SWEAK relies on logistic regression, search, and an industrial scheduling simulator, enabling it to analyze more complex systems involving uncertainties in task arrivals, context switching time, multiple processing cores, and advanced scheduling policies such as APS.
The complex relationships among these system characteristics are difficult to capture in a (mixed) ILP formulation.

\textbf{Adaptive partitioning scheduling (APS).} 
Due to the increasing complexity of real-time systems, several APS approaches have been proposed~\cite{Bletsas2009,Massa2016,Abeni2020}. 
These approaches combine global scheduling, which allows for the free migration of tasks, and static partitioned scheduling policies. 
\citet{Bletsas2009} proposed a semi-adaptive partitioning scheduling approach that first allocates heavy-load tasks with utilization higher than 50\% to processors. The scheduling approach then assigns the remaining tasks to the available time slots on each processor.
\citet{Massa2016} designed an online adaptive partitioning scheduling approach for a selected group of tasks based on offline analysis of their requirements. The groups are scheduled by global-like or partitioned scheduling methods depending on system load. 
\citet{Abeni2020} introduced an approach that harmonizes global scheduling and static partitioned scheduling policies. The approach operates in partitioned mode and, if a system workload is over a specified threshold, it switches to global mode and then rapidly returns to the partitioned mode.
Unlike these studies, BlackBerry provides an RTOS (i.e., QNX Neutrino) equipped with APS that supports the dynamic partition budget management for industrial use.

Even though the usage of APS for QNX Neutrino has been increasing, only a few schedulability analysis studies have recently been conducted for APS~\cite{Dasari2021,Dasari2022}.
Dasari et al.~\cite{Dasari2021} found that APS has drawbacks in partition configurations. Hence, they provided some guidelines to help engineers properly configure partitions.
Dasari et al.~\cite{Dasari2022} investigated the APS behavior in practice and proposed a technique that verifies the end-to-end delay (i.e., schedulability) of event chains, which are the sequences of tasks activated by events, on a real-time system using APS. 
The technique employs a response time analysis of the chains at later development stages of systems containing periodic tasks with fixed WCET values.
SWEAK complements these APS-related research strands as an analysis tool for estimating safe WCET ranges at early design stages. It helps engineers design their systems with safe WCET ranges as objectives with a certain level of confidence.

\textbf{Search-based analysis in real-time systems.} 
In real-time systems, search-based techniques are used in many prior approaches that aim at testing the systems~\cite{Wegener1997,Wegener1998,Lin2009,Arcuri2010,Shin2018}.
For example, \citet{Wegener1997} introduced a testing approach that relies on a genetic algorithm aiming at checking whether or not the system violates its timing constraints.
Specifically, this prior work checks the timing constraints with regard to maximum and minimum execution times measured in processor cycles.
\citet{Lin2009} proposed a search-based approach to check whether a real-time system satisfies both its deadline and security constraints.
Regarding security constraints, their work aims at optimally using security services for confidentiality, integrity, and authentication, while ensuring the schedulability of real-time tasks.
\citet{Shin2018} presented a test case prioritization approach based on a multi-objective search algorithm.
This prior work, specifically, focused on studying the test case prioritization problem in the context of acceptance testing for cyber-physical real-time systems.
In this context, their approach accounts for not only the criticality levels of test cases but also the risk of hardware damage posed by executing the test cases.

Beyond testing real-time systems, recently, \citet{Lee2022b} proposed OPAM, an optimal task priority assignment method for real-time systems.
OPAM uses a multi-objective, competitive coevolutionary search algorithm to find near-optimal priority assignments that maximize the magnitude of safety margins and the extent to which engineering constraints are satisfied.
\citet{Lee2022a} introduced SAFE, a tool that provides probabilistic estimation of safe WCET ranges for real-time systems.
As described in Section~\ref{sec:approach}, however, SAFE is not applicable to weakly hard real-time systems, as SAFE accounts for hard deadline constraints that do not allow any occurrence of a deadline miss and relies on a simple task model. 
SWEAK extends SAFE to account for weakly hard deadline constraints, context switching times, and the APS policy.

In contrast to these prior works, SWEAK is the first attempt to address the problem of probabilistically estimating safe WCET ranges for weakly hard real-time systems. 
Furthermore, SWEAK accounts for multiple objectives to generate test cases that likely violate weakly hard deadline constraints and maximize the magnitude of deadline misses.
SWEAK relies on logistic regression, adapted from SAFE, to infer safe WCET ranges in the form of safe WCET border with a probability of violating weakly hard deadline constraints. This enables engineers to investigate suitable WCET values by analyzing trade-offs within the safe ranges.

 \section{Conclusion}
\label{sec:conclusion}
This article introduced SWEAK, which probabilistically estimates the safe WCET ranges of tasks for weakly hard real-time systems at early design stages.
SWEAK employs a multi-objective search algorithm to find test cases (i.e., worst-case task arrivals and context switching times), aiming to maximize the magnitude of deadline misses and the degree of consecutive deadline misses.
Based on the search results, SWEAK infers safe WCET ranges, for a probability of violating deadline constraints, using logistic regression.
We evaluated SWEAK on a mission-critical real-time satellite system and several synthetic systems created by following the guidelines provided by BlackBerry.
The results indicate that SWEAK provides high flexibility in selecting WCET ranges for practitioners. We further evaluated SWEAK through 600 synthetic systems with different characteristics, varying their degree of complexity. The results show that SWEAK scales to complex systems. Furthermore, SWEAK completed all experiments within at most 22.1h. Hence, SWEAK is acceptable in practice as an offline analysis technique for estimating safe WCET ranges given a probability of violating deadline constraints.

For future directions of this research, we plan to develop a real-time task modeling language that facilitates schedulability analysis and represents task constraints and system behaviors.
In addition, the usefulness of SWEAK should be further validated with other case studies from different domains and through user studies.
In particular, it is important to assess the usefulness of SWEAK from the perspective of practitioners. We plan to investigate how SWEAK assists engineers by providing safe WCET estimates, which, in turn, guide engineers in making appropriate decisions during the development process.

 \begin{acks}
This work was supported by Mitacs through the Mitacs Accelerate program (IT23234), the European Research Council (ERC) under the European Union's Horizon 2020 research and innovation programme (grant agreement No 694277), and NSERC of Canada under the Discovery and CRC programs. The experiments presented in this paper were carried out using the HPC facilities of the University of Luxembourg ~\cite{Varrette2014}{\small -- see \href{http://hpc.uni.lu}{hpc.uni.lu}}. We thank Chris Hobbs and Alexandre Koppel from BlackBerry for their help in identifying the problem based on practitioners' needs and conducting realistic case studies.
\end{acks} 

\bibliographystyle{ACM-Reference-Format}

\begin{thebibliography}{77}



\ifx \showCODEN    \undefined \def \showCODEN     #1{\unskip}     \fi
\ifx \showDOI      \undefined \def \showDOI       #1{#1}\fi
\ifx \showISBNx    \undefined \def \showISBNx     #1{\unskip}     \fi
\ifx \showISBNxiii \undefined \def \showISBNxiii  #1{\unskip}     \fi
\ifx \showISSN     \undefined \def \showISSN      #1{\unskip}     \fi
\ifx \showLCCN     \undefined \def \showLCCN      #1{\unskip}     \fi
\ifx \shownote     \undefined \def \shownote      #1{#1}          \fi
\ifx \showarticletitle \undefined \def \showarticletitle #1{#1}   \fi
\ifx \showURL      \undefined \def \showURL       {\relax}        \fi
\providecommand\bibfield[2]{#2}
\providecommand\bibinfo[2]{#2}
\providecommand\natexlab[1]{#1}
\providecommand\showeprint[2][]{arXiv:#2}

\bibitem[Abella et~al\mbox{.}(2017)]{Abella2017}
\bibfield{author}{\bibinfo{person}{Jaume Abella}, \bibinfo{person}{Maria
  Padilla}, \bibinfo{person}{Joan~Del Castillo}, {and}
  \bibinfo{person}{Francisco~J. Cazorla}.} \bibinfo{year}{2017}\natexlab{}.
\newblock \showarticletitle{Measurement-Based Worst-Case Execution Time
  Estimation Using the Coefficient of Variation}.
\newblock \bibinfo{journal}{\emph{ACM Transactions Design Automation of
  Electronic Systems}} \bibinfo{volume}{22}, \bibinfo{number}{4}, Article
  \bibinfo{articleno}{72} (\bibinfo{date}{Jun} \bibinfo{year}{2017}),
  \bibinfo{numpages}{29}~pages.
\newblock


\bibitem[Abeni and Cucinotta(2020)]{Abeni2020}
\bibfield{author}{\bibinfo{person}{Luca Abeni} {and} \bibinfo{person}{Tommaso
  Cucinotta}.} \bibinfo{year}{2020}\natexlab{}.
\newblock \showarticletitle{{Adaptive partitioning of real-time tasks on
  multiple processors}}. In \bibinfo{booktitle}{\emph{Proceedings of the ACM
  Symposium on Applied Computing (SAC'17)}}. \bibinfo{publisher}{{ACM}},
  \bibinfo{address}{New York, NY, USA}, \bibinfo{pages}{572--579}.
\newblock


\bibitem[Akesson et~al\mbox{.}(2020)]{Akesson2020}
\bibfield{author}{\bibinfo{person}{Benny Akesson}, \bibinfo{person}{Mitra
  Nasri}, \bibinfo{person}{Geoffrey Nelissen}, \bibinfo{person}{Sebastian
  Altmeyer}, {and} \bibinfo{person}{Robert~I. Davis}.}
  \bibinfo{year}{2020}\natexlab{}.
\newblock \showarticletitle{{An Empirical Survey-based Study into Industry
  Practice in Real-time Systems}}. In \bibinfo{booktitle}{\emph{Proceedings of
  the 2020 IEEE Real-Time Systems Symposium (RTSS'20)}},
  Vol.~\bibinfo{volume}{2020-Decem}. \bibinfo{pages}{3--11}.
\newblock


\bibitem[AlEnawy and Aydin(2005)]{Alenawy2005}
\bibfield{author}{\bibinfo{person}{T.A. AlEnawy} {and} \bibinfo{person}{H.
  Aydin}.} \bibinfo{year}{2005}\natexlab{}.
\newblock \showarticletitle{Energy-constrained scheduling for weakly-hard
  real-time systems}. In \bibinfo{booktitle}{\emph{Proceedings of the 26th IEEE
  International Real-Time Systems Symposium (RTSS'05)}}.
  \bibinfo{pages}{376--385}.
\newblock


\bibitem[Altenbernd et~al\mbox{.}(2016)]{Altenbernd2016}
\bibfield{author}{\bibinfo{person}{Peter Altenbernd}, \bibinfo{person}{Jan
  Gustafsson}, \bibinfo{person}{Bj{\"{o}}rn Lisper}, {and}
  \bibinfo{person}{Friedhelm Stappert}.} \bibinfo{year}{2016}\natexlab{}.
\newblock \showarticletitle{Early Execution Time-Estimation Through
  Automatically Generated Timing Models}.
\newblock \bibinfo{journal}{\emph{Real-Time Systems}} \bibinfo{volume}{52},
  \bibinfo{number}{6} (\bibinfo{year}{2016}), \bibinfo{pages}{731--760}.
\newblock


\bibitem[Altmeyer and Gebhard(2008)]{Altmeyer2008}
\bibfield{author}{\bibinfo{person}{Sebastian Altmeyer} {and}
  \bibinfo{person}{Gernot Gebhard}.} \bibinfo{year}{2008}\natexlab{}.
\newblock \showarticletitle{WCET Analysis for Preemptive Scheduling}. In
  \bibinfo{booktitle}{\emph{{Proceeding of the 8th International Workshop
  Worst-Case Execution Time Analysis (WCET'08)}}}. \bibinfo{publisher}{Schloss
  Dagstuhl}, \bibinfo{address}{Dagstuhl, Germany}, \bibinfo{pages}{105--112}.
\newblock


\bibitem[Arcuri and Briand(2014)]{Arcuri2014}
\bibfield{author}{\bibinfo{person}{Andrea Arcuri} {and}
  \bibinfo{person}{Lionel~C. Briand}.} \bibinfo{year}{2014}\natexlab{}.
\newblock \showarticletitle{A Hitchhiker's Guide to Statistical Tests for
  Assessing Randomized Algorithms in Software Engineering}.
\newblock \bibinfo{journal}{\emph{Software Testing, Verification and
  Reliability}} \bibinfo{volume}{24}, \bibinfo{number}{3}
  (\bibinfo{year}{2014}), \bibinfo{pages}{219--250}.
\newblock


\bibitem[Arcuri et~al\mbox{.}(2010)]{Arcuri2010}
\bibfield{author}{\bibinfo{person}{Andrea Arcuri},
  \bibinfo{person}{Muhammad~Zohaib Iqbal}, {and} \bibinfo{person}{Lionel~C.
  Briand}.} \bibinfo{year}{2010}\natexlab{}.
\newblock \showarticletitle{Black-box system testing of real-time embedded
  systems using random and search-based testing}. In
  \bibinfo{booktitle}{\emph{Proceedings of the IFIP International Conference on
  Testing Software and Systems (ICTSS'10)}}, Vol.~\bibinfo{volume}{6435}.
  \bibinfo{pages}{95--110}.
\newblock


\bibitem[Baruah et~al\mbox{.}(2011)]{Baruah2011}
\bibfield{author}{\bibinfo{person}{S.K. Baruah}, \bibinfo{person}{A. Burns},
  {and} \bibinfo{person}{R.I. Davis}.} \bibinfo{year}{2011}\natexlab{}.
\newblock \showarticletitle{{Response-Time Analysis for Mixed Criticality
  Systems}}. In \bibinfo{booktitle}{\emph{Proceedings of the IEEE 32nd
  Real-Time Systems Symposium (RTSS'11)}} (Vienna, Austria).
  \bibinfo{publisher}{{IEEE}}, \bibinfo{pages}{34--43}.
\newblock


\bibitem[Berezovskyi et~al\mbox{.}(2014)]{Berezovskyi2014}
\bibfield{author}{\bibinfo{person}{Kostiantyn Berezovskyi},
  \bibinfo{person}{Luca Santinelli}, \bibinfo{person}{Konstantinos Bletsas},
  {and} \bibinfo{person}{Eduardo Tovar}.} \bibinfo{year}{2014}\natexlab{}.
\newblock \showarticletitle{WCET Measurement-Based and Extreme Value Theory
  Characterisation of CUDA Kernels}. In \bibinfo{booktitle}{\emph{Proceedings
  of the 22nd International Conference on Real-Time Networks and Systems
  (RTNS'14)}} (Versaille, France). \bibinfo{publisher}{ACM},
  \bibinfo{address}{New York, NY, USA}, \bibinfo{pages}{279–288}.
\newblock


\bibitem[Bernat et~al\mbox{.}(2001)]{Bernat2001}
\bibfield{author}{\bibinfo{person}{Guillem Bernat}, \bibinfo{person}{Alan
  Burns}, {and} \bibinfo{person}{Albert Llamos{\'{\i}}}.}
  \bibinfo{year}{2001}\natexlab{}.
\newblock \showarticletitle{Weakly Hard Real-Time Systems}.
\newblock \bibinfo{journal}{\emph{IEEE Trans. Comput.}} \bibinfo{volume}{50},
  \bibinfo{number}{4} (\bibinfo{year}{2001}), \bibinfo{pages}{308--321}.
\newblock


\bibitem[Bernat et~al\mbox{.}(2002)]{Bernat2002}
\bibfield{author}{\bibinfo{person}{Guillem Bernat}, \bibinfo{person}{Antoine
  Colin}, {and} \bibinfo{person}{Stefan~M. Petters}.}
  \bibinfo{year}{2002}\natexlab{}.
\newblock \showarticletitle{{WCET} Analysis of Probabilistic Hard Real-Time
  System}. In \bibinfo{booktitle}{\emph{Proceedings of the 23rd {IEEE}
  Real-Time Systems Symposium (RTSS'02)}} (Austin, TX, USA).
  \bibinfo{publisher}{{IEEE}}, \bibinfo{pages}{279--288}.
\newblock


\bibitem[Bini and Buttazzo(2004)]{Bini2004}
\bibfield{author}{\bibinfo{person}{Enrico Bini} {and}
  \bibinfo{person}{Giorgio~C. Buttazzo}.} \bibinfo{year}{2004}\natexlab{}.
\newblock \showarticletitle{{Schedulability analysis of periodic fixed priority
  systems}}.
\newblock \bibinfo{journal}{\emph{IEEE Trans. Comput.}} \bibinfo{volume}{53},
  \bibinfo{number}{11} (\bibinfo{year}{2004}), \bibinfo{pages}{1462--1473}.
\newblock


\bibitem[Bini and Buttazzo(2005)]{bini2005measuring}
\bibfield{author}{\bibinfo{person}{Enrico Bini} {and}
  \bibinfo{person}{Giorgio~C Buttazzo}.} \bibinfo{year}{2005}\natexlab{}.
\newblock \showarticletitle{Measuring the Performance of Schedulability Tests}.
\newblock \bibinfo{journal}{\emph{Real-Time Systems}}  \bibinfo{volume}{30}
  (\bibinfo{year}{2005}), \bibinfo{pages}{129--154}.
\newblock


\bibitem[{BlackBerry QNX}(2022a)]{APS}
\bibfield{author}{\bibinfo{person}{{BlackBerry QNX}}.}
  \bibinfo{year}{2022}\natexlab{a}.
\newblock \bibinfo{title}{Adaptive Partitioning Scheduler ({APS})}.
\newblock
\newblock
\urldef\tempurl \url{https://www.qnx.com/developers/docs/7.1/#com.qnx.doc.neutrino.sys_arch/topic/adaptive.html}
\showURL{Retrieved October 6, 2022 from \tempurl}


\bibitem[{BlackBerry QNX}(2022b)]{QNXNeutrino}
\bibfield{author}{\bibinfo{person}{{BlackBerry QNX}}.}
  \bibinfo{year}{2022}\natexlab{b}.
\newblock \bibinfo{title}{{QNX} Neutrino 7.1}.
\newblock
\newblock
\urldef\tempurl \url{https://blackberry.qnx.com/en/products/foundation-software/qnx-rtos}
\showURL{Retrieved October 6, 2022 from \tempurl}


\bibitem[Bletsas and Andersson(2009)]{Bletsas2009}
\bibfield{author}{\bibinfo{person}{Konstantinos Bletsas} {and}
  \bibinfo{person}{Björn Andersson}.} \bibinfo{year}{2009}\natexlab{}.
\newblock \showarticletitle{Notional Processors: An Approach for Multiprocessor
  Scheduling}. In \bibinfo{booktitle}{\emph{Proceedings of the 15th IEEE
  Real-Time and Embedded Technology and Applications Symposium (RTAS'09)}} (San
  Francisco, CA, USA). \bibinfo{publisher}{{IEEE}}, \bibinfo{pages}{3--12}.
\newblock


\bibitem[Bonenfant et~al\mbox{.}(2017)]{Bonenfant2017}
\bibfield{author}{\bibinfo{person}{Armelle Bonenfant}, \bibinfo{person}{Denis
  Claraz}, \bibinfo{person}{Marianne~De Michiel}, {and} \bibinfo{person}{Pascal
  Sotin}.} \bibinfo{year}{2017}\natexlab{}.
\newblock \showarticletitle{Early {WCET} Prediction Using Machine Learning}. In
  \bibinfo{booktitle}{\emph{Proceedings of the 17th International Workshop on
  Worst-Case Execution Time Analysis (WCET'17)}}, Vol.~\bibinfo{volume}{57}.
  \bibinfo{publisher}{{Schloss Dagstuhl}}, \bibinfo{address}{Dagstuhl,
  Germany}, Article \bibinfo{articleno}{5}, \bibinfo{numpages}{9}~pages.
\newblock


\bibitem[Breiman(2001)]{Breiman2001}
\bibfield{author}{\bibinfo{person}{Leo Breiman}.}
  \bibinfo{year}{2001}\natexlab{}.
\newblock \showarticletitle{Random Forests}.
\newblock \bibinfo{journal}{\emph{Machine Learning}} \bibinfo{volume}{45},
  \bibinfo{number}{1} (\bibinfo{year}{2001}), \bibinfo{pages}{5--32}.
\newblock


\bibitem[Burns and Edgar(2000)]{Burns2000}
\bibfield{author}{\bibinfo{person}{A. Burns} {and} \bibinfo{person}{S. Edgar}.}
  \bibinfo{year}{2000}\natexlab{}.
\newblock \showarticletitle{Predicting computation time for advanced processor
  architectures}. In \bibinfo{booktitle}{\emph{Proceedings of the 12th
  Euromicro Conference on Real-Time Systems (ECRTS'00)}}.
  \bibinfo{pages}{89--96}.
\newblock


\bibitem[Byers et~al\mbox{.}(2015)]{ByersCD15}
\bibfield{author}{\bibinfo{person}{Chad~M. Byers}, \bibinfo{person}{Betty H.~C.
  Cheng}, {and} \bibinfo{person}{Kalyanmoy Deb}.}
  \bibinfo{year}{2015}\natexlab{}.
\newblock \showarticletitle{Unwanted Feature Interactions Between the Problem
  and Search Operators in Evolutionary Multi-objective Optimization}. In
  \bibinfo{booktitle}{\emph{Proceedings of the 8th International Conference on
  Evolutionary Multi-Criterion Optimization}}. \bibinfo{pages}{19--33}.
\newblock


\bibitem[Chen et~al\mbox{.}(2018)]{Chen2018a}
\bibfield{author}{\bibinfo{person}{Jian~Jia Chen}, \bibinfo{person}{Georg {Von
  Der Br{\"{u}}ggen}}, {and} \bibinfo{person}{Niklas Ueter}.}
  \bibinfo{year}{2018}\natexlab{}.
\newblock \showarticletitle{{Push forward: Global fixed-priority scheduling of
  arbitrary-deadline sporadic task systems}}.
\newblock \bibinfo{journal}{\emph{Leibniz International Proceedings in
  Informatics, LIPIcs}}  \bibinfo{volume}{106} (\bibinfo{year}{2018}),
  \bibinfo{pages}{1--24}.
\newblock


\bibitem[Cheng(2003)]{Cheng2003}
\bibfield{author}{\bibinfo{person}{Albert M.~K. Cheng}.}
  \bibinfo{year}{2003}\natexlab{}.
\newblock \bibinfo{booktitle}{\emph{{Real-Time Systems: Scheduling, Analysis,
  and Verification}}}.
\newblock \bibinfo{publisher}{Wiley}. 552 pages.
\newblock


\bibitem[Cucu-Grosjean et~al\mbox{.}(2012)]{Cucu2012}
\bibfield{author}{\bibinfo{person}{Liliana Cucu-Grosjean},
  \bibinfo{person}{Luca Santinelli}, \bibinfo{person}{Michael Houston},
  \bibinfo{person}{Code Lo}, \bibinfo{person}{Tullio Vardanega},
  \bibinfo{person}{Leonidas Kosmidis}, \bibinfo{person}{Jaume Abella},
  \bibinfo{person}{Enrico Mezzetti}, \bibinfo{person}{Eduardo Quiñones}, {and}
  \bibinfo{person}{Francisco~J. Cazorla}.} \bibinfo{year}{2012}\natexlab{}.
\newblock \showarticletitle{Measurement-Based Probabilistic Timing Analysis for
  Multi-path Programs}. In \bibinfo{booktitle}{\emph{Proceedings of the 24th
  Euromicro Conference on Real-Time Systems (ECRTS'12)}}.
  \bibinfo{pages}{91--101}.
\newblock


\bibitem[Dasari et~al\mbox{.}(2022)]{Dasari2022}
\bibfield{author}{\bibinfo{person}{Dakshina Dasari}, \bibinfo{person}{Matthias
  Becker}, \bibinfo{person}{Daniel Casini}, {and} \bibinfo{person}{Tobias
  Blas}.} \bibinfo{year}{2022}\natexlab{}.
\newblock \showarticletitle{{End-to-End Analysis of Event Chains under the QNX
  Adaptive Partitioning Scheduler}}. In \bibinfo{booktitle}{\emph{Proceedings
  of the IEEE 28th Real-Time and Embedded Technology and Applications Symposium
  (RTAS'22)}}. \bibinfo{publisher}{IEEE}, \bibinfo{pages}{214--227}.
\newblock


\bibitem[Dasari et~al\mbox{.}(2021)]{Dasari2021}
\bibfield{author}{\bibinfo{person}{Dakshina Dasari}, \bibinfo{person}{Arne
  Hamann}, \bibinfo{person}{Holger Broede}, \bibinfo{person}{Michael Pressler},
  {and} \bibinfo{person}{Dirk Ziegenbein}.} \bibinfo{year}{2021}\natexlab{}.
\newblock \showarticletitle{{Brief Industry Paper: Dissecting the QNX Adaptive
  Partitioning Scheduler}}. In \bibinfo{booktitle}{\emph{Proceedings of the
  IEEE 27th Real-Time and Embedded Technology and Applications Symposium
  (RTAS'21)}}. \bibinfo{publisher}{IEEE}, \bibinfo{pages}{477--480}.
\newblock


\bibitem[Davis and Cucu-Grosjean(2019)]{Davis2019}
\bibfield{author}{\bibinfo{person}{Robert Davis} {and} \bibinfo{person}{Liliana
  Cucu-Grosjean}.} \bibinfo{year}{2019}\natexlab{}.
\newblock \showarticletitle{{A Survey of Probabilistic Timing Analysis
  Techniques for Real-Time Systems}}.
\newblock \bibinfo{journal}{\emph{LITES: Leibniz Transactions on Embedded
  Systems}} (\bibinfo{year}{2019}), \bibinfo{pages}{1--60}.
\newblock


\bibitem[Davis and Burns(2011)]{Davis2011}
\bibfield{author}{\bibinfo{person}{Robert~I. Davis} {and} \bibinfo{person}{Alan
  Burns}.} \bibinfo{year}{2011}\natexlab{}.
\newblock \showarticletitle{Improved Priority Assignment for Global Fixed
  Priority Pre-Emptive Scheduling in Multiprocessor Real-Time Systems}.
\newblock \bibinfo{journal}{\emph{Real-Time Systems}} \bibinfo{volume}{47},
  \bibinfo{number}{1} (\bibinfo{date}{Jan} \bibinfo{year}{2011}),
  \bibinfo{pages}{1–40}.
\newblock


\bibitem[Davis et~al\mbox{.}(2008)]{Davis2008}
\bibfield{author}{\bibinfo{person}{Robert~I. Davis}, \bibinfo{person}{Attila
  Zabos}, {and} \bibinfo{person}{Alan Burns}.} \bibinfo{year}{2008}\natexlab{}.
\newblock \showarticletitle{{Efficient exact schedulability tests for fixed
  priority real-time systems}}.
\newblock \bibinfo{journal}{\emph{IEEE Trans. Comput.}} \bibinfo{volume}{57},
  \bibinfo{number}{9} (\bibinfo{year}{2008}), \bibinfo{pages}{1261--1276}.
\newblock


\bibitem[D{\"{u}}rr et~al\mbox{.}(2019)]{Durr2019}
\bibfield{author}{\bibinfo{person}{Marco D{\"{u}}rr}, \bibinfo{person}{Georg
  Von~Der Br{\"{u}}ggen}, \bibinfo{person}{Kuan~Hsun Chen}, {and}
  \bibinfo{person}{Jian-Jia Chen}.} \bibinfo{year}{2019}\natexlab{}.
\newblock \showarticletitle{{End-to-End Timing Analysis of Sporadic
  Cause-Effect Chains in Distributed Systems}}.
\newblock \bibinfo{journal}{\emph{ACM Transactions on Embedded Computing
  Systems}} \bibinfo{volume}{18}, \bibinfo{number}{5s} (\bibinfo{date}{Oct}
  \bibinfo{year}{2019}), \bibinfo{pages}{1--24}.
\newblock


\bibitem[Emberson et~al\mbox{.}(2010)]{Emberson2010}
\bibfield{author}{\bibinfo{person}{Paul Emberson}, \bibinfo{person}{Roger
  Stafford}, {and} \bibinfo{person}{Robert~I. Davis}.}
  \bibinfo{year}{2010}\natexlab{}.
\newblock \showarticletitle{Techniques for the synthesis of multiprocessor
  tasksets}. In \bibinfo{booktitle}{\emph{Proceedings of the 1st International
  Workshop on Analysis Tools and Methodologies for Embedded and Real-time
  Systems (WATERS'10)}}. \bibinfo{pages}{6--11}.
\newblock


\bibitem[Ferdinand and Wilhelm(1998)]{Ferdinand1998}
\bibfield{author}{\bibinfo{person}{Christian Ferdinand} {and}
  \bibinfo{person}{Reinhard Wilhelm}.} \bibinfo{year}{1998}\natexlab{}.
\newblock \showarticletitle{{On predicting data cache behavior for real-time
  systems}}.
\newblock \bibinfo{journal}{\emph{Lecture Notes in Computer Science (including
  subseries Lecture Notes in Artificial Intelligence and Lecture Notes in
  Bioinformatics)}}  \bibinfo{volume}{1474} (\bibinfo{year}{1998}),
  \bibinfo{pages}{16--30}.
\newblock


\bibitem[Gamma et~al\mbox{.}(1994)]{gamma1994design}
\bibfield{author}{\bibinfo{person}{Erich Gamma}, \bibinfo{person}{Richard
  Helm}, \bibinfo{person}{Ralph Johnson}, {and} \bibinfo{person}{John
  Vlissides}.} \bibinfo{year}{1994}\natexlab{}.
\newblock \bibinfo{booktitle}{\emph{Design Patterns: Elements of Reusable
  Object-Oriented Software}}.
\newblock \bibinfo{publisher}{Addison-Wesley Professional}.
\newblock


\bibitem[Gendreau and Potvin(2010)]{Gendreau2010}
\bibfield{author}{\bibinfo{person}{Michel Gendreau} {and}
  \bibinfo{person}{Jean-Yves Potvin}.} \bibinfo{year}{2010}\natexlab{}.
\newblock \bibinfo{booktitle}{\emph{Handbook of Metaheuristics}
  (\bibinfo{edition}{2nd} ed.)}.
\newblock \bibinfo{publisher}{Springer}.
\newblock


\bibitem[Gettings et~al\mbox{.}(2015)]{Gettings2015}
\bibfield{author}{\bibinfo{person}{Oliver Gettings}, \bibinfo{person}{Sophie
  Quinton}, {and} \bibinfo{person}{Robert~I. Davis}.}
  \bibinfo{year}{2015}\natexlab{}.
\newblock \showarticletitle{Mixed Criticality Systems with Weakly-Hard
  Constraints}. In \bibinfo{booktitle}{\emph{Proceedings of the 23rd
  International Conference on Real Time and Networks Systems (RTNS'15)}}.
  \bibinfo{publisher}{ACM}, \bibinfo{pages}{237–246}.
\newblock


\bibitem[Grass and Nguyen(2018)]{Grass2018}
\bibfield{author}{\bibinfo{person}{Werner Grass} {and} \bibinfo{person}{Thi
  Huyen~Chau Nguyen}.} \bibinfo{year}{2018}\natexlab{}.
\newblock \showarticletitle{Improved response-time bounds in fixed priority
  scheduling with arbitrary deadlines}.
\newblock \bibinfo{journal}{\emph{Real-Time Systems}} \bibinfo{volume}{54},
  \bibinfo{number}{1} (\bibinfo{year}{2018}), \bibinfo{pages}{1--30}.
\newblock


\bibitem[Gustafsson et~al\mbox{.}(2009)]{Gustafsson2009}
\bibfield{author}{\bibinfo{person}{Jan Gustafsson}, \bibinfo{person}{Peter
  Altenbernd}, \bibinfo{person}{Andreas Ermedahl}, {and}
  \bibinfo{person}{Bj{\"{o}}rn Lisper}.} \bibinfo{year}{2009}\natexlab{}.
\newblock \showarticletitle{Approximate Worst-Case Execution Time Analysis for
  Early Stage Embedded Systems Development}. In
  \bibinfo{booktitle}{\emph{Proceedings of the 7th {IFIP} {WG} 10.2
  International Workshop on Software Technologies for Embedded and Ubiquitous
  Systems (SEUS'09)}}. \bibinfo{publisher}{{Springer}},
  \bibinfo{address}{Berlin, Heidelberg}, \bibinfo{pages}{308--319}.
\newblock


\bibitem[Hansen et~al\mbox{.}(2009)]{Hansen2009}
\bibfield{author}{\bibinfo{person}{Jeffery~P. Hansen},
  \bibinfo{person}{Scott~A. Hissam}, {and} \bibinfo{person}{Gabriel~A.
  Moreno}.} \bibinfo{year}{2009}\natexlab{}.
\newblock \showarticletitle{Statistical-Based {WCET} Estimation and
  Validation}. In \bibinfo{booktitle}{\emph{Proceedings of the 9th
  International Workshop on Worst-Case Execution Time Analysis (WCET'09)}}.
  \bibinfo{publisher}{{Schloss Dagstuhl}}, \bibinfo{address}{Dagstuhl,
  Germany}, \bibinfo{pages}{1--11}.
\newblock


\bibitem[Hardy and Puaut(2011)]{Hardy2011}
\bibfield{author}{\bibinfo{person}{Damien Hardy} {and}
  \bibinfo{person}{Isabelle Puaut}.} \bibinfo{year}{2011}\natexlab{}.
\newblock \showarticletitle{{WCET analysis of instruction cache hierarchies}}.
\newblock \bibinfo{journal}{\emph{Journal of Systems Architecture}}
  \bibinfo{volume}{57}, \bibinfo{number}{7} (\bibinfo{date}{Aug}
  \bibinfo{year}{2011}), \bibinfo{pages}{677--694}.
\newblock


\bibitem[Harman et~al\mbox{.}(2012)]{Harman2012}
\bibfield{author}{\bibinfo{person}{Mark Harman}, \bibinfo{person}{S.~Afshin
  Mansouri}, {and} \bibinfo{person}{Yuanyuan Zhang}.}
  \bibinfo{year}{2012}\natexlab{}.
\newblock \showarticletitle{Search-based Software Engineering: Trends,
  Techniques and Applications}.
\newblock \bibinfo{journal}{\emph{{ACM} Computing Survey}}
  \bibinfo{volume}{45}, \bibinfo{number}{1}, Article \bibinfo{articleno}{11}
  (\bibinfo{year}{2012}), \bibinfo{numpages}{61}~pages.
\newblock


\bibitem[Hastie et~al\mbox{.}(2009)]{Hastie2009}
\bibfield{author}{\bibinfo{person}{Trevor Hastie}, \bibinfo{person}{Robert
  Tibshirani}, {and} \bibinfo{person}{Jerome~H. Friedman}.}
  \bibinfo{year}{2009}\natexlab{}.
\newblock \bibinfo{booktitle}{\emph{The Elements of Statistical Learning: Data
  Mining, Inference, and Prediction} (\bibinfo{edition}{2nd} ed.)}.
\newblock \bibinfo{publisher}{Springer}. 745 pages.
\newblock


\bibitem[Haupt and Haupt(1998)]{Haupt1988}
\bibfield{author}{\bibinfo{person}{Randy~L. Haupt} {and}
  \bibinfo{person}{Sue~Ellen Haupt}.} \bibinfo{year}{1998}\natexlab{}.
\newblock \bibinfo{booktitle}{\emph{Practical Genetic Algorithms}}.
\newblock \bibinfo{publisher}{John Wiley \& Sons, Inc.} 288 pages.
\newblock
\showISBNx{047-1188735}


\bibitem[Hideko and Hiroaki(2012)]{Hideko2012}
\bibfield{author}{\bibinfo{person}{Kawakubo Hideko} {and}
  \bibinfo{person}{Yoshida Hiroaki}.} \bibinfo{year}{2012}\natexlab{}.
\newblock \showarticletitle{Rapid Feature Selection Based on Random Forests for
  High-Dimensional Data}.
\newblock \bibinfo{journal}{\emph{Expert Systems with Applications}}
  \bibinfo{volume}{40}, \bibinfo{number}{1} (\bibinfo{year}{2012}),
  \bibinfo{pages}{6241--6252}.
\newblock


\bibitem[{International Organization for Standardization}(2018)]{ISO26262}
\bibfield{author}{\bibinfo{person}{{International Organization for
  Standardization}}.} \bibinfo{year}{2018}\natexlab{}.
\newblock \showarticletitle{{ISO} 26262: Road vehicles-functional safety}.
\newblock  (\bibinfo{year}{2018}).
\newblock


\bibitem[Jr. et~al\mbox{.}(2013)]{Hosmer2013}
\bibfield{author}{\bibinfo{person}{David W.~Hosmer Jr.},
  \bibinfo{person}{Stanley Lemeshow}, {and} \bibinfo{person}{Rodney~X.
  Sturdivant}.} \bibinfo{year}{2013}\natexlab{}.
\newblock \bibinfo{booktitle}{\emph{Applied Logistic Regression}
  (\bibinfo{edition}{3rd} ed.)}.
\newblock \bibinfo{publisher}{John Wiley \& Sons, Inc.} 528 pages.
\newblock


\bibitem[Khuri and Mukhopadhyay(2010)]{Khuri2010}
\bibfield{author}{\bibinfo{person}{Andr{\'e}~I Khuri} {and}
  \bibinfo{person}{Siuli Mukhopadhyay}.} \bibinfo{year}{2010}\natexlab{}.
\newblock \showarticletitle{Response surface methodology}.
\newblock \bibinfo{journal}{\emph{Wiley Interdisciplinary Reviews:
  Computational Statistics}} \bibinfo{volume}{2}, \bibinfo{number}{2}
  (\bibinfo{year}{2010}), \bibinfo{pages}{128--149}.
\newblock


\bibitem[Le~Moigne et~al\mbox{.}(2004)]{le2004generic}
\bibfield{author}{\bibinfo{person}{Rocco Le~Moigne}, \bibinfo{person}{Olivier
  Pasquier}, {and} \bibinfo{person}{J-P Calvez}.}
  \bibinfo{year}{2004}\natexlab{}.
\newblock \showarticletitle{A Generic {RTOS} Model for Real-Time Systems
  Simulation with {SystemC}}. In \bibinfo{booktitle}{\emph{Proceedings of the
  2004 Design, Automation and Test in Europe Conference and Exhibition}}.
  \bibinfo{pages}{82--87}.
\newblock


\bibitem[Lee et~al\mbox{.}(2022a)]{Lee2022b}
\bibfield{author}{\bibinfo{person}{Jaekwon Lee}, \bibinfo{person}{Seung~Yeob
  Shin}, \bibinfo{person}{Shiva Nejati}, {and} \bibinfo{person}{Lionel~C.
  Briand}.} \bibinfo{year}{2022}\natexlab{a}.
\newblock \showarticletitle{{Optimal Priority Assignment for Real-Time Systems:
  A Coevolution-Based Approach}}.
\newblock \bibinfo{journal}{\emph{Empirical Software Engineering}}
  \bibinfo{volume}{27}, \bibinfo{number}{6} (\bibinfo{year}{2022}),
  \bibinfo{pages}{142:1--49}.
\newblock


\bibitem[Lee et~al\mbox{.}(2023)]{Artifacts3}
\bibfield{author}{\bibinfo{person}{Jaekwon Lee}, \bibinfo{person}{Seung~Yeob
  Shin}, \bibinfo{person}{Shiva Nejati}, {and} \bibinfo{person}{Lionel~C.
  Briand}.} \bibinfo{year}{2023}\natexlab{}.
\newblock \bibinfo{title}{[Case study data] Probabilistic Safe WCET Estimation
  for Weakly Hard Real-Time Systems at Design Stages}.
\newblock \bibinfo{howpublished}{\url{https://github.com/SNTSVV/SWEAK}}.
\newblock


\bibitem[Lee et~al\mbox{.}(2022b)]{Lee2022a}
\bibfield{author}{\bibinfo{person}{Jaekwon Lee}, \bibinfo{person}{Seung~Yeob
  Shin}, \bibinfo{person}{Shiva Nejati}, \bibinfo{person}{Lionel~C. Briand},
  {and} \bibinfo{person}{Yago~Isasi Parache}.}
  \bibinfo{year}{2022}\natexlab{b}.
\newblock \showarticletitle{Estimating Probabilistic Safe WCET Ranges of
  Real-Time Systems at Design Stages}.
\newblock \bibinfo{journal}{\emph{ACM Transactions on Software Engineering and
  Methodology}} (\bibinfo{date}{Jun} \bibinfo{year}{2022}).
\newblock
\newblock
\shownote{Just Accepted}.


\bibitem[Lin et~al\mbox{.}(2009)]{Lin2009}
\bibfield{author}{\bibinfo{person}{Man Lin}, \bibinfo{person}{Li Xu},
  \bibinfo{person}{Laurence~T. Yang}, \bibinfo{person}{Xiao Qin},
  \bibinfo{person}{Nenggan Zheng}, \bibinfo{person}{Zhaohui Wu}, {and}
  \bibinfo{person}{Meikang Qiu}.} \bibinfo{year}{2009}\natexlab{}.
\newblock \showarticletitle{Static security optimization for real-time
  systems}.
\newblock \bibinfo{journal}{\emph{IEEE Transactions on Industrial Informatics}}
  \bibinfo{volume}{5}, \bibinfo{number}{1} (\bibinfo{year}{2009}),
  \bibinfo{pages}{22--37}.
\newblock


\bibitem[Liu and Layland(1973)]{Liu1973}
\bibfield{author}{\bibinfo{person}{Chang Liu} {and} \bibinfo{person}{James~W.
  Layland}.} \bibinfo{year}{1973}\natexlab{}.
\newblock \showarticletitle{Scheduling Algorithms for Multiprogramming in a
  Hard-Real-Time Environment}.
\newblock \bibinfo{journal}{\emph{Journal of the {ACM} (JACM)}}
  \bibinfo{volume}{20}, \bibinfo{number}{1} (\bibinfo{year}{1973}),
  \bibinfo{pages}{46--61}.
\newblock


\bibitem[Liu(2000)]{Liu2000}
\bibfield{author}{\bibinfo{person}{Jane W.~S. Liu}.}
  \bibinfo{year}{2000}\natexlab{}.
\newblock \bibinfo{booktitle}{\emph{Real-Time Systems} (\bibinfo{edition}{1st}
  ed.)}.
\newblock \bibinfo{publisher}{Prentice Hall PTR}.
\newblock


\bibitem[Luke(2013)]{Luke2013}
\bibfield{author}{\bibinfo{person}{Sean Luke}.}
  \bibinfo{year}{2013}\natexlab{}.
\newblock \bibinfo{booktitle}{\emph{Essentials of Metaheuristics}
  (\bibinfo{edition}{2nd} ed.)}.
\newblock \bibinfo{publisher}{Lulu}.
\newblock
\urldef\tempurl \url{http://cs.gmu.edu/~sean/book/metaheuristics/}
\showURL{\tempurl}


\bibitem[Mann and Whitney(1947)]{Mann1947}
\bibfield{author}{\bibinfo{person}{Henry~B. Mann} {and}
  \bibinfo{person}{Donald~R. Whitney}.} \bibinfo{year}{1947}\natexlab{}.
\newblock \showarticletitle{On a Test of Whether one of Two Random Variables is
  Stochastically Larger than the Other}.
\newblock \bibinfo{journal}{\emph{The Annals of Mathematical Statistics}}
  \bibinfo{volume}{18}, \bibinfo{number}{1} (\bibinfo{year}{1947}),
  \bibinfo{pages}{50--60}.
\newblock


\bibitem[Massa et~al\mbox{.}(2016)]{Massa2016}
\bibfield{author}{\bibinfo{person}{Ernesto Massa}, \bibinfo{person}{George
  Lima}, \bibinfo{person}{Paul Regnier}, \bibinfo{person}{Greg Levin}, {and}
  \bibinfo{person}{Scott Brandt}.} \bibinfo{year}{2016}\natexlab{}.
\newblock \showarticletitle{{Quasi-partitioned scheduling: optimality and
  adaptation in multiprocessor real-time systems}}.
\newblock \bibinfo{journal}{\emph{Real-Time Systems}} \bibinfo{volume}{52},
  \bibinfo{number}{5} (\bibinfo{year}{2016}), \bibinfo{pages}{566--597}.
\newblock


\bibitem[Moallemi and Wainer(2013)]{moallemi2013modeling}
\bibfield{author}{\bibinfo{person}{Mohammad Moallemi} {and}
  \bibinfo{person}{Gabriel Wainer}.} \bibinfo{year}{2013}\natexlab{}.
\newblock \showarticletitle{Modeling and Simulation-Driven Development of
  Embedded Real-Time Systems}.
\newblock \bibinfo{journal}{\emph{Simulation Modelling Practice and Theory}}
  \bibinfo{volume}{38} (\bibinfo{year}{2013}), \bibinfo{pages}{115--131}.
\newblock


\bibitem[Mueller(2000)]{Mueller2000}
\bibfield{author}{\bibinfo{person}{Frank Mueller}.}
  \bibinfo{year}{2000}\natexlab{}.
\newblock \showarticletitle{{Timing analysis for instruction caches}}.
\newblock \bibinfo{journal}{\emph{Real-Time Systems}} \bibinfo{volume}{18},
  \bibinfo{number}{2} (\bibinfo{year}{2000}), \bibinfo{pages}{217--247}.
\newblock


\bibitem[Nguyen et~al\mbox{.}(2015)]{Nguyen2015}
\bibfield{author}{\bibinfo{person}{Thanh-Tung Nguyen},
  \bibinfo{person}{Joshua~Zhexue Huang}, {and} \bibinfo{person}{Thuy~Thi
  Nguyen}.} \bibinfo{year}{2015}\natexlab{}.
\newblock \showarticletitle{Unbiased Feature Selection in Learning Random
  Forests for High-Dimensional Data}.
\newblock \bibinfo{journal}{\emph{The Scientific World Journal}}
  \bibinfo{volume}{2015} (\bibinfo{year}{2015}), \bibinfo{pages}{1--18}.
\newblock


\bibitem[Pazzaglia et~al\mbox{.}(2021)]{Pazzaglia2021}
\bibfield{author}{\bibinfo{person}{Paolo Pazzaglia}, \bibinfo{person}{Youcheng
  Sun}, {and} \bibinfo{person}{Marco~Di Natale}.}
  \bibinfo{year}{2021}\natexlab{}.
\newblock \showarticletitle{Generalized Weakly Hard Schedulability Analysis for
  Real-Time Periodic Tasks}.
\newblock \bibinfo{journal}{\emph{{ACM} Transactions on Embedded Computing
  Systems}} \bibinfo{volume}{20}, \bibinfo{number}{1}, Article
  \bibinfo{articleno}{3} (\bibinfo{year}{2021}), \bibinfo{numpages}{26}~pages.
\newblock


\bibitem[Ralph et~al\mbox{.}(2020)]{Ralph2020}
\bibfield{author}{\bibinfo{person}{Paul Ralph}, \bibinfo{person}{Nauman bin
  Ali}, \bibinfo{person}{Sebastian Baltes}, \bibinfo{person}{Domenico
  Bianculli}, \bibinfo{person}{Jessica Diaz}, \bibinfo{person}{Yvonne
  Dittrich}, \bibinfo{person}{Neil Ernst}, \bibinfo{person}{Michael Felderer},
  \bibinfo{person}{Robert Feldt}, \bibinfo{person}{Antonio Filieri},
  \bibinfo{person}{Breno Bernard~Nicolau de França},
  \bibinfo{person}{Carlo~Alberto Furia}, \bibinfo{person}{Greg Gay},
  \bibinfo{person}{Nicolas Gold}, \bibinfo{person}{Daniel Graziotin},
  \bibinfo{person}{Pinjia He}, \bibinfo{person}{Rashina Hoda},
  \bibinfo{person}{Natalia Juristo}, \bibinfo{person}{Barbara Kitchenham},
  \bibinfo{person}{Valentina Lenarduzzi}, \bibinfo{person}{Jorge Martínez},
  \bibinfo{person}{Jorge Melegati}, \bibinfo{person}{Daniel Mendez},
  \bibinfo{person}{Tim Menzies}, \bibinfo{person}{Jefferson Molleri},
  \bibinfo{person}{Dietmar Pfahl}, \bibinfo{person}{Romain Robbes},
  \bibinfo{person}{Daniel Russo}, \bibinfo{person}{Nyyti Saarimäki},
  \bibinfo{person}{Federica Sarro}, \bibinfo{person}{Davide Taibi},
  \bibinfo{person}{Janet Siegmund}, \bibinfo{person}{Diomidis Spinellis},
  \bibinfo{person}{Miroslaw Staron}, \bibinfo{person}{Klaas Stol},
  \bibinfo{person}{Margaret-Anne Storey}, \bibinfo{person}{Davide Taibi},
  \bibinfo{person}{Damian Tamburri}, \bibinfo{person}{Marco Torchiano},
  \bibinfo{person}{Christoph Treude}, \bibinfo{person}{Burak Turhan},
  \bibinfo{person}{Xiaofeng Wang}, {and} \bibinfo{person}{Sira Vegas}.}
  \bibinfo{year}{2020}\natexlab{}.
\newblock \bibinfo{title}{Empirical Standards for Software Engineering
  Research}.
\newblock
\newblock
\showeprint[arxiv]{2010.03525}~[cs.SE]


\bibitem[Russell and Norvig(2010)]{Russell2010}
\bibfield{author}{\bibinfo{person}{Stuart~J. Russell} {and}
  \bibinfo{person}{Peter Norvig}.} \bibinfo{year}{2010}\natexlab{}.
\newblock \bibinfo{booktitle}{\emph{Artificial Intelligence - {A} Modern
  Approach} (\bibinfo{edition}{3rd} ed.)}.
\newblock \bibinfo{publisher}{Pearson Education}. 1132 pages.
\newblock


\bibitem[Santinelli et~al\mbox{.}(2017)]{Santinelli2017}
\bibfield{author}{\bibinfo{person}{Luca Santinelli}, \bibinfo{person}{Fabrice
  Guet}, {and} \bibinfo{person}{Jerome Morio}.}
  \bibinfo{year}{2017}\natexlab{}.
\newblock \showarticletitle{Revising Measurement-Based Probabilistic Timing
  Analysis}. In \bibinfo{booktitle}{\emph{Proceedings of the 2017 IEEE
  Real-Time and Embedded Technology and Applications Symposium (RTAS'17)}}.
  \bibinfo{pages}{199--208}.
\newblock


\bibitem[Schaffer and Reid(2011)]{Schaffer2011}
\bibfield{author}{\bibinfo{person}{Jeff Schaffer} {and} \bibinfo{person}{Steve
  Reid}.} \bibinfo{year}{2011}\natexlab{}.
\newblock \showarticletitle{The joy of scheduling}.
\newblock \bibinfo{journal}{\emph{QNX Software Systems}}
  (\bibinfo{year}{2011}).
\newblock


\bibitem[Shin et~al\mbox{.}(2018)]{Shin2018}
\bibfield{author}{\bibinfo{person}{Seung~Yeob Shin}, \bibinfo{person}{Shiva
  Nejati}, \bibinfo{person}{Mehrdad Sabetzadeh}, \bibinfo{person}{Lionel~C.
  Briand}, {and} \bibinfo{person}{Frank Zimmer}.}
  \bibinfo{year}{2018}\natexlab{}.
\newblock \showarticletitle{Test Case Prioritization for Acceptance Testing of
  Cyber Physical Systems: {A} Multi-Objective Search-Based Approach}. In
  \bibinfo{booktitle}{\emph{Proceedings of the 27th ACM SIGSOFT International
  Symposium on Software Testing and Analysis (ISSTA'18)}}.
  \bibinfo{pages}{49--60}.
\newblock


\bibitem[Theiling et~al\mbox{.}(2000)]{Theiling2000}
\bibfield{author}{\bibinfo{person}{Henrik Theiling}, \bibinfo{person}{Christian
  Ferdinand}, {and} \bibinfo{person}{Reinhard Wilhelm}.}
  \bibinfo{year}{2000}\natexlab{}.
\newblock \showarticletitle{{Fast and Precise WCET Prediction by Separated
  Cache and Path Analyses}}.
\newblock \bibinfo{journal}{\emph{Real-Time Systems}} \bibinfo{volume}{18},
  \bibinfo{number}{2} (\bibinfo{year}{2000}), \bibinfo{pages}{157--179}.
\newblock


\bibitem[Vargha and Delaney(2000)]{Vargha2000}
\bibfield{author}{\bibinfo{person}{Andr\'as Vargha} {and}
  \bibinfo{person}{Harold~D. Delaney}.} \bibinfo{year}{2000}\natexlab{}.
\newblock \showarticletitle{A Critique and Improvement of the {CL} Common
  Language Effect Size Statistics of McGraw and Wong}.
\newblock \bibinfo{journal}{\emph{Journal of Educational and Behavioral
  Statistics}} \bibinfo{volume}{25}, \bibinfo{number}{2}
  (\bibinfo{year}{2000}), \bibinfo{pages}{101--132}.
\newblock


\bibitem[Varrette et~al\mbox{.}(2014)]{Varrette2014}
\bibfield{author}{\bibinfo{person}{S{\'{e}}bastien Varrette},
  \bibinfo{person}{Pascal Bouvry}, \bibinfo{person}{Hyacinthe Cartiaux}, {and}
  \bibinfo{person}{Fotis Georgatos}.} \bibinfo{year}{2014}\natexlab{}.
\newblock \showarticletitle{Management of an Academic {HPC} Cluster: The {UL}
  Experience}. In \bibinfo{booktitle}{\emph{Proceedings of the 2014
  International Conference on High Performance Computing \& Simulation
  (HPCS'14)}} (Bologna, Italy). \bibinfo{publisher}{{IEEE}},
  \bibinfo{pages}{959--967}.
\newblock


\bibitem[von~der Br{\"u}ggen et~al\mbox{.}(2018)]{Vonder2018}
\bibfield{author}{\bibinfo{person}{Georg von~der Br{\"u}ggen},
  \bibinfo{person}{Nico Piatkowski}, \bibinfo{person}{Kuan-Hsun Chen},
  \bibinfo{person}{Jian-Jia Chen}, {and} \bibinfo{person}{Katharina Morik}.}
  \bibinfo{year}{2018}\natexlab{}.
\newblock \showarticletitle{{Efficiently Approximating the Probability of
  Deadline Misses in Real-Time Systems}}. In
  \bibinfo{booktitle}{\emph{Proceedings of the 30th Euromicro Conference on
  Real-Time Systems (ECRTS'18)}}, Vol.~\bibinfo{volume}{106}.
  \bibinfo{publisher}{Schloss Dagstuhl}, \bibinfo{address}{Dagstuhl, Germany},
  Article \bibinfo{articleno}{6}, \bibinfo{numpages}{22}~pages.
\newblock


\bibitem[Wang(2017)]{Wang2017}
\bibfield{author}{\bibinfo{person}{K.~C. Wang}.}
  \bibinfo{year}{2017}\natexlab{}.
\newblock \bibinfo{booktitle}{\emph{Embedded and Real-Time Operating Systems}
  (\bibinfo{edition}{1st} ed.)}.
\newblock \bibinfo{publisher}{Springer}. 481 pages.
\newblock


\bibitem[Wegener and Grochtmann(1998)]{Wegener1998}
\bibfield{author}{\bibinfo{person}{Joachim Wegener} {and}
  \bibinfo{person}{Matthias Grochtmann}.} \bibinfo{year}{1998}\natexlab{}.
\newblock \showarticletitle{Verifying timing constraints of real-time systems
  by means of evolutionary testing}.
\newblock \bibinfo{journal}{\emph{Real-Time Systems}} \bibinfo{volume}{15},
  \bibinfo{number}{3} (\bibinfo{year}{1998}), \bibinfo{pages}{275--298}.
\newblock


\bibitem[Wegener et~al\mbox{.}(1997)]{Wegener1997}
\bibfield{author}{\bibinfo{person}{Joachim Wegener}, \bibinfo{person}{Harmen
  Sthamer}, \bibinfo{person}{Bryan~F. Jones}, {and} \bibinfo{person}{David~E.
  Eyres}.} \bibinfo{year}{1997}\natexlab{}.
\newblock \showarticletitle{Testing real-time systems using genetic
  algorithms}.
\newblock \bibinfo{journal}{\emph{Software Quality Journal}}
  \bibinfo{volume}{6}, \bibinfo{number}{2} (\bibinfo{year}{1997}),
  \bibinfo{pages}{127--135}.
\newblock


\bibitem[Wenzel et~al\mbox{.}(2005)]{Wenzel2005}
\bibfield{author}{\bibinfo{person}{I. Wenzel}, \bibinfo{person}{R. Kirner},
  \bibinfo{person}{B. Rieder}, {and} \bibinfo{person}{P. Puschner}.}
  \bibinfo{year}{2005}\natexlab{}.
\newblock \showarticletitle{Measurement-based worst-case execution time
  analysis}. In \bibinfo{booktitle}{\emph{Proceedings of the Third IEEE
  Workshop on Software Technologies for Future Embedded and Ubiquitous Systems
  (SEUS'05)}}. \bibinfo{pages}{7--10}.
\newblock


\bibitem[Witten et~al\mbox{.}(2011)]{Witten2011}
\bibfield{author}{\bibinfo{person}{Ian~H. Witten}, \bibinfo{person}{Eibe
  Frank}, {and} \bibinfo{person}{Mark~A. Hall}.}
  \bibinfo{year}{2011}\natexlab{}.
\newblock \bibinfo{booktitle}{\emph{Data Mining: Practical Machine Learning
  Tools and Techniques} (\bibinfo{edition}{3rd} ed.)}.
\newblock \bibinfo{publisher}{Morgan Kaufmann Publishers Inc.} 665 pages.
\newblock


\bibitem[Xu et~al\mbox{.}(2015)]{Xu2015}
\bibfield{author}{\bibinfo{person}{Wenbo Xu}, \bibinfo{person}{Zain
  Alabedin~Haj Hammadeh}, \bibinfo{person}{Alexander Kr{\"{o}}ller},
  \bibinfo{person}{Rolf Ernst}, {and} \bibinfo{person}{Sophie Quinton}.}
  \bibinfo{year}{2015}\natexlab{}.
\newblock \showarticletitle{Improved Deadline Miss Models for Real-Time Systems
  Using Typical Worst-Case Analysis}. In \bibinfo{booktitle}{\emph{Proceedings
  of the 27th Euromicro Conference on Real-Time Systems (ECRTS'15)}} (Lund,
  Sweden). \bibinfo{publisher}{{IEEE}}, \bibinfo{pages}{247--256}.
\newblock


\bibitem[Yamashita et~al\mbox{.}(2007)]{Yamashita2007}
\bibfield{author}{\bibinfo{person}{Toshie Yamashita}, \bibinfo{person}{Keizo
  Yamashita}, {and} \bibinfo{person}{Ryotaro Kamimura}.}
  \bibinfo{year}{2007}\natexlab{}.
\newblock \showarticletitle{A Stepwise AIC Method for Variable Selection in
  Linear Regression}.
\newblock \bibinfo{journal}{\emph{Communications in Statistics - Theory and
  Methods}} \bibinfo{volume}{36}, \bibinfo{number}{13} (\bibinfo{year}{2007}),
  \bibinfo{pages}{2395--2403}.
\newblock


\bibitem[Zhang and Burns(2009)]{Zhang2009}
\bibfield{author}{\bibinfo{person}{Fengxiang Zhang} {and} \bibinfo{person}{Alan
  Burns}.} \bibinfo{year}{2009}\natexlab{}.
\newblock \showarticletitle{Schedulability analysis for real-time systems with
  {EDF} scheduling}.
\newblock \bibinfo{journal}{\emph{IEEE Trans. Comput.}} \bibinfo{volume}{58},
  \bibinfo{number}{9} (\bibinfo{year}{2009}), \bibinfo{pages}{1250--1258}.
\newblock


\end{thebibliography}
\balance

\vfill

\end{document}